\documentclass[11pt]{article}
\usepackage{jheppub}
\usepackage{tikz}

\newcommand{\ri}{{\rm i}}

\def\th{\theta}

\newcommand{\hf}{\frac{1}{2}}
\newcommand{\qu}{\frac{1}{4}}
\def\til#1{\widetilde{#1}}

\def\b#1{\overline{#1}}
\def\del{\partial}

\def\lap{\Delta}

\def\h#1{\widehat{#1}}

\def\Ga{\Gamma}
\def\al{\alpha}

\newcommand{\Om}{\Omega}
\def\rt#1{\sqrt{#1}}
\newcommand{\Ztop}{\mathcal{Z}^{\text{top}}}

\newcommand{\nn}{\nonumber}
\renewcommand{\[}{\begin{eqnarray}}
\renewcommand{\]}{\end{eqnarray}}

\newcommand{\varth}{\vartheta}

\newcommand{\bbZ}{{\mathbb Z}}

\newcommand{\tZ}{{\widetilde Z}}
\newcommand{\tPhi}{{\widetilde\Phi}}

\newcommand{\Zefull}{{\cal Z}^{\text{full}}}
\newcommand{\Zofull}{{\widetilde{\cal Z}}^{\text{full}}}
\newcommand{\Zeven}{{\cal Z}}
\newcommand{\Zodd}{{\widetilde{\cal Z}}}
\newcommand{\Feven}{{\cal F}}
\newcommand{\Fodd}{{\widetilde{\cal F}}}
\newcommand{\Thetaeven}{\Theta}
\newcommand{\Thetaodd}{{\widetilde{\Theta}}}
\newcommand{\htop}{h^{\text{top}}}
\newcommand{\heven}{h}

\newcommand{\Atop}{A}
\newcommand{\Aeven}{{\cal A}}
\newcommand{\Aodd}{{\tilde{\cal A}}}
\newcommand{\btop}{b}
\newcommand{\beven}{\mathfrak{b}}
\newcommand{\bodd}{{\tilde{\beven}}}

\newcommand{\ftop}{f}
\newcommand{\feven}{\mathfrak{f}}
\newcommand{\fodd}{{\tilde{\feven}}}
\newcommand{\mutop}{\mu}
\newcommand{\mueven}{\mathfrak{m}}
\newcommand{\muodd}{\tilde{\mueven}}

\newcommand{\DTeven}{D_\Thetaeven}
\newcommand{\DTodd}{D_\Thetaodd}

\newcommand{\hh}[1]{\hspace{.2em}\widehat{\hspace{-.2em}#1}}

\begin{document}

\title{Resurgence analysis of 2d Yang-Mills theory on a torus}

\author[a]{Kazumi Okuyama}
\author[b]{and Kazuhiro Sakai}

\affiliation[a]{Department of Physics, 
Shinshu University, Matsumoto 390-8621, Japan}
\affiliation[b]{Institute of Physics, Meiji Gakuin University, Yokohama 244-8539, Japan}

\emailAdd{kazumi@azusa.shinshu-u.ac.jp, kzhrsakai@gmail.com}

\abstract{We study the large $N$ 't Hooft expansion of the
partition function of 2d $U(N)$ Yang-Mills theory on a torus.
We compute the  $1/N$  genus expansion of both the chiral and the full
partition function of 2d Yang-Mills using the recursion relation found
by Kaneko and Zagier with a slight modification.
Then we 
study the large order behavior of this genus expansion,
from which we extract the non-perturbative correction
using the resurgence relation.
It turns out that the genus expansion is not Borel summable and the coefficient
of 1-instanton correction, the so-called Stokes parameter, is pure imaginary. 
We find that the non-perturbative correction obtained from the resurgence
is reproduced from a certain analytic continuation of
the grand partition function of a system of non-relativistic fermions on a circle.
Our analytic continuation is different from that considered in hep-th/0504221.  
}

\maketitle

\section{Introduction \label{sec:intro}}

The holographic large $N$ duality between certain Yang-Mills theory and
string theory provides us with an important clue for understanding the still mysterious
quantum  gravity and the behavior of spacetime at the Planck scale.
In particular, in a certain situation some quantities on the Yang-Mills side can be 
computed exactly at finite $N$ and we expect that one can extract some interesting
quantum gravity effects
on the string theory side from the analysis of Yang-Mills side. 

As discussed in \cite{Vafa:2004qa}, this expectation is realized in a concrete example
of the equality between
the partition function $Z_N$ of 2d $U(N)$ Yang-Mills theory on a torus and that of four dimensional
BPS black holes. The black holes in question appear as bound states of D-branes
wrapping some cycles in a certain local Calabi-Yau threefold.
The partition function of the field theory on the branes
reduces to that of the 2d Yang-Mills theory due to the supersymmetric localization 
\cite{Vafa:2004qa}, which in turn is related to
the norm-squared $|\psi^{\text{top}}|^2$ of the topological string partition function 
$\psi^{\text{top}}$ according to the OSV conjecture \cite{Ooguri:2004zv}. 
It is argued in \cite{Vafa:2004qa} that 
the factorized structure $Z_N=|\psi^{\text{top}}|^2$ of the OSV relation has a natural
interpretation as the chiral factorization of
2d Yang-Mills  studied 
by Gross and Taylor \cite{Gross:1992tu,Gross:1993hu,Gross:1993yt}\footnote{See e.g. \cite{Moore:1994dk,Cordes:1994fc} for a review of 2d Yang-Mills theory and its large $N$ limit.}.
This factorized structure is also consistent with 
the existence of two boundaries of $AdS_2$ spacetime
in the near horizon $AdS_2\times S^2$ geometry of BPS black hole. 
Also, this factorized structure
naturally arises in the
free fermion representation of the partition function 
of 2d Yang-Mills \cite{Minahan:1993np}, where the two chiral factors
correspond to the positive and negative Fermi levels. 

It is further argued \cite{Vafa:2004qa} that this factorization is 
valid only in the perturbative $1/N$ expansion and if we include the non-perturbative
$\mathcal{O}(e^{-N})$ effects the exact factorization no longer holds.
In the free fermion picture, this corresponds to the entanglement of 
two Fermi levels at finite $N$.
In \cite{Dijkgraaf:2005bp} an interesting spacetime picture for this failure of factorization
was put forward:
the non-perturbative
$\mathcal{O}(e^{-N})$ corrections come from multi-center black holes
and the 2d Yang-Mills theory is actually dual to a coherent ensemble of black holes. This 
in particular implies that
the partition function of 2d Yang-Mills includes the effect of creation of baby universes
on the dual gravity side.

In this paper, we will revisit this problem from the viewpoint of resurgence. 
According to the theory of resurgence, 
non-perturbative corrections are encoded in the 
large order behavior of the perturbative series and one can 
``decode'' the non-perturbative effects from the information 
of perturbative computation alone (see e.g. \cite{Aniceto:2018bis,Dunne:2015eaa,Marino:2012zq} 
for review of resurgence).
For this purpose, we will compute the genus $g$ free energy $F_g(t)$ 
of 2d Yang-Mills theory on $T^2$ in the large $N$ limit with fixed 't Hooft coupling $t$. 
We will consider the $1/N$ expansion of both the chiral part and the full partition function 
of 2d Yang-Mills theory on $T^2$.

The chiral part of free energy $F_g(t)$ is identified as the genus $g$
topological string free energy counting the holomorphic maps from genus $g$ Riemann surface
to $T^2$, and it has interesting mathematical
properties.
In particular, as shown in  \cite{zagier,dijk},
$F_g(t)$ is a quasi-modular form of weight $6g-6$
given by a combination of Eisenstein series.
After the first computation of genus-one free energy in \cite{Gross:1992tu},
$F_g(t)$ has been computed up to $g=2$ by Douglas in 1993 \cite{Douglas:1993wy}
and up to $g=8$ by Rudd in 1994 \cite{Rudd:1994ta}.

In this paper, we have computed $F_g(t)$ up to $g=60$ using 
the recursion relation found by Kaneko and Zagier \cite{zagier}
with a slight modification.
It turns out that the $1/N$ expansion of 2d Yang-Mills on $T^2$ is not Borel summable
and there is a pole on the positive real axis on the Borel plane when $t>0$.
From the large genus behavior of $F_g(t)$ we find that the non-perturbative correction scales
as $e^{-A(t)/g_s}$ where $g_s$ denotes the topological string coupling and the ``instanton action''
$A(t)$ is given by $A(t)=t^2/2$.\footnote{See 
also \cite{Matsuo:2004nn,deMelloKoch:2005rq,Dhar:2006ru,Lelli:2002gr}
for the study of nonperturbative $\mathcal{O}(e^{-N})$
effects in 2d Yang-Mills theory.}
We also find that after including the fluctuation around
the 1-instanton $e^{-A(t)/g_s}\sum_n f_n(t)g_s^n$, 
it is proportional to $\psi^{\text{top}}(t+g_s)$, i.e.,
the 1-instanton correction is given by the topological string partition
function $\psi^{\text{top}}(t)$ with a shift of 't Hooft coupling $t\to t+g_s$.\footnote{Note 
that such a shift of $t$ naturally appears as
an effect of D-brane insertion \cite{Aganagic:2003qj,Aganagic:2011mi}.}  
Moreover, it turns out that 
the overall coefficient of 1-instanton, the so-called Stokes
parameter,
is pure imaginary and this imaginary contribution is
exactly canceled by the imaginary part of Borel resummation
coming from the contour deformation to avoid the pole on the positive real axis
of the Borel plane.
Interestingly, we find that the 1-instanton correction obtained from
this resurgence analysis is reproduced from 
a certain analytic continuation of
the grand partition function of fermions. 

We also study the genus expansion of the full partition function $Z_N$
when the topological $\th$-angle of 2d Yang-Mills is zero.
We derive a set of recursion relations
that determine the $\mathcal{O}(g_s^{2n})$ term in the
genus expansion and elucidate its modular properties.
We then obtain the 1-instanton correction of full
partition function from the large order behavior
of genus expansion, which we have computed up to $n=60$.
This is again reproduced from
our prescription of the analytic continuation. 
We find that there appear two types of partition functions
in the instanton expansion of $Z_N$ at $\th=0$, which we denote
as $\Zefull(t)$ and $\Zofull(t)$.
It turns out that $\Zefull(t)$ is the perturbative part of the
$1/N$ expansion of $Z_N$, while $\Zofull(t)$
corresponds to the perturbative part of another partition function,
$\tZ_N$. The difference between $Z_N$ and $\tZ_N$
is the boundary condition of $N$
free fermions on a circle: these fermions obey periodic boundary condition
in  $Z_N$, while
in $\tZ_N$ they obey anti-periodic boundary condition.
In the large $N$ expansion of $Z_N$,
on top of the perturbative part
$\Zefull(t)$,  we find that $Z_N$ receives a 1-instanton correction
proportional to
$\Zofull(t+g_s/2)$. On the other hand, in the large $N$ expansion of $\tZ_N$
this relation is reversed:
$\Zofull(t)$ is the perturbative part and $\Zefull(t+g_s/2)$ appears as a 1-instanton
correction.

In \cite{Dijkgraaf:2005bp} a similar analytic continuation of
the grand partition function of fermions was considered 
in order to rewrite the partition function in the form of
a sum of binary branching trees, which was interpreted as the creation of baby universes.
Our analytic continuation is different from that in \cite{Dijkgraaf:2005bp}.
In particular, the pure imaginary Stokes parameter naturally
arises in our prescription 
and this imaginary contribution is necessary
for the cancellation of the non-perturbative ambiguity of Borel resummation.
On the other hand, there is no such imaginary contribution in the 
analytic continuation considered in \cite{Dijkgraaf:2005bp}.
We should stress that our prescription 
of analytic continuation is strongly
supported by the explicit computation of the genus expansion
up to very high genera and 
the resurgence analysis of the large genus behavior.

This paper is organized as follows.
In section~\ref{sec:gen}, 
we first review the fact that the
partition function of 2d $U(N)$ Yang-Mills on $T^2$
is identified as a system of $N$ non-relativistic fermions
on a circle. Then we argue 
that the non-perturbative corrections
to the large $N$ expansion of 
the partition function can be systematically obtained 
by a certain analytic continuation of the grand partition 
function of non-relativistic fermions.
Along the way, we propose a non-perturbative completion
of $\psi^{\text{top}}$. 
In section~\ref{sec:genus},
we compute the genus expansion of both the chiral partition function
$\psi^{\text{top}}$ and the full partition function
$Z_N$ when the $\th$-angle is zero.
We find that  
the recursion relation of Kaneko and Zagier
can be slightly modified so that the modular properties
of $F_g(t)$ become more transparent.
We also write down the recursion relations
for the genus expansion of full partition functions
$\Zefull$ and $\Zofull$.
In section~\ref{sec:large}, we study the large order behavior 
of genus expansion numerically, and we extract the
1-instanton correction from this large genus behavior.
We find that the 1-instanton correction obtained in this way 
is consistent with our prescription of analytic continuation
considered in section~\ref{sec:gen}.
In section~\ref{sec:borel},
we consider the Borel-Pad\'{e} resummation
of the genus expansion.
It turns out that 
the genus expansion is not Borel summable and
the imaginary part of lateral Borel resummation is precisely
canceled by the imaginary contribution 
coming from the 1-instanton correction.
In section~\ref{sec:theta}, we briefly comment on the case of
non-zero $\th$-angle. We show that when $\th=\pi$
the full partition function is equal to 
the chiral partition function up to a rescaling of the coupling.
We conclude in section~\ref{sec:discussion}
with some discussion for future directions.
In appendix~\ref{app:notation} we summarize our convention of Jacobi theta functions.
In appendix~\ref{app:Zfullproof} we present a proof of some nontrivial
identities used in the main text.

\section{Generating function of partition function\label{sec:gen}}

\subsection{Partition function of Yang-Mills on $T^2$}

Let us first review the partition function of
2d Yang-Mills on a torus and its connection to topological string.
As explained in \cite{Vafa:2004qa},
the worldvolume theory on $N$ D4-branes in the Type IIA theory on a local Calabi-Yau
threefold $X$
\begin{align}
X:~ \mathcal{O}(-m)\oplus \mathcal{O}(m)\to T^2
\label{eq:CY}
\end{align}
reduces to the 2d $U(N)$ Yang-Mills on $T^2$ thanks to the supersymmetric localization.
The $N$ D4-branes in question are wrapping around the total space
of $\mathcal{O}(-m)\to T^2$ with $m$ being a positive integer. 
The D4-branes with gauge fluxes threading the worldvolume  can be thought of as a bound state
of D4, D2, and D0-branes, which in turn can be seen as 
a black hole in the 4-dimensional spacetime
after a compactification of Type IIA theory on the 6-dimensional space
$X$ in \eqref{eq:CY}.
Then the partition function $Z_N$ of $U(N)$ Yang-Mills on $T^2$
is identified as the partition function $Z_{\text{BH}}$
of black hole microstates, which is further 
related to the partition function $\psi^{\text{top}}$
of topological string on $X$
via the OSV conjecture \cite{Ooguri:2004zv}
\begin{align}
 Z_N=Z_{\text{BH}}=|\psi^{\text{top}}|^2.
\label{eq:OSV}
\end{align} 
The topological string coupling $g_s$
and the 2d Yang-Mills coupling $g_{\text{YM}}$ are related by
\begin{align}
 g_s=mg_{\text{YM}}^2A
\end{align} 
where $A$ is the area of the torus.

It is well-known that the 2d Yang-Mills partition function is given by
a sum over $U(N)$ representations $R$ \cite{Migdal:1975zg,Rusakov:1990rs}
\begin{align}
 Z_N=\sum_R q^{\hf C_2(R)}e^{\ri \th C_1(R)}
\label{eq:sumR}
\end{align}
where $C_1(R)$ and $C_2(R)$ denote the first and second Casimir of $R$, respectively, 
and $q$ is 
given by
\begin{align}
 q:=e^{-g_s}.
\end{align}
The partition function \eqref{eq:sumR} has a nice interpretation as a system of $N$
non-relativistic free fermions on a circle \cite{Minahan:1993np}.
The Casimirs $C_1(R)$ and $C_2(R)$ correspond to the total momentum and total
energy of $N$ fermions, respectively.
A single fermion with momentum $p$ has an 
energy
$E=\frac{1}{2}p^2$, 
and the momentum $p$ is quantized by the condition
\begin{align}
 e^{2\pi\ri p}=(-1)^{N-1}.
\label{eq:pcond}
\end{align}
This quantization condition of $p$
has a simple physical interpretation \cite{Minahan:1993np}: when a fermion is transported once around the circle
it passes through other $N-1$ fermions and picks up $N-1$ minus signs. 
This condition \eqref{eq:pcond}
implies that $p$ is half-integer for even $N$ and integer for odd $N$.
This free fermion picture allows us to write down the partition function
as
\begin{align}
\begin{aligned}
 Z_N&=\oint\frac{dx}{2\pi\ri x^{N+1}}
\prod_{p\in\mathbb{Z}+\frac{N-1}{2} }(1+xe^{\ri p\th}q^{\hf p^2}).
\end{aligned}
\label{eq:ZN-mode}
\end{align}
In this paper we will assume $N$ is even
for simplicity.
When $N$ is even, $p$ runs over the half-integers and 
\eqref{eq:ZN-mode} is rewritten as
\begin{equation}
\begin{aligned}
 Z_N&=\oint\frac{dx}{2\pi\ri x^{N+1}}\exp\left[\sum_{\ell=1}^\infty
\frac{(-1)^{\ell-1}x^\ell}{\ell}\vartheta_2(e^{\ri\ell\th},q^\ell)\right] ,
\end{aligned} 
\label{eq:ZN-int-th}
\end{equation}
where $\vartheta_2$ denotes the Jacobi theta function (see appendix \ref{app:notation} for
our definition of theta functions).
For instance, the partition function of $U(2)$ Yang-Mills is given by
\begin{align}
 Z_2=\hf \vartheta_2(e^{\ri\th},q)^2-\hf \vartheta_2(e^{2\ri\th},q^2).
\label{eq:Z2}
\end{align} 

We are interested in the 
behavior of $Z_N$ in 
the large $N$ 't Hooft limit
\begin{align}
 N\to\infty,~g_s\to 0,~~ \text{with}~~ t=\hf Ng_s-\ri\th~~\text{fixed}.
\label{eq:thooft}
\end{align}
Then the OSV relation \eqref{eq:OSV}
is expected to hold  at least perturbatively in $1/N$ expansion 
under the identification of $t$ as the K\"{a}hler parameter of the base $T^2$
of $X$. The topological string
free energy $F=\log\psi^{\text{top}}$ has a genus expansion
in the small $g_s$ limit
\begin{align}
 F=\sum_{g=0}^\infty g_s^{2g-2}F_g(t)
\label{eq:free-g}
\end{align}
and the first two terms are given by
\begin{align}
 F_0(t)=-\frac{t^3}{6},\quad F_1(t)=-\log\eta(Q).
\label{eq:free-01}
\end{align}
Here $\eta(Q):=Q^{\frac{1}{24}}\prod_{n=1}^\infty (1-Q^n)$ denotes the Dedekind eta-function
and $Q$ is defined by
\begin{align}
 Q:=e^{-t}.
\end{align}
In the relation  $Z_N=\psi^{\text{top}}\b{\psi}^{\text{top}}$ \eqref{eq:OSV},
the anti-topological partition function $\b{\psi}^{\text{top}}$ is obtained from 
$\psi^{\text{top}}$ by reversing the sign of $\th$ in
\eqref{eq:thooft}: 
\begin{align}
 \b{\psi}^{\text{top}}(t)=\psi^{\text{top}}(\b{t}),\qquad
\b{t}=\hf Ng_s+\ri\th.
\end{align}
In this paper we will be mostly focusing on the $\th=0$ case,
in which the 't Hooft coupling $t$ reduces to
\begin{align}
 t=\hf  Ng_s.
\label{eq:thooft-t}
\end{align} 
We will briefly comment on the non-zero $\th$ case in section \ref{sec:theta}.

The partition function $\psi^{\text{top}}(t)$ of topological
string also 
has a simple expression in the free fermion picture.
It is
given by a formal power series \cite{Douglas:1993wy} 
\begin{align}
\psi^{\text{top}}(t)
 &= e^{F^{\text{cl}}(t)}
 \oint\frac{dx}{2\pi\ri x}
 \prod_{p>0}\left(1+xQ^p q^{\hf p^2}\right)
            \left(1+x^{-1}Q^p q^{-\hf p^2}\right),
\label{eq:Ztop-fermi}
\end{align}
where  $p$ runs over positive half-integers
and $F^{\text{cl}}(t)$ is a polynomial of $t$ 
\begin{align}
 F^{\text{cl}}(t)=\frac{1}{g_s^2}F_0(t)+\frac{t}{24}=-\frac{t^3}{6g_s^2}+\frac{t}{24}.
\label{eq:Fcl}
\end{align}
This classical part of free energy comes from 
the ground state of $N$ fermions where
the momentum modes between $p=-\frac{N-1}{2}$ and $p=\frac{N-1}{2}$ are occupied
\cite{Vafa:2004qa}
\begin{align}
 \begin{aligned}
  F^{\text{cl}}(t)+F^{\text{cl}}(\b{t})&=-g_sE_0,\qquad
E_0=\hf \sum_{p=-\frac{N-1}{2}}^{\frac{N-1}{2}}p^2
=\frac{N^3-N}{24}.
 \end{aligned}
\label{eq:Egnd}
\end{align} 
From the expression \eqref{eq:Ztop-fermi}, we can easily find the small $Q$ expansion of 
$\psi^{\text{top}}(t)$
\begin{align}
\begin{aligned}
\psi^{\text{top}}(t)
 &=e^{F^{\text{cl}}(t)}\Big[1+Q+(q+q^{-1})Q^2+(1+q^3+q^{-3})Q^3+\cdots\Big],
\end{aligned}
\label{eq:Ztop-Q}
\end{align}
from which one can extract the Gromov-Witten 
and Gopakumar-Vafa invariants of $X$.

As observed in \cite{Vafa:2004qa},
the OSV relation $Z_N=|\psi^{\text{top}}|^2$
has a natural interpretation 
as the chiral factorization of the 2d Yang-Mills theory
studied by Gross and Taylor \cite{Gross:1992tu,Gross:1993hu}.
This norm-squared form $Z_N=|\psi^{\text{top}}|^2$ is in accord with
the interpretation of the topological string partition function as
a wavefunction \cite{Witten:1993ed,Ooguri:2005vr}.
Moreover, 
this is consistent with the black hole picture \cite{Vafa:2004qa}:
the near horizon geometry of 4d charged black hole is $AdS_2\times S^2$,
and the two boundaries of Lorentzian $AdS_2$ naturally correspond to the two factors
$\psi^{\text{top}}$ and $\b{\psi}^{\text{top}}$.

However, this relation \eqref{eq:OSV} is only schematic;
we have to sum over the $U(1)$ charge of representation $R$
\begin{align}
 Z_N=\sum_{l\in\mathbb{Z}}\psi^{\text{top}}(t+g_sl)
\bar{\psi}^{\text{top}}(t-g_sl)
\label{eq:RRsum}
\end{align}
which corresponds to the sum over RR fluxes on the topological string side
\cite{Vafa:2004qa}.
In  \cite{Dijkgraaf:2005bp} it is further
argued that this is not the end of the story: 
the chiral factorization is valid only approximately 
and if we include the non-perturbative $\mathcal{O}(e^{-N})$ effects the expansion \eqref{eq:RRsum}
is modified to
\begin{align}
 Z_N=\sum_{n=1}^\infty (-1)^{n-1}C_{n-1}
\sum_{\sum_{i=1}^n N_+^i+N_-^i=N}\prod_{i=1}^n\psi^{\text{top}}_{N_+^i}
\b{\psi}^{\text{top}}_{N_-^i},
\label{eq:baby-expansion}
\end{align}
where 
$C_{n}$ denotes the Catalan number
\begin{align}
 C_n=\frac{(2n)!}{n!(n+1)!},
\end{align}
and $\psi_{N_+}^{\text{top}}$ in \eqref{eq:baby-expansion}
is equal to the topological string partition function $\psi^{\text{top}}(t)$
with the identification $t=N_+g_s$
\begin{align}
 \psi_{N_+}^{\text{top}}=\psi^{\text{top}}(t=N_+g_s).
\label{eq:psitop-N}
\end{align}
This expansion \eqref{eq:baby-expansion}
is interpreted in \cite{Dijkgraaf:2005bp}
as the creation of baby universes and the Catalan number counts the
number of ways that the baby universes are created. 
This seems to be also consistent with the black hole
picture that there is a quantum tunneling
from single-center to multi-center black holes \cite{Brill:1991rw}
due to a peculiar nature of $AdS_2$ spacetime \cite{Maldacena:1998uz,Sen:2011cn}.

However, it is not obvious in what sense the expansion \eqref{eq:baby-expansion} holds.
The Yang-Mills partition function $Z_N$ on the left hand side (LHS) of \eqref{eq:baby-expansion}
is non-perturbatively well-defined while the topological string partition function
$\psi^{\text{top}}$ on the right hand side (RHS) of \eqref{eq:baby-expansion}
is only defined perturbatively,
and the non-perturbative
completion of $\psi^{\text{top}}$ still remains as a problem.

In this paper we will propose a non-perturbative
completion of $\psi^{\text{top}}$
which makes sense at finite $N$. We will also show that our non-perturbative
definition of $\psi^{\text{top}}$ is consistent with the 
large genus behavior of free energy $F_g(t)$ and the resurgence analysis.

\subsection{Non-perturbative completion of $\psi^{\text{top}}$}

The expression of $\psi^{\text{top}}(t)$ in \eqref{eq:Ztop-fermi}
is not  non-perturbatively complete {\it per se}, since it involves
the power series in both $q$ and $q^{-1}$
and hence the infinite product
in \eqref{eq:Ztop-fermi} is not convergent.
Here we would like to propose a simple candidate
of the non-perturbative completion of $\psi^{\text{top}}(t)$.

We start with the free fermion description of
the partition function \eqref{eq:ZN-int-th}
\begin{align}
 Z_N=\oint\frac{dx}{2\pi\ri x^{N+1}}
\prod_{p\in\mathbb{Z}+\hf}(1+xq^{\hf p^2}),
\label{eq:ZN-int}
\end{align}
where we have set $\theta=0$ for simplicity.
The integrand of \eqref{eq:ZN-int}
can be thought of as a grand partition function of fermions
\begin{align}
 Z(x,g_s):=\prod_{p\in\mathbb{Z}+\frac{1}{2}}(1+xq^{\hf p^2})=\sum_{N=0}^\infty Z_Nx^N.
\label{eq:grandZ}
\end{align} 
One can naturally decompose this grand partition function
into two parts according to the sign of momentum $p$
\begin{align}
\begin{aligned}
 \prod_{p>0}(1+xq^{\hf p^2})&=:\sum_{N_+=0}^\infty \psi_{N_+}x^{N_+},\\
\prod_{p<0}(1+xq^{\hf p^2})&=:\sum_{N_-=0}^\infty \b{\psi}_{N_-}x^{N_-}.
\end{aligned}
\label{eq:psi-gen-func}
\end{align}
In other words, $\psi_{N_+}$ is the canonical partition function of $N_+$ fermions with
positive momentum, while 
$\b{\psi}_{N_-}$ is the canonical partition function of $N_-$ fermions with
negative momentum.
When $\th=0$, $\psi_{k}$ and $\b{\psi}_{k}$ are actually equal: $\psi_{k}=\b{\psi}_{k}$.
If $\th\ne0$ they are related by the sign flip of $\th$
\begin{align}
 \b{\psi}_{k}(\th)=\psi_{k}(-\th).
\end{align}
From the obvious relation
\begin{align}
 \prod_{p\in\mathbb{Z}+\frac{1}{2}}(1+xq^{\hf p^2})=\prod_{p>0}(1+xq^{\hf p^2})
\prod_{p<0}(1+xq^{\hf p^2})
\end{align}
it follows that the full partition function $Z_N$
is decomposed as
\begin{align}
 Z_N=\sum_{N_++N_-=N}\psi_{N_+}\b{\psi}_{N_-}=\sum_{k=0}^N \psi_{k}\b{\psi}_{N-k}.
\label{eq:ZN-decomp}
\end{align}
We propose that $\psi_{N_+}$ in \eqref{eq:psi-gen-func}
gives a natural non-perturbative completion
of the topological string partition function 
$\psi^{\text{top}}_{N_+}$ in \eqref{eq:psitop-N},
in the sense that
$\psi_{N_+}$ is equal to 
 $\psi^{\text{top}}_{N_+}$ in the asymptotic $1/N_{+}$ 
expansion up to exponentially small corrections
\begin{align}
 \psi_{N_+}=\psi^{\text{top}}_{N_+}+\mathcal{O}(e^{-N_+}).
\label{eq:psiapprox}
\end{align}
We should stress that our definition of $\psi_{N_+}$ is well-defined at finite $N_+$
\begin{align}
\begin{aligned}
 \psi_{N_+}&=\oint\frac{dx}{2\pi\ri x^{N_++1}}\prod_{p>0}(1+xq^{\hf p^2})\\
&=\oint\frac{dx}{2\pi\ri x^{N_++1}}
\exp\left[\hf\sum_{\ell=1}^\infty\frac{(-1)^{\ell-1}x^\ell}{\ell}
\vartheta_2\bigl(q^\ell\bigr)
\right], 
\end{aligned}
\label{eq:psi-int}
\end{align}
where $\vartheta_2\bigl(q^\ell\bigr)=\vartheta_2\bigl(1,q^\ell\bigr)$.
For instance, the first few terms are given by
\begin{align}
 \psi_0=1,\qquad
\psi_1=\hf\vartheta_2(q),\qquad
\psi_2=\frac{1}{8} \vartheta_2(q)^2-\qu \vartheta_2(q^2). 
\end{align}

To see that $\psi_{N_+}$ is a non-perturbative completion of $\psi^{\text{top}}_{N_+}$, 
we notice that $\psi_{N_+}$ can be also written as
\begin{align}
 \psi_{N_+}=e^{F^{\text{cl}}(t)}\oint \frac{dx}{2\pi\ri x}\prod_{p>0}(1+xQ^pq^{\hf p^2})
\prod_{N_+>p>0}(1+x^{-1}Q^p q^{-\hf p^2})
\end{align}
which indeed becomes $\psi^{\text{top}}(t)$ in \eqref{eq:Ztop-fermi}
in the large $N_+$ limit. We will also see in the next subsection that the difference
between $\psi_{N_+}$ and $\psi^{\text{top}}(t)$ is indeed exponentially small
in the large $N_+$ limit.
The identification $t=N_+g_s$ in \eqref{eq:psitop-N}
is consistent with the definition of 't Hooft coupling in
\eqref{eq:thooft-t} since the sum \eqref{eq:ZN-decomp} is peaked around
$N_+=N_-=\hf N$ and hence the two definition of the 't Hooft parameter agree:
$t=N_+g_s=\hf Ng_s$.

Some comments are in order here:
\begin{itemize}
 \item[$(i)$] 
By our definition of $\psi_{N_+}$, the chiral factorization in \eqref{eq:ZN-decomp}
is {\it exact}. There are only bi-linear terms of $\psi_{N_+}$ in \eqref{eq:ZN-decomp};
there are no multi-linear terms of $\psi_{N_+}$
which appeared in the baby universe expansion \eqref{eq:baby-expansion} in \cite{Dijkgraaf:2005bp}.
\item[$(ii)$]
In our expansion \eqref{eq:ZN-decomp} both sides of the equation are well-defined
at finite $N$.
\end{itemize}

\subsection{Analytic continuation}
One can systematically compute the non-perturbative $\mathcal{O}(e^{-N_+})$ 
correction in \eqref{eq:psiapprox}
using the technique of generating function 
as in \cite{Dijkgraaf:2005bp}.
For this purpose, we first rewrite the integral representation of $\psi_{N_+}$ in \eqref{eq:psi-int} 
as
\begin{align}
 \psi_{N_+}=\oint\frac{dx}{2\pi\ri x^{N_++1}}\prod_{p>0}\frac{1}{1+x^{-1}q^{-\hf p^2}}
\prod_{p>0}(1+xq^{\hf p^2})(1+x^{-1}q^{-\hf p^2}). 
\label{eq:psiN-formal}
\end{align}
Here we have multiplied the integrand of \eqref{eq:psi-int}
by the factor
$\prod_{p>0}(1+x^{-1}q^{-\hf p^2})$ and divided it by the same factor.
On the other hand $\psi^{\text{top}}_{N_+}$ is written as \cite{Dijkgraaf:2005bp}
\begin{align}
 \psi_{N_+}^{\text{top}}=\oint\frac{dx}{2\pi\ri x^{N_++1}}
 \prod_{p>0}\left(1+x q^{\hf p^2}\right)
            \left(1+x^{-1} q^{-\hf p^2}\right),
\label{eq:top-int}
\end{align}
which can be formally inverted as
\begin{align}
\prod_{p>0}(1+xq^{\hf p^2})(1+x^{-1}q^{-\hf p^2}) =
\sum_{N_+} x^{N_+} \psi_{N_+}^{\text{top}}.
\end{align} 
By expanding the first factor of \eqref{eq:psiN-formal}
\begin{align}
 \prod_{p>0}\frac{1}{1+x^{-1}q^{-\hf p^2}}=:\sum_{k=0}^\infty \phi_kx^{-k},
\label{eq:phi-gen}
\end{align}
we find that \eqref{eq:psiN-formal} becomes
\begin{align}
\begin{aligned}
 \psi_{N_+}
&=\sum_{k=0}^\infty \phi_k\psi_{N_++k}^{\text{top}}
= \sum_{k=0}^\infty \phi_k\psi^{\text{top}}(t+kg_s).
\label{eq:phiN-exp}
\end{aligned}
\end{align}
In the last equality we have used \eqref{eq:psitop-N}. 
However, the above expansion \eqref{eq:phi-gen}
of the denominator of \eqref{eq:psiN-formal}
is merely a formal expression and $\phi_k$ is not a well-defined function of
$q$ as it stands.
We will argue below that we can define 
$\phi_k$ by an analytic continuation.
To do this, we  rewrite \eqref{eq:phi-gen} as
\begin{align}
\begin{aligned}
\prod_{p>0}\frac{1}{1+x^{-1}q^{-\hf p^2}}
&=\exp\left[\hf\sum_{\ell=1}^\infty \frac{(-x^{-1})^\ell}{\ell}
\vartheta_2(q^{-\ell})\right].
\end{aligned}
\label{eq:hole-phi}
\end{align}
For the physical value of string coupling $g_s>0$,
the parameter $q=e^{-g_s}$ satisfies $|q|<1$, which implies $|q^{-1}|>1$.
However, the theta function $\vartheta_2(q^{-1})$ 
is not well-defined in the region $|q^{-1}|>1$ and it should be defined by 
a certain analytic continuation.
We define $\vartheta_2(q^{-1})$
by using the zeta-function regularization as follows:
\begin{align}
 \begin{aligned}
  \vartheta_2(q^{-1})&=2q^{-\frac{1}{8}}\prod_{n=1}^\infty(1-q^{-n})(1+q^{-n})^2\\
&=2q^{-\frac{1}{8}}\prod_{n=1}^\infty(-1)q^{-3n}(1-q^{n})(1+q^{n})^2\\
&=2q^{-\frac{1}{8}}(-1)^{\zeta(0)}q^{-3\zeta(-1)}\prod_{n=1}^\infty(1-q^{n})(1+q^{n})^2.
 \end{aligned}
\label{eq:theta-zeta}
\end{align}
Plugging the value of the zeta-function $\zeta(0)=-\hf$ and $\zeta(-1)=-\frac{1}{12}$
into \eqref{eq:theta-zeta}, we find
\begin{align}
 \vartheta_2(q^{-1})=\ri \vartheta_2(q).
\label{eq:acon}
\end{align}
Here, for definiteness we have chosen a branch of the square-root
$(-1)^{\zeta(0)}=\ri$.
We will see in section \ref{sec:borel} that
the existence of the other branch $(-1)^{\zeta(0)}=-\ri$
is related to the Stokes phenomenon. 
Via this analytic continuation, $\phi_k$ in \eqref{eq:phi-gen} becomes a well-defined 
function of~$q$
\begin{align}
 \sum_{k=0}^\infty \phi_kx^{-k}=
\exp\left[\frac{\ri}{2}\sum_{\ell=1}^\infty \frac{(-x^{-1})^\ell}{\ell}
\vartheta_2(q^{\ell})\right].
\end{align}
In particular, $\phi_1$ is imaginary
\begin{align}
 \phi_1=-\frac{\ri}{2}\vartheta_2(q),
\end{align}
and the expansion of $\psi_{N_+}$ in \eqref{eq:phiN-exp} becomes
\begin{align}
 \begin{aligned}
  \psi_{N_+}
=\psi^{\text{top}}(t)-\frac{\ri}{2}\vartheta_2(q)\psi^{\text{top}}(t+g_s)+\cdots.
 \end{aligned}
\label{eq:psiN-acon}
\end{align}
The second term and the ellipses of \eqref{eq:psiN-acon}
correspond to the non-perturbative $\mathcal{O}(e^{-N_+})$
correction in \eqref{eq:psiapprox}.
This can be seen by taking the ratio of the two terms
$\psi^{\text{top}}(t)$ and $\psi^{\text{top}}(t+g_s)$
\begin{align}
\frac{\psi^{\text{top}}(t+g_s)}{\psi^{\text{top}}(t)}
\sim e^{F^{\text{cl}}(t+g_s)-F^{\text{cl}}(t)}
=e^{-\frac{t^2}{2g_s}-\frac{t}{2}-\frac{g_s}{8}}
\sim e^{-\frac{t}{2}N_+},
\end{align}
where we approximated $\psi^{\text{top}}(t)$ and $\psi^{\text{top}}(t+g_s)$
by their leading terms $e^{F^{\text{cl}}(t)}$ and 
$e^{F^{\text{cl}}(t+g_s)}$ with $F^{\text{cl}}(t)$
given by \eqref{eq:Fcl}.
One might think that the appearance of the imaginary term
in \eqref{eq:psiN-acon} looks strange since $\psi_{N_+}$ on the LHS
of \eqref{eq:psiN-acon} is real.
However,
as we will see in section \ref{sec:borel}, 
the second term of \eqref{eq:psiN-acon} is precisely
canceled by the imaginary part coming from the Borel resummation
of $\psi^{\text{top}}(t)$
in accord with the theory of resurgence.

A similar expansion of $Z_N$ is obtained by plugging the expansion 
\eqref{eq:phiN-exp} into \eqref{eq:ZN-decomp}
\begin{align}
 Z_N=
\sum_{N_++N_-=N}\sum_{k,l=0}^\infty
\phi_k\b{\phi}_l
\psi^{\text{top}}_{N_++k}\b{\psi}^{\text{top}}_{N_-+l}.
\label{eq:ZN-psi-exp}
\end{align}
Here $\b{\phi}_l$ is not the complex conjugate of
$\phi_l$ but it is defined by $\b{\phi}_l(\th)=\phi_l(-\th)$.
In particular, when $\th=0$ they are equal: $\b{\phi}_l=\phi_l$.

When $\th=0$, we can write down another useful expansion of $Z_N$.
To do this, let us introduce 
the perturbative part $\Zefull_N$ of the full partition function
$Z_N$ in the $1/N$ expansion
\begin{align}
\Zefull_N := \Zefull\left(t=\tfrac{1}{2}Ng_s\right).
\label{eq:ZefullN-def}
\end{align}
One can show that $\Zefull(t)$
is obtained by
squaring the integrand of $\psi^{\text{top}}(t)$ in \eqref{eq:Ztop-fermi}
\begin{align}
\begin{aligned}
\Zefull(t)
 =&\ e^{-\frac{t^3}{3g_s^2}}Q^{-\frac{1}{12}}
 \oint\frac{dx}{2\pi\ri x}
 \prod_{p\in\mathbb{Z}_{\geq0}+\hf}\left(1+xQ^p q^{\hf p^2}\right)^2
            \left(1+x^{-1}Q^p q^{-\hf p^2}\right)^2.
 \end{aligned}
\label{eq:Zfull}
\end{align}
Note that $Z_N$ can be thought of as a
non-perturbative completion of $\Zefull_N$
\begin{align}
Z_N = \Zefull_N+\mathcal{O}(e^{-N}),
\label{eq:ZNapprox}
\end{align}
which is an analogue of the relation between
$\psi_{N_+}$ and $\psi^{\text{top}}_{N_+}$ in \eqref{eq:psiapprox}. 
One can also show  that $\Zefull(t)$ in \eqref{eq:Zfull} 
given by the product over half-integer $p$
is the perturbative part of $Z_N$ for both even $N$ and odd $N$,
although it is not so obvious from the
definition of  $Z_N$
in \eqref{eq:ZN-mode} with $\th=0$.
To see this, we notice that $Z_N$
can also be written (for both even and odd $N$) as
\begin{align}
\begin{aligned}
Z_N
 = e^{-\frac{t^3}{3g_s^2}}Q^{-\frac{1}{12}}
 \oint\frac{dx}{2\pi\ri x}
 \prod_{p>0}\left(1+xQ^{p}q^{\hf p^2}\right)^2
 &\prod_{0<p<\frac{N}{2}}
  \left(1+x^{-1}Q^{p}q^{-\hf p^2}\right)\\
\times
 &\prod_{0<p<\frac{N+1}{2}}
  \left(1+x^{-1}Q^{p}q^{-\hf p^2}\right),
\end{aligned}
\end{align}
where products are over half-integer $p$
and we identify $t=\hf Ng_s$ (i.e.~$Q=q^{N/2}$). This indeed becomes 
$\Zefull(t)$ in \eqref{eq:Zfull} in the large $N$ limit.
In the rest of this section we will assume $N$ is even
for simplicity.

One can systematically compute 
the non-perturbative $\mathcal{O}(e^{-N})$ corrections
in \eqref{eq:ZNapprox} in a similar manner as
the expansion of $\psi_{N_+}$ in \eqref{eq:phiN-exp}.
It turns out that the non-perturbative corrections
in \eqref{eq:ZNapprox} involve not only $\Zefull_N$
but also another type of
partition function, which we denote
by $\Zofull_N$
\begin{align}
\Zofull_N:=\Zofull\left(t=\tfrac{1}{2}Ng_s\right),
\label{eq:ZofullN-def}
\end{align}
where
\begin{align}
\begin{aligned}
\Zofull(t):=&\ e^{-\frac{t^3}{3g_s^2}}Q^{\frac{1}{6}}\oint\frac{dx}{2\pi\ri x}
 (1+x)(1+x^{-1})
 \prod_{n=1}^\infty
 \left(1+xQ^n q^{\hf n^2}\right)^2
 \left(1+x^{-1}Q^n q^{-\hf n^2}\right)^2.
\end{aligned}
\label{eq:tZfull}
\end{align}
One might think that the 
introduction of $\Zofull(t)$ seems 
ad hoc, but it actually
has a 
clear physical interpretation as we mentioned in section~\ref{sec:intro}:
it can be regarded as the perturbative part of
another partition function 
\begin{align}
\tZ_N:=\oint\frac{dx}{2\pi\ri x^{N+1}}
 \prod_{p\in\bbZ+\frac{N}{2}}\left(1+xq^{\hf p^2}\right),
\label{eq:tZN-def}
\end{align}
which is the partition function of
$N$ non-relativistic free fermions on a circle
with {\it anti-periodic} boundary condition.
Here, notice that $p\in\bbZ$ for even $N$ and $p\in\bbZ+\frac{1}{2}$
for odd $N$.
This is in contrast to the case of $Z_N$,
in which periodic boundary condition \eqref{eq:pcond}
is imposed.
We should stress that $\Zofull(t)$ in \eqref{eq:tZfull}
is {\em not} the large $N$ limit of $Z_N$ with odd $N$.

Now we are ready to consider the expansion of $Z_N$
in \eqref{eq:ZN-int}.
By rewriting \eqref{eq:ZN-int}
as
\begin{equation}
\begin{aligned}
 Z_N=\oint\frac{dx}{2\pi\ri x^{N+1}}
\prod_{p\in\mathbb{Z}_{\geq0}+\hf}\frac{1}{\left(1+x^{-1}q^{-\hf p^2}\right)^2}
\prod_{p\in\mathbb{Z}_{\geq0}+\hf}
\left(1+xq^{\hf p^2}\right)^2\left(1+x^{-1}q^{-\hf p^2}\right)^2,
\end{aligned} 
\label{eq:ZN-dn}
\end{equation}
we find that $Z_N$ is written as
\begin{align}
\begin{aligned}
 Z_N
&=\sum_{k=0}^\infty \Phi_k\mathcal{W}_{N+k},
\label{eq:ZN-exp}
\end{aligned}
\end{align}
where  $\Phi_k$ 
is the expansion coefficient of the first factor of
\eqref{eq:ZN-dn}
\begin{align}
\sum_{k=0}^\infty \Phi_kx^{-k}
:= \prod_{p\in\mathbb{Z}_{\geq0}+\hf}\frac{1}{\left(1+x^{-1}q^{-\hf p^2}\right)^2}
=\exp\left[\sum_{\ell=1}^\infty \frac{(-x^{-1})^\ell}{\ell}
\vartheta_2(q^{-\ell})\right],
\label{eq:CapPhi-gen}
\end{align}
while $\mathcal{W}_K$ comes from
the second factor of
\eqref{eq:ZN-dn}
\begin{align}
\mathcal{W}_K:=
\oint\frac{dx}{2\pi\ri x^{K+1}}
 \prod_{p\in\mathbb{Z}_{\geq0}+\hf}\left(1+x q^{\hf p^2}\right)^2
            \left(1+x^{-1} q^{-\hf p^2}\right)^2.
\end{align}
As we anticipated, $\mathcal{W}_K$
is equal to either $\Zefull_K$ or $\Zofull_K$
depending on the parity of $K$
\begin{align}
\mathcal{W}_K
 =\left\{\begin{array}{l}
 \Zefull_K, \qquad \mbox{($K$: even)}, \\[1ex]
 \Zofull_K, \qquad \mbox{($K$: odd)}.
         \end{array}\right.
\label{eq:WK-cond}
\end{align}%
We present a proof of this relation in appendix~\ref{app:Zfullproof}.
As in the case of $\phi_k$ appearing in 
\eqref{eq:phiN-exp}, 
$\Phi_k$ in \eqref{eq:CapPhi-gen} is merely a formal expression
and thus we apply our prescription of the analytic continuation \eqref{eq:acon}
\begin{align}
\sum_{k=0}^\infty \Phi_kx^{-k}
=\exp\left[\ri \sum_{\ell=1}^\infty \frac{(-x^{-1})^\ell}{\ell}
\vartheta_2(q^\ell)\right].
\end{align}
Finally, the expansion of $Z_N$ in \eqref{eq:ZN-exp} becomes\footnote{
Here, the sum is divided merely for appearances' sake;
it should be taken in ascending order of $k$.}
\begin{equation}
\begin{aligned}
 Z_N&=\sum_{k:\,\text{even}}\Phi_k\Zefull_{N+k}+
\sum_{k:\,\text{odd}}\Phi_k\Zofull_{N+k}\\
&=\sum_{k:\,\text{even}}\Phi_k\Zefull\left(t+\tfrac{k}{2}g_s\right)+
\sum_{k:\,\text{odd}}\Phi_k\Zofull\left(t+\tfrac{k}{2}g_s\right).
\end{aligned} 
\label{eq:ZN-instexp}
\end{equation}
More explicitly, the first two terms of this expansion read
\begin{align}
Z_N
 =\Zefull(t)
 -\ri\varth_2(q)\Zofull\left(t+\tfrac{1}{2}g_s\right) + \cdots.
\label{eq:ZN-acon}
\end{align}
Again, the second term of \eqref{eq:ZN-acon} is imaginary but it is exactly 
canceled by the imaginary part coming from the Borel resummation
of the first term of \eqref{eq:ZN-acon}
as we will see in section \ref{sec:borel}.

In a similar manner as above, we can find the expansion of $\tZ_N$.
When $N$ is even,  \eqref{eq:tZN-def}
is written as
\begin{equation}
\begin{aligned}
 \tZ_N&=\oint\frac{dx}{2\pi\ri x^{N+1}}
(1+x)
 \prod_{n=1}^\infty\left(1+x q^{\hf n^2}\right)^2\\
&=\oint\frac{dx}{2\pi\ri x^{N+1}} \frac{1}{1+x^{-1}}
  \prod_{n=1}^\infty\frac{1}{\left(1+x^{-1}q^{-\hf n^2}\right)^2}\\
&\qquad\times 
(1+x)(1+x^{-1})
 \prod_{n=1}^\infty\left(1+x q^{\hf n^2}\right)^2
            \left(1+x^{-1} q^{-\hf n^2}\right)^2,
\end{aligned}  
\label{eq:tZ-intPW}
\end{equation}
and this can be expanded in a similar form as 
\eqref{eq:ZN-exp}
\begin{align}
\tZ_N
&=\sum_{k=0}^\infty \tPhi_k\til{\mathcal{W}}_{N+k},
\label{eq:tZN-exp}
\end{align}
where $\til{\mathcal{W}}_{K}$
comes from the last factor of \eqref{eq:tZ-intPW}
\begin{equation}
\begin{aligned}
 \til{\mathcal{W}}_{K}:=\oint\frac{dx}{2\pi\ri x^{K+1}}
 (1+x)(1+x^{-1})
 \prod_{n=1}^\infty\left(1+x q^{\hf n^2}\right)^2
            \left(1+x^{-1} q^{-\hf n^2}\right)^2.
\end{aligned} 
\end{equation}
One can show that (see appendix~\ref{app:Zfullproof})
$\til{\mathcal{W}}_{K}$ is equal to
$\Zofull_K$ or $\Zefull_K$
in the opposite ordering of
$\mathcal{W}_K$ in \eqref{eq:WK-cond}
\begin{equation}
\begin{aligned}
 \til{\mathcal{W}}_{K}=\left\{\begin{array}{l}
 \Zofull_K, \qquad \mbox{($K$: even)}, \\[1ex]
 \Zefull_K, \qquad \mbox{($K$: odd)}.
         \end{array}\right.
\end{aligned}
\label{eq:tWK-cond}
\end{equation}
The coefficient $\tPhi_k$ in \eqref{eq:tZN-exp} is formally given by
\begin{align}
\sum_{k=0}^\infty \tPhi_kx^{-k}
:= \frac{1}{1+x^{-1}}
  \prod_{n=1}^\infty\frac{1}{\left(1+x^{-1}q^{-\hf n^2}\right)^2}
=\exp\left[\sum_{\ell=1}^\infty \frac{(-x^{-1})^\ell}{\ell}
\vartheta_3(q^{-\ell})\right],
\label{eq:tPhi-gen}
\end{align}
which should be defined by a certain analytic continuation.
We define $\varth_3(q^{-1})$ by using the zeta-function
regularization, in a similar manner as we did for $\vartheta_2(q^{-1})$
in \eqref{eq:theta-zeta}
\begin{align}
\begin{aligned}
\varth_3\left(q^{-1}\right)
 &=\prod_{n=1}^\infty\left(1-q^{-n}\right)
  \left(1+q^{-\left(n-\hf\right)}\right)^2\\
 &=(-1)^{\zeta(0)}q^{-\zeta(-1)}q^{-2\zeta\left(-1,\hf\right)}
   \prod_{n=1}^\infty\left(1-q^n\right)
  \left(1+q^{n-\hf}\right)^2\\
 &=\ri \vartheta_3(q),
\end{aligned} 
\end{align}
where we have used
$\zeta(0)=-\hf,\,\zeta(-1)=-\frac{1}{12},\,\zeta(-1,\hf)=\frac{1}{24}$
and $\zeta(z,a):=\sum_{n=0}^\infty(a+n)^{-z}$
is the Hurwitz zeta function.
Then $\tPhi_k$ becomes a well-defined
function of $q$
\begin{align}
\sum_{k=0}^\infty \tPhi_kx^{-k}
=\exp\left[\ri \sum_{\ell=1}^\infty \frac{(-x^{-1})^\ell}{\ell}
\vartheta_3(q^\ell)\right].
\end{align}
Finally,  the expansion of $\tZ_N$ in \eqref{eq:tZN-exp} becomes
\begin{equation}
\begin{aligned}
 \tZ_N
=\sum_{k:\,\text{even}}\tPhi_k\Zofull\left(t+\tfrac{k}{2}g_s\right)+
\sum_{k:\,\text{odd}}\tPhi_k\Zefull\left(t+\tfrac{k}{2}g_s\right),
\end{aligned} 
\label{eq:tZN-instexp}
\end{equation}
and the first two terms of this expansion read
\begin{align}
\tZ_N
 =\Zofull(t)
 -\ri\varth_3(q)\Zefull\left(t+\tfrac{1}{2} g_s\right) + \cdots.
\label{eq:tZN-acon}
\end{align}

To summarize, $\Zefull(t)$ and $\Zofull(t)$
are the perturbative part of $Z_N$ and $\tZ_N$, respectively,
and $\Zefull(t+\tfrac{k}{2} g_s)$ and $\Zofull(t+\tfrac{k}{2} g_s)$ appear alternatingly
as non-perturbative $k$-instanton corrections
in the expansion of $Z_N$ \eqref{eq:ZN-instexp}
and $\tZ_N$ \eqref{eq:tZN-instexp}.
In other words,
each time one instanton is added, $\Zefull$ and $\Zofull$
are exchanged and $t$ is shifted with
a unit $\lap t=g_s/2$.
This reminds us of the effect of adding D-branes 
discussed in \cite{Aganagic:2003qj,Aganagic:2011mi}.
It would be interesting to understand this relation further.

\subsection{Comparison with Dijkgraaf-Gopakumar-Ooguri-Vafa \cite{Dijkgraaf:2005bp}} 

Let us compare our expansion \eqref{eq:ZN-psi-exp}
with the baby universe expansion \eqref{eq:baby-expansion} in \cite{Dijkgraaf:2005bp}.
In \cite{Dijkgraaf:2005bp}, the expansion \eqref{eq:baby-expansion} 
of $Z_N$ was obtained starting from the following
relation
\begin{align}
Z(x,g_s)Z(x^{-1},-g_s)= \psi^{\text{top}}(x)\b{\psi}^{\text{top}}(x), 
\label{eq:Zxrel}
\end{align}
where $Z(x,g_s)$ is defined in \eqref{eq:grandZ} and 
$\psi^{\text{top}}(x)$ and $\b{\psi}^{\text{top}}(x)$ are given by
\begin{align}
 \begin{aligned}
 \psi^{\text{top}}(x)&= \sum_{N_+}x^{N_+}\psi^{\text{top}}_{N_+}=\prod_{p>0}(1+xq^{\hf p^2})(1+x^{-1}q^{-\hf p^2}),\\
\b{\psi}^{\text{top}}(x)&=\sum_{N_-}x^{N_-}\b{\psi}^{\text{top}}_{N_-}=\prod_{p<0}(1+xq^{\hf p^2})
(1+x^{-1}q^{-\hf p^2}).
 \end{aligned}
\end{align}
In \cite{Dijkgraaf:2005bp} it is argued that under a certain 
analytic continuation $Z(x^{-1},-g_s)$
can be identified with $Z(x,g_s)$
\begin{align}
 Z(x^{-1},-g_s)= Z(x,g_s).
\label{eq:acon-baby}
\end{align}
Then \eqref{eq:Zxrel} becomes $Z(x,g_s)^2=\psi^{\text{top}}(x)\b{\psi}^{\text{top}}(x)$,
which implies that $Z_N$ obeys
\begin{align}
 \sum_{k=0}^N Z_k Z_{N-k}=\sum_{N_{+}+N_{-}=N} \psi^{\text{top}}_{N_+}
\b{\psi}^{\text{top}}_{N_-}.
\end{align}
Solving this relation iteratively, $Z_N$ is written 
as \eqref{eq:baby-expansion} and it was interpreted as creation of baby universes 
in \cite{Dijkgraaf:2005bp}.

However, our resurgence analysis suggests that 
we should consider different analytic continuation \eqref{eq:acon}
in order to cancel the non-perturbative ambiguity (imaginary part)
in the Borel resummation of $\psi^{\text{top}}$.
Our analytic continuation \eqref{eq:acon} is different from that in \cite{Dijkgraaf:2005bp}
\begin{align}
 Z(x^{-1},-g_s)= 
\exp\left[-\sum_{\ell=1}^\infty
\frac{(-x^{-1})^{\ell}}{\ell}\vartheta_2\bigl(q^{-\ell}\bigr)\right]=\exp\left[-\ri\sum_{\ell=1}^\infty
\frac{(-x^{-1})^{\ell}}{\ell}\vartheta_2\bigl(q^\ell\bigr)\right].
\label{eq:our-acon}
\end{align}
In particular, $Z(x^{-1},-g_s)$ is not equal to $Z(x,g_s)$
\begin{align}
 Z(x^{-1},-g_s)\ne Z(x,g_s).
\end{align}
In our approach, $Z(x^{-1},-g_s)$
corresponds to the denominator appeared in  \eqref{eq:CapPhi-gen}
\begin{align}
\begin{aligned}
 Z_N&=\oint\frac{dx}{2\pi\ri x^{N+1}}Z(x,g_s)
=\oint\frac{dx}{2\pi\ri x^{N+1}}\frac{\psi^{\text{top}}(x)\b{\psi}^{\text{top}}(x)}{Z(x^{-1},-g_s)}\\
&=\oint\frac{dx}{2\pi\ri x^{N+1}}
\exp\left[\ri\sum_{\ell=1}^\infty
\frac{(-x^{-1})^{\ell}}{\ell}\vartheta_2\bigl(q^\ell\bigr)\right]\psi^{\text{top}}(x)\b{\psi}^{\text{top}}(x) 
\end{aligned}
\end{align}
which leads to the expansion \eqref{eq:ZN-psi-exp}.

We think that there is no clear justification for
the analytic continuation \eqref{eq:acon-baby} used in \cite{Dijkgraaf:2005bp}. 
On the other hand, our analytic continuation  \eqref{eq:our-acon}
is supported by the resurgence analysis as we will see in the rest of this paper.

\section{Genus expansion of partition function\label{sec:genus}}

In this section we consider the genus expansion of
$\psi^{\text{top}}$, $\Zefull$, and $\Zofull$.
In subsection \ref{subsec:genustop}
we study the genus expansion of $\psi^{\text{top}}$
following the approach of Kaneko and Zagier in \cite{zagier}
with a slight modification.
In subsection \ref{subsec:genus-full} we consider
the genus expansion of $\Zefull$ and $\Zofull$.
We derive two different methods for obtaining the genus expansion:
by using the chiral factorization relation \eqref{eq:RRsum}
or by using recursion relations similar to that of Kaneko and Zagier.
We also elucidate the modular properties of $\Zefull$ and $\Zofull$.

\subsection{Genus expansion of $\psi^{\text{top}}$\label{subsec:genustop}}

We first consider the genus expansion of 
topological string partition function $\psi^{\text{top}}$ \eqref{eq:free-g}.
On general grounds, one can in principle compute the genus $g$ free
energy $F_g(t)$ recursively by solving the holomorphic anomaly equation \cite{Bershadsky:1993cx},
up to a holomorphic ambiguity.
The holomorphic anomaly equation in the case of 2d Yang-Mills on $T^2$
was studied in \cite{Dijkgraaf:1996iy,dijk}\footnote{
See also \cite{Griguolo:2004uz,Griguolo:2004jp} for 
the genus expansion of chiral partition function and its double scaling limit.}.
However,
it turns out that to compute the genus expansion of $\psi^{\text{top}}$
it is more efficient to use a different recursion relation
found by Kaneko and Zagier \cite{zagier}. Their relation determines
the higher genus 
amplitudes completely without holomorphic ambiguity.
We also find a slight modification
of the recursion relation of \cite{zagier},
which makes the modular property of $F_g(t)$ more transparent than the original 
one in \cite{zagier}.

In this section, we will often use the
rescaled topological string partition function
$\psi(t)$ and $\h{\psi}(t)$ defined by  
removing the genus-zero (and genus-one) part from $\psi^{\text{top}}(t)$
\begin{align}
 \psi^{\text{top}}(t)=\psi_0(t)\psi(t)
=\psi_{01}(t)\h{\psi}(t),
\label{eq:psi01}
\end{align}
where
\begin{align}
 \begin{aligned}
  \psi_0(t)&=\exp\Biggl(\frac{1}{g_s^2}F_0(t)\Biggr)=\exp\Biggl(-\frac{t^3}{6g_s^2}\Biggr),\\
 \psi_{01}(t)&=\exp\Biggl(\frac{1}{g_s^2}F_0(t)+F_1(t)\Biggr)
 =\frac{1}{\eta(Q)}\exp\Biggl(-\frac{t^3}{6g_s^2}\Biggr).  
 \end{aligned}
 \label{eq:def-psi01}
\end{align}
In other words,
$\psi(t)$ and $\h{\psi}(t)$ are given by
the sum of $F_g(t)$ for $g\geq1$ and $g\geq2$, respectively
\begin{align}
\begin{aligned}
 \psi(t)&=\exp\Biggl(\sum_{g=1}^\infty g_s^{2g-2}F_g(t)\Biggr),\\
\h{\psi}(t)&=\exp\Biggl(\sum_{g=2}^\infty g_s^{2g-2}F_g(t)\Biggr) =\eta(Q)\psi(t).
\label{eq:psihat}
\end{aligned}
\end{align}
Now we want to find the genus expansion of $\h{\psi}(t)$
\begin{align}
 \h{\psi}(t)=\sum_{n=0}^\infty \Ztop_n(t)g_s^{2n}.
\label{eq:Zn-expand}
\end{align}
From the definition \eqref{eq:psihat}, 
one can  see that  $\Ztop_0(t)=1$.
As we will show below, starting from $\Ztop_0(t)=1$
we can compute $\Ztop_n(t)$ recursively.
Once we know $\Ztop_n(t)$, the genus $g$ free energy 
$F_{g}(t)$ is  obtained 
from the relation
\begin{align}
 F_{g+1}(t)=\Ztop_g(t)-\frac{1}{g}\sum_{h=1}^{g-1}hF_{h+1}(t)\Ztop_{g-h}(t),\quad(g\geq1),
\label{eq:ZtoF}
\end{align}
which is easily derived by taking the $g_s$-derivative of the both sides of
\eqref{eq:Zn-expand}.

Let us first recall the approach in \cite{zagier}.
By dropping the genus-zero part of $\psi^{\text{top}}$ in \eqref{eq:Ztop-fermi},
$\psi(t)$ is written as
\begin{align}
 \psi(t)=\oint\frac{dx}{2\pi\ri x}H(q,Q,-x),
\end{align}
where $H(q,Q,z)$ is a function introduced in \cite{zagier}\footnote{In \cite{zagier}
$H(q,Q,z)$ is denoted as $H(w,q,\zeta)$.}
\begin{align}
 H(q,Q,z):=Q^{-\frac{1}{24}}\prod_{p\in\bbZ_{\geq0}+\hf}(1-zQ^pq^{\hf p^2})(1-z^{-1}Q^pq^{-\hf p^2}).
\end{align}
As shown in \cite{zagier}, $H(q,Q,z)$ is related to
$\psi(t)$ as
\begin{align}
 H(q,Q,z)
=\sum_{n\in\bbZ}
  \psi(t+n g_s)q^{\frac{n^3}{6}}Q^{\frac{n^2}{2}}(-z)^n.
\label{eq:prop-KZ}
\end{align}
Expanding the both sides of \eqref{eq:prop-KZ} in $g_s$,
it is proved in \cite{zagier}
that $F_g(t)$ is a quasi-modular form
of weight $6g-6$ for $\Ga=PSL(2,\bbZ)$.
The relation obtained from \eqref{eq:prop-KZ} by expanding in $g_s$
can be thought of as a recursion relation for $\Ztop_n(t)$.
However, this relation involves the quasi-modular forms of 
both $\Ga$ and $\Ga^0(2)$,\footnote{$\Gamma^0(2)$ is
a subgroup of $\Gamma$ which consists of matrices of the form
$
\begin{pmatrix}
 a&b\\c&d
\end{pmatrix}\in\Gamma
$
with $b\equiv 0$ mod $2$.} and it is not straightforward to
see that
$\Ztop_n(t)$ is a quasi-modular form of $\Ga$.

It turns out that we can modify the relation \eqref{eq:prop-KZ}
in such a way that it becomes manifest that
$\Ztop_n(t)$ is a quasi-modular form of weight $6n$ for $\Ga$.
To see this, let us introduce a new generating function
$\Xi(q,Q,z)$
\begin{align}
\begin{aligned}
 \Xi(q,Q,z):=&\ z^{-\frac{1}{2}}q^{-\frac{1}{48}}Q^{\frac{1}{8}}
  H\left(q,q^{-\frac{1}{2}}Q,q^{\frac{1}{8}}Q^{-\frac{1}{2}}z\right)\\
=&\ Q^{\frac{1}{12}}
 \left(z^{-\frac{1}{2}}-z^{\frac{1}{2}}\right)
 \prod_{n\in\bbZ_{> 0}}
 \left(1-zQ^{n}q^{\hf n^2}\right)
 \left(1-z^{-1}Q^{n}q^{-\hf n^2}\right),
\label{eq:Xi-def} 
\end{aligned}
\end{align}
which is related to $\psi(t)$ as
 \begin{align}
 \Xi(q,Q,z)=\sum_{p\in\mathbb{Z}+\hf}(-1)^{p-\hf}
\psi(t-pg_s)q^{-\frac{p^3}{6}}Q^{\frac{p^2}{2}}z^{-p}.
\label{eq:Xi-sum}
\end{align}
This is just a rescaled version of the original relation \eqref{eq:prop-KZ} in \cite{zagier}.
The above relation \eqref{eq:Xi-sum} leads to infinitely many relations when expanded in terms of 
the chemical 
potential $\mu=-\log z$.
Here we focus on the linear term in the small $\mu$ expansion 
\begin{align}
  \mathcal{K}(q,Q)=\sum_{p\in\mathbb{Z}+\hf}(-1)^{p-\hf}
p\psi(t-pg_s)q^{-\frac{p^3}{6}}Q^{\frac{p^2}{2}},
\label{eq:xi-psi}
\end{align}
where 
 $\mathcal{K}(q,Q)$
is given by
\begin{align}
\mathcal{K}(q,Q):=\lim_{\mu\to0}\frac{1}{\mu}\Xi(q,Q,e^{-\mu})=
Q^{\frac{1}{12}}\prod_{n=1}^\infty (1-Q^nq^{\hf n^2})(1-Q^nq^{-\hf n^2}).
\label{eq:hat-xi} 
\end{align}
By comparing the small $g_s$ expansion of the both sides of \eqref{eq:hat-xi},
we can write down a recursion relation for $\Ztop_n$.

Let us first consider the LHS of \eqref{eq:hat-xi}. 
It turns out that it is useful
to normalize $\mathcal{K}(q,Q)$ by
$\mathcal{K}(1,Q)=\eta(Q)^2$ 
\begin{equation}
\begin{aligned}
 \h{\mathcal{K}}(q,Q):=
\frac{\mathcal{K}(q,Q)}{\eta(Q)^2}=\prod_{n=1}^\infty
\frac{(1-Q^nq^{\hf n^2})(1-Q^nq^{-\hf n^2})}{(1-Q^n)^2}.
\end{aligned} 
\label{eq:hxi-def}
\end{equation}
As we will see below, this function
plays an important role in the recursion relation 
of $\Ztop_n$. 
Let us introduce $h_n^{\text{top}}$ and 
$e_l$ as the coefficients 
in the small $g_s$ expansion of $\h{\mathcal{K}}(q,Q)$
\begin{align}
 \h{\mathcal{K}}=:\sum_{n=0}^\infty h_n^{\text{top}}g_s^{2n}
=:\exp\left(\sum_{l=1}^\infty \frac{e_l}{(2l)!}g_s^{2l}\right).
\label{eq:hxi-exp}
\end{align}
Here we suppressed the argument of $\h{\mathcal{K}}(q,Q)$
for brevity.
As we will show below, 
$e_l$ is given by the derivative of Eisenstein series
\begin{align}
 e_l=\frac{B_{2l+2}}{2l+2}2^{-2l}D^{2l-1}E_{2l+2}(Q).
\label{eq:el}
\end{align}
Here $B_{2k}$ denotes the Bernoulli number
and the Eisenstein series $E_{2k}(Q)$
of weight $2k$ is defined by
\begin{align}
 E_{2k}(Q):=1-\frac{4k}{B_{2k}}\sum_{n=1}^\infty  \frac{n^{2k-1}Q^n}{1-Q^n},
\label{eq:def-Eisen}
\end{align}
and $D$ in \eqref{eq:el} is a differential operator defined by
\begin{align}
 D:=Q\del_Q=-\del_t.
\end{align}
The derivation of \eqref{eq:el}
is almost parallel to the similar computation of $H(q,Q,z)$ in \cite{zagier}.
Taking the log of $\h{\mathcal{K}}$ in \eqref{eq:hat-xi}
\begin{align}
 \begin{aligned}
  \log\h{\mathcal{K}}=-\sum_{r,n=1}^\infty
\frac{Q^{rn}}{r}\bigl(q^{\hf rn^2}+q^{-\hf rn^2}-2\bigr)=
\sum_{l=1}^\infty \frac{e_l}{(2l)!}g_s^{2l},
 \end{aligned}
\end{align}
$e_l$ is given by
\begin{align}
 \begin{aligned}
  e_l&=-2\sum_{r,n=1}^\infty\biggl(\frac{rn^2}{2}\biggr)^{2l}\frac{Q^{rn}}{r}=-2^{1-2l} \sum_{r,n=1}^\infty r^{2l-1}n^{4l}Q^{rn}\\
&=-2^{1-2l}D^{2l-1}\sum_{r,n=1}^\infty n^{2l+1}Q^{rn}=-2^{1-2l}D^{2l-1}\sum_{n=1}^\infty\frac{n^{2l+1}Q^n}{1-Q^n}.
 \end{aligned}
\end{align} 
Comparing this with the definition of Eisenstein series in \eqref{eq:def-Eisen},
we arrive at the expression of $e_l$  in \eqref{eq:el}.

On the other hand, the $g_s$-expansion of the RHS of \eqref{eq:xi-psi}
is given by
\begin{align}
 \begin{aligned}
 \text{RHS~of~}\eqref{eq:xi-psi}&=\sum_{p\in\mathbb{Z}+\hf}(-1)^{p-\hf}p
\sum_{l,m\geq0} \frac{(pg_s)^l}{l!}
D^l\psi\frac{p^{3m}g_s^m}{6^m m!}Q^{\hf p^2}\\
&=\sum_{l,m\geq0} \frac{g_s^{l+m}}{l!6^mm!}
D^l\psi\frac{1+(-1)^{l+m}}{2}(2D)^{\frac{l+3m}{2}}\sum_{p\in\mathbb{Z}+\hf}(-1)^{p-\hf}pQ^{\hf p^2}\\
&=\sum_{l,m\geq0} \frac{g_s^{l+m}}{l!6^mm!}
D^l\psi\frac{1+(-1)^{l+m}}{2}(2D)^{\frac{l+3m}{2}}\eta(Q)^3,
 \end{aligned}
\label{eq:rhs-comp}
\end{align}
where we have used the identity
\begin{equation}
\begin{aligned}
 \sum_{p\in\mathbb{Z}+\hf}(-1)^{p-\hf}pQ^{\hf p^2}=\eta(Q)^3.
\end{aligned} 
\end{equation}
When going from the first line to the second line of \eqref{eq:rhs-comp}, 
we replaced
$p^{l+3m}Q^{\frac{p^2}{2}}$
$\to$
\linebreak
$(2D)^{\frac{l+3m}{2}}Q^{\frac{p^2}{2}}$
and inserted the projection $\frac{1+(-1)^{l+m}}{2}$ to even $l+m$, 
since the contribution of odd $l+m$ vanishes
by the cancellation between $p$ and $-p$.
\eqref{eq:rhs-comp} can be further simplified as follows.
Introducing the notation $D_k$ by
\begin{align}
 D_{k}:=\eta(Q)^{-k}D\eta(Q)^k=D+kD\log\eta(Q)=D+\frac{kE_2(Q)}{24},
\label{eq:Dk}
\end{align}
and using the relation 
\begin{equation}
\begin{aligned}
 D^l\psi=\eta(Q)^{-1}D_{-1}^l\h{\psi},\qquad
D^n\eta(Q)^3=\eta(Q)^3D_3^n1,
\end{aligned} 
\end{equation}
we can formally perform the summation in \eqref{eq:rhs-comp}
\begin{align}
 \begin{aligned}
  \text{RHS~of~}\eqref{eq:xi-psi}&=\eta(Q)^2\sum_{l,m\geq0} \frac{g_s^{l+m}}{l!6^mm!}
D_{-1}^l\h{\psi}\frac{1+(-1)^{l+m}}{2}(2D_{3})^{\frac{l+3m}{2}}1\\
&=\eta(Q)^2\cosh\Bigl[g_s\rt{2D_{3}}\Bigl(D_{-1}+\frac{1}{3}D_{3}\Bigr)\Bigr]\h{\psi}\cdot 1.
\label{eq:rhs-op}
 \end{aligned}
\end{align}
Here it should be understood that $D_{-1}$ and $D_{3}$ act on $\h{\psi}$ and $1$, respectively.
From \eqref{eq:hxi-exp} and \eqref{eq:rhs-op}, we arrive at
our ``master equation'' for the genus expansion of $\h{\psi}$
\begin{align}
\h{\mathcal{K}}=
\cosh\Bigl[g_s\rt{2D_{3}}\Bigl(D_{-1}+\frac{1}{3}D_{3}\Bigr)\Bigr]\h{\psi}\cdot 1.
\label{eq:el-to-DZ}
\end{align}
Finally, comparing the $\mathcal{O}(g_s^{2n})$ term of \eqref{eq:el-to-DZ}, we arrive at the
desired recursion relation
of $\Ztop_n(t)$
\begin{align}
 \Ztop_n=\htop_n-\sum_{m=1}^n\frac{\bigl[D_{-1}+\frac{1}{3}D_{3}\bigr]^{2m}}{(2m)!}\Ztop_{n-m}\cdot (2D_{3})^{m}1.
\label{eq:rec-1}
\end{align}
The explicit form of $\htop_n$ is obtained from $e_l$ in \eqref{eq:el}
by expanding the exponential in \eqref{eq:hxi-exp}.

Now one can easily compute $\Ztop_n$ using our recursion relation \eqref{eq:rec-1}
with the initial condition 
$\Ztop_0=1$.
We emphasize that our recursion relation \eqref{eq:rec-1}
determines $\Ztop_n$  unambiguously. This is in contrast to
the case of the holomorphic anomaly equation that determines the derivative of 
$\Ztop_n$: there is an ambiguity in the integration constant
which should be fixed by some other conditions.\footnote{
From the $\mathcal{O}(\mu^0)$ term of \eqref{eq:Xi-sum}, one can write another relation.
After a similar computation as above, we find
\begin{align}
\begin{aligned}
0&=\sum_{p\in\mathbb{Z}+\hf}(-1)^{p-\hf}
\psi(t-pg_s)q^{-\frac{p^3}{6}}Q^{\frac{p^2}{2}}= 
\frac{\sinh\Bigl[g_s\rt{2D_{3}}(D_{-1}+\frac{1}{3}D_{3})\Bigr]}{\rt{2D_{3}}}
\h{\psi}\cdot 1
\end{aligned}
\label{eq:p-sum}
\end{align}
Using the relation $(D_{-1}+\frac{1}{3}D_{3})\Ztop_n\cdot1=D\Ztop_n$,
we find the recursion relation without the inhomogeneous term $\htop_n$
\eqref{eq:rec-1}
\begin{align}
 D\Ztop_n=-\sum_{m=1}^n \frac{\bigl[D_{-1}+\frac{1}{3}D_{3}\bigr]^{2m+1}}{(2m+1)!}\Ztop_{n-m}\cdot (2D_{3})^m1
\label{eq:rec-2}
\end{align}
which determines the derivative $D\Ztop_n$.
This recursion 
relation was also considered in \cite{Zagier-BO}.
}

We also note that the $g_s$-expansion of \eqref{eq:prop-KZ} originally considered in \cite{zagier}
involves $E_{2k}(Q)$ and $E_{2k}(Q^{1/2})$, while in our case
the expansion \eqref{eq:hxi-exp}
of $\h{\mathcal{K}}$ involves $E_{2k}(Q)$ only and $E_{2k}(Q^{1/2})$ does not show up. 
This is the advantage of the use of $\h{\mathcal{K}}$ 
over the original $H(q,Q,z)$ in \cite{zagier}
and it is clear from our recursion relation \eqref{eq:rec-1}
that $\Ztop_n$ is written as a combination of $E_{2k}(Q)$ only.

As is well known, $E_{2k}(Q)$~$(k\geq2)$ is a modular form for $\Gamma$ of
weight $2k$ in $\tau=\frac{1}{2\pi\ri}\ln Q$
and thus can be expressed as a polynomial of $E_4(Q)$ and $E_6(Q)$.
This can be done easily by using the recursion relation
\begin{align}
\frac{B_{2k}E_{2k}(Q)}{(2k)!}
 =\frac{3}{(3-k)(4k^2-1)}\sum_{\substack{p+q=k\\ p,q\ge 2}}
(2p-1)(2q-1)\frac{B_{2p}E_{2p}(Q)}{(2p)!}\frac{B_{2q}E_{2q}(Q)}{(2q)!}
\quad (k\ge 4).
\end{align}
From our recursion relation \eqref{eq:rec-1} it is manifest that
$\Ztop_n$ is a quasi-modular form of weight $6n$ for $\Ga$,
i.e.~it can be expressed as a polynomial of $E_2(Q),\,E_4(Q),$ and $E_6(Q)$.

Using the recursion relation \eqref{eq:rec-1}, we have computed $\Ztop_n$ up to $n=60$.\footnote{
The data
of $\Ztop_n~(n=1,\ldots,60)$ are available upon request to the authors.}
The first few terms read
\begin{align}
\begin{aligned}
\Ztop_1&=\frac{5 E_2^3-3 E_2 E_4 -2 E_6}{51840},\\
\Ztop_2&= \frac{-875 E_2^6+2220 E_2^4 E_4 +580 E_2^3 E_6 -1791
   E_2^2 E_4^2 -1788 E_2 E_4 E_6 +1050 E_4^3+604
   E_6^2}{5374771200},\\
\Ztop_3&=\frac{1}{835884417024000}
\left(
 625625E_2^9-2469375E_2^7E_4-1065750E_2^6E_6+3079485E_2^5E_4^2\right.\\
&\left.\hspace{4em}
 +7892280E_2^4E_4E_6-3829077E_2^3E_4^3-3342540E_2^3E_6^2
 -11313054E_2^2E_4^2E_6\right.\\
&\left.\hspace{4em}
 +6470550E_2E_4^4+8753364E_2E_4E_6^2
 -4034700E_4^3E_6-766808E_6^3
\right),
\end{aligned}
\end{align}
where we abbreviated $E_{2k}=E_{2k}(Q)$.
One can check that the genus-$g$ free energy $F_{g}(t)$ obtained from 
\eqref{eq:ZtoF} reproduces the known result in \cite{Douglas:1993wy,Rudd:1994ta}.

\subsection{Genus expansion of $\Zefull$ and $\Zofull$ \label{subsec:genus-full}}

In this subsection we will compute the genus expansion of
the full partition functions $\Zefull(t)$ \eqref{eq:Zfull} and $\Zofull(t)$
\eqref{eq:tZfull}.
Throughout this subsection we set $\theta=0$ for simplicity.

We can define
the genus-$g$ free energies
$\Feven_g(t),\,\Fodd_g(t)$ of
$\Zefull(t),\,\Zofull(t)$ in the usual way
\begin{align}
\Zefull(t)=:\exp\left(\sum_{g=0}^\infty g_s^{2g-2}\Feven_g(t)\right),
\quad
\Zofull(t)=:\exp\left(\sum_{g=0}^\infty g_s^{2g-2}\Fodd_g(t)\right).
\label{eq:Zeogenusexp}
\end{align}
The first few terms are found as
\begin{align}
\begin{aligned}
\Feven_0(t)=\Fodd_0(t)=2F_0(t)=-\frac{t^3}{3},\quad
\Feven_1(t)=\ln\frac{\Thetaeven}{\eta(Q)^2},\quad
\Fodd_1(t)=\ln\frac{\Thetaodd}{\eta(Q)^2},
\end{aligned}
\label{eq:Feo-01}
\end{align}
where
\begin{align}
\begin{aligned}
\Thetaeven&:=\sum_{l\in\mathbb{Z}}Q^{l^2}=\vartheta_3(Q^2)
 =\frac{\eta(Q^2)^5}{\eta(Q)^2\eta(Q^4)^2},\\
\Thetaodd&:=\sum_{p\in\mathbb{Z}+\hf}Q^{p^2}=\vartheta_2(Q^2)
 =\frac{2\eta(Q^4)^2}{\eta(Q^2)}.
\end{aligned}
\label{eq:defTh}
\end{align}
The genus-one free energy in \eqref{eq:Feo-01}
can be obtained by setting $q=1$ in \eqref{eq:Zfull} and \eqref{eq:tZfull}. 
The appearance of $\Thetaeven,\,\Thetaodd$
can be also understood from the relation
 in \eqref{eq:RRsum},
as we will see shortly.

It is convenient to introduce
the rescaled partition functions
$\Zeven(t)$, $\Zodd(t)$, $\h{\Zeven}(t)$, $\hh{\Zodd}(t)$ by
stripping off the genus-zero (and genus-one) pieces
in the same way as $\psi^{\text{top}}$ in \eqref{eq:psi01}.
More specifically, the rescaled partition functions are given by
(see \eqref{eq:Zfull} and \eqref{eq:tZfull})
\begin{align}
\Zeven(t) &= \frac{\Thetaeven}{\eta(Q)^2}\h{\Zeven}(t) =
 Q^{-\frac{1}{12}}
 \oint\frac{dx}{2\pi\ri x}
 \prod_{p\in\bbZ_{\ge 0}+\frac{1}{2}}
 \left(1+xQ^p q^{\hf p^2}\right)^2
 \left(1+x^{-1}Q^pq^{-\hf p^2}\right)^2,\nn\\
\Zodd(t) &= \frac{\Thetaodd}{\eta(Q)^2}\hh{\Zodd}(t) =
 Q^{\frac{1}{6}}\oint\frac{dx}{2\pi\ri x}
 (1+x)(1+x^{-1})
 \prod_{n=1}^\infty
 \left(1+xQ^n q^{\hf n^2}\right)^2
 \left(1+x^{-1}Q^n q^{-\hf n^2}\right)^2.
\label{eq:Z-tZ}
\end{align}
We would like to find the $g_s$-expansion of free energy \eqref{eq:Zeogenusexp}
as well as the $g_s$-expansion of partition function itself
\begin{align}
\h{\Zeven}(t)=:\sum_{n=0}^\infty g_s^{2n}\Zeven_n(t),\qquad
\hh{\Zodd}(t)=:\sum_{n=0}^\infty g_s^{2n}\Zodd_n(t).
\label{eq:Zn-exp}
\end{align}
Note that from the definition of $\h{\Zeven}(t)$
and $\hh{\Zodd}(t)$, the $\mathcal{O}(g_s^0)$ term is unity:
$\Zeven_0(t)=\Zodd_0(t)=1$.

One way to find the above expansion is to make use of the
factorization relation \eqref{eq:RRsum},
which holds exactly at the perturbative level
\begin{equation}
\begin{aligned}
 \Zefull(t)=\sum_{l\in\mathbb{Z}}\psi^{\text{top}}(t+lg_s)
\psi^{\text{top}}(t-lg_s),
\end{aligned} 
\label{eq:pert-RR}
\end{equation}
and also the data of $\Ztop_n$ obtained in the last subsection.
Note that there is no distinction between $\psi^{\text{top}}$ and
$\b{\psi}^{\text{top}}$ when $\th=0$.
We can rewrite the relation \eqref{eq:pert-RR}
in terms of $\Zeven(t)$ and $\psi(t)$
by removing the genus-zero part.
By using the relation
\begin{equation}
\begin{aligned}
 e^{g_s^{-2}\bigl[F_0(t+lg_s)+F_0(t-lg_s)-\mathcal{F}_0(t)\bigr]}=Q^{l^2},
\end{aligned} 
\end{equation}
\eqref{eq:pert-RR} becomes
\begin{align}
\begin{aligned}
 \Zeven&=\sum_{l\in\mathbb{Z}}Q^{l^2}\psi(t-lg_s)\psi(t+lg_s)\\
&=\sum_{l\in\mathbb{Z}}Q^{l^2} \sum_{n,m=0}^\infty \frac{1+(-1)^{n+m}}{2}
D^n\psi D^m\psi\frac{(-1)^n(lg_s)^{n+m}}{n!m!}\\
&=\sum_{n,m=0}^\infty \frac{(-1)^n+(-1)^{m}}{2}
D^n\psi D^m\psi\frac{(\rt{D}g_s)^{n+m}}{n!m!}\Thetaeven\\
&=\cosh \Bigl[g_s(D^{(1)}-D^{(2)})\rt{D^{(3)}}\Bigr]\psi\cdot \psi\cdot \Thetaeven,
\end{aligned}
\label{eq:Zfull-pert}
\end{align}
where $D^{(i)}$ act on the $i$-th factor of
$\psi\cdot \psi\cdot \Thetaeven$.
As advertised, $\Thetaeven$ in \eqref{eq:defTh}
naturally arises from the 
sum over $U(1)$ charges \eqref{eq:pert-RR}.
We can further rewrite \eqref{eq:Zfull-pert}  by
performing the conjugation
with respect to the genus-one part
\begin{equation}
\begin{aligned}
 \h{\mathcal{Z}}&= \eta^2 \Thetaeven^{-1}
\cosh \Bigl[g_s(D^{(1)}-D^{(2)})\rt{D^{(3)}}\Bigr]\eta^{-1}\h{\psi}\cdot \eta^{-1}\h{\psi}
\cdot \Thetaeven\\
&=\cosh \Bigl[g_s(D^{(1)}_{-1}-D^{(2)}_{-1})\rt{D^{(3)}_\Thetaeven}\Bigr]\h{\psi}\cdot \h{\psi}
\cdot 1,
\end{aligned} 
\end{equation}
where $D_{-1}$ is defined in \eqref{eq:Dk}
and 
$D_{\Thetaeven}$ is given by
\begin{equation}
\begin{aligned}
 D_{\Thetaeven}:=\Thetaeven^{-1}D\Thetaeven
=D+(D\ln\Thetaeven)
=D-\frac{E_2(Q)}{12}+\frac{5E_2(Q^2)}{12}-\frac{E_2(Q^4)}{3}.
\end{aligned} 
\label{eq:DTe}
\end{equation}
Finally, the  coefficient $\Zeven_n$ in the $g_s$-expansion
of $\h{\mathcal{Z}}$ in
\eqref{eq:Zn-exp} is given by
\begin{align}
 \Zeven_n=\sum_{k+l+m=n}\frac{(D_{-1}^{(1)}-D_{-1}^{(2)})^{2k}}{(2k)!}
\Ztop_l\cdot\Ztop_m \cdot D^k_{\Thetaeven}1.
\label{eq:Zfull-genus}
\end{align}

We find that $\Zodd$ has a similar expansion as $\Zeven$
\begin{align}
 \begin{aligned}
  \Zodd&=\sum_{p\in\mathbb{Z}+\hf}
Q^{p^2}\psi(t-pg_s)\psi(t+pg_s)\\
&=\cosh \Bigl[g_s(D^{(1)}-D^{(2)})\rt{D^{(3)}}\Bigr]\psi\cdot \psi\cdot
\Thetaodd.
 \end{aligned}
\end{align}
The  $\mathcal{O}(g_s^{2n})$ term $\Zodd_n$ in the $g_s$-expansion
\eqref{eq:Zn-exp} is then given by
\begin{align}
  \Zodd_n=\sum_{k+l+m=n}\frac{(D_{-1}^{(1)}-D_{-1}^{(2)})^{2k}}{(2k)!}
\Ztop_l\cdot\Ztop_m\cdot \DTodd^k1,
\label{eq:Zodd-genus}
\end{align}
where $\DTodd$ is defined by
\begin{equation}
\begin{aligned}
  \DTodd:=\Thetaodd^{-1}D\Thetaodd
=D+(D\ln\Thetaodd)
=D-\frac{E_2(Q^2)}{12}+\frac{E_2(Q^4)}{3}.
\end{aligned} 
\label{eq:DTo}
\end{equation}

There is another way to find $\Zeven_n$ and $\Zodd_n$,
which is based on a set of recursion relations similar to
\eqref{eq:rec-1}.
This method also elucidates the modular properties
of $\Zeven_n$ and $\Zodd_n$.
To see this, let us first point out that
$\Zeven_n$ and $\Zodd_n$ have an interesting structure:
they are expressed as
\begin{align}
\label{ZeoinXY}
\Zeven_n=X_n+(D\ln\Thetaeven)Y_n,\qquad
\Zodd_n=X_n+(D\ln\Thetaodd)Y_n,
\end{align}
with $X_n,Y_n$ being quasi-modular forms
in $\tau=\frac{1}{2\pi\ri}\ln Q$
of weight $6n,\,6n-2$, respectively,
for $\Gamma=PSL(2,\bbZ)$.
We will prove this after deriving a set of recursion relations
for $X_n, Y_n$.
Note here that $D\ln\Thetaeven$ and $D\ln\Thetaodd$
are not quasi-modular forms for $\Gamma$,
but rather for the subgroup $\Gamma_0(4)$ of $\Gamma$.\footnote{
$\Gamma_0(4)$ is a subgroup of $\Gamma$
which consists of matrices of the form
$
\begin{pmatrix}
 a&b\\c&d
\end{pmatrix}\in\Gamma
$
with $c\equiv 0$ mod $4$.
It is generated by
$T=
\begin{pmatrix}
 1&1\\0&1
\end{pmatrix}
$
and
$ST^4S=
\begin{pmatrix}
 -1&0\\4&-1
\end{pmatrix}
$.}
This can be seen directly from their expression
appearing in \eqref{eq:DTe} and \eqref{eq:DTo}.
Alternatively, one can rewrite them as
\begin{align}
D\ln\Thetaeven
 =\frac{1}{24}\left[
 E_2(Q)-\Thetaeven^4+5\Thetaodd^4\right],\qquad
D\ln\Thetaodd
 =\frac{1}{24}\left[
 E_2(Q)-\Thetaodd^4+5\Thetaeven^4\right].
\end{align}
This expression clarifies the modular anomaly:
$E_2$ is a quasi-modular form (i.e.~anomalous) for $\Gamma$
while $\Thetaeven^4$ and $\Thetaodd^4$ are
modular forms (i.e.~non-anomalous) for $\Gamma_0(4)$.
All these are of weight two.
Consequently,
$\Zeven_n$ and $\Zodd_n$
are quasi-modular forms of weight $6n$
for $\Gamma_0(4)$.

Let us now derive the recursion relations for $X_n$ and $Y_n$.
We start with the relation between the generating function $\Xi$
in \eqref{eq:Xi-def} 
and the rescaled partition functions $\Zeven,\,\Zodd$ in \eqref{eq:Z-tZ}
\begin{align}
\Xi^2(q,Q,z)
&=\sum_{p\in\bbZ+\frac{1}{2}}
  \Zeven(t+p g_s)q^{\frac{p^3}{3}}Q^{p^2}z^{2p}
 -\sum_{n\in\bbZ}
  \Zodd(t+n g_s)q^{\frac{n^3}{3}}Q^{n^2}z^{2n}.
\end{align}
This is analogous to \eqref{eq:Xi-sum} and is derived
in the same way as \eqref{eq:Xi-sum} from the definitions of 
$\Xi,\,\Zeven,$ and $\Zodd$.
Recall that $\Xi$ is expanded in the chemical potential $\mu=-\log z$
as $\Xi=\mu\mathcal{K}+{\cal O}(\mu^2)$, so that
\begin{align}
\Xi^2\Big|_{\mu=0}=0,\qquad
\frac{\partial^2}{\partial\mu^2}\Xi^2\Big|_{\mu=0}=2\mathcal{K}^2,
\end{align}
where $\mathcal{K}$ is given by \eqref{eq:hat-xi}.
From these one obtains
\begin{align}
0&=
\sum_{p\in\bbZ+\frac{1}{2}}
  \Zeven(t+pg_s)q^{\frac{p^3}{3}}Q^{p^2}
 -\sum_{n\in\bbZ}
  \Zodd(t+ng_s)q^{\frac{n^3}{3}}Q^{n^2},\\
\frac{1}{2}\mathcal{K}^2&=
\sum_{p\in\bbZ+\frac{1}{2}}
  p^2\Zeven(t+pg_s)q^{\frac{p^3}{3}}Q^{p^2}
 -\sum_{n\in\bbZ}
  n^2\Zodd(t+ng_s)q^{\frac{n^3}{3}}Q^{n^2}.
\end{align}
These relations are rewritten as
\begin{align}
\label{ZeoRecExp1}
0&=
 \sum_{k,l,m\ge 0}\frac{g_s^{2k+l+m}}{l!m!3^m}
 \frac{1+(-1)^{l+m}}{2}\left(
  D^l\frac{\Thetaeven\Zeven_k}{\eta^2}
  D^{\frac{l+3m}{2}}\Thetaodd
 -D^l\frac{\Thetaodd\Zodd_k}{\eta^2}
  D^{\frac{l+3m}{2}}\Thetaeven
 \right),\\
\frac{\eta^4}{2}\h{\mathcal{K}}^2
\label{ZeoRecExp2}
&=
 \sum_{k,l,m\ge 0}\frac{g_s^{2k+l+m}}{l!m!3^m}
 \frac{1+(-1)^{l+m}}{2}\left(
  D^l\frac{\Thetaeven\Zeven_k}{\eta^2}
  D^{\frac{l+3m+2}{2}}\Thetaodd
 -D^l\frac{\Thetaodd\Zodd_k}{\eta^2}
  D^{\frac{l+3m+2}{2}}\Thetaeven
 \right),
\end{align}
where $\h{\mathcal{K}}$ is defined in \eqref{eq:hxi-def}.
Let us now plug (\ref{ZeoinXY}) into the above relations
and compare the ${\cal O}(g_s^{2n})$ parts.
After a bit of algebra, one obtains
\begin{align}
\label{RecRelX}
X_n &= \heven_n 
 -2\sum_{j=1}^n\ \sum_{\substack{k+l+m=2j\\ k,l,m\ge 0}}
 \frac{1}{k!l!m!3^m}
 \left(a_{k,m+j+1}D_{-2}^l X_{n-j}
      +a_{k+1,m+j+1}D_{-2}^l Y_{n-j}\right),\\
\label{RecRelY}
Y_n &=
  2\sum_{j=1}^n\ \sum_{\substack{k+l+m=2j\\ k,l,m\ge 0}}
 \frac{1}{k!l!m!3^m}
 \left(a_{k,m+j}D_{-2}^l X_{n-j}
      +a_{k+1,m+j}D_{-2}^l Y_{n-j}\right),
\end{align}
where we have introduced
\begin{align}
a_{i,j}:=
 \frac{D^i\Thetaeven D^j\Thetaodd-D^j\Thetaeven D^i\Thetaodd}{\eta^6},
\end{align}
and $\heven_n$
in \eqref{RecRelX} is obtained from $e_l$ in \eqref{eq:el} by the relation
\begin{align}
\label{GenRelheven}
 \h{\mathcal{K}}^2=:\sum_{n=0}^\infty \heven_n g_s^{2n}
=\exp\left(2\sum_{l=1}^\infty \frac{e_l}{(2l)!}g_s^{2l}\right).
\end{align}
The above relations, together with the initial data
\begin{align}
X_0=1,\qquad Y_0=0,
\end{align}
determine $X_n,Y_n$ recursively.

We are now in a position to prove that $X_n,Y_n$ are quasi-modular forms
for $\Gamma$.
First, it is obvious from the relation (\ref{GenRelheven})
that $\heven_n$ is a quasi-modular form of weight $6n$. 
Next, notice that $a_{i,j}$ can also be obtained from the relation
\begin{align}
\begin{aligned}
&\hspace{-2em}
\lefteqn{
 \sum_{i,j\ge 0}\frac{(-1)^{i+j}(2x)^{2i}(2y)^{2j}}{(2i)!(2j)!}a_{i,j}
        }\\
 &=\frac{\varth_3(e^{2\ri x},Q^2)\varth_2(e^{2\ri y},Q^2)
        -\varth_2(e^{2\ri x},Q^2)\varth_3(e^{2\ri y},Q^2)}{\eta(Q)^6}\\
 &=\frac{\varth_1(e^{\ri(x+y)},Q)\varth_1(e^{\ri(x-y)},Q)}{\eta(Q)^6}\\
 &=(x^2-y^2)\exp\left[\sum_{k=1}^\infty
  \frac{(-1)^kB_{2k}}{2k(2k)!}E_{2k}(Q)
  \left((x+y)^{2k}+(x-y)^{2k}\right)\right].
\end{aligned}
\end{align}
From this it is clear that $a_{i,j}$ is a quasi-modular form
of weight $2i+2j-2$.
Third, if $A_n$ is a quasi-modular form of weight $n$,
$D_{-2}A_n=(D-\frac{1}{12}E_2)A_n$ is a quasi-modular form
of weight $n+2$.
Hence, the recursion relations (\ref{RecRelX}), (\ref{RecRelY})
ensure that $X_n,Y_n$ are quasi-modular forms of weight $6n,6n-2$,
respectively.
First few $\heven_n$ and $a_{i,j}$ read
\begin{align}
h_0&=1,&
h_1&=\frac{-E_2E_4+E_6}{1440},\quad
h_2=\frac{50E_2^3E_6-147E_2^2E_4^2+144E_2E_4E_6-25E_4^3-22E_6^2}{12441600},\\
a_{0,1}&=\frac{1}{2},&
a_{0,2}&=\frac{E_2}{8},\quad
a_{0,3}=\frac{5E_2^2-E_4}{128},\quad
a_{1,2}=\frac{E_2^2-E_4}{384}.
\end{align}
Note that by definition $a_{i,j}=-a_{j,i}$.
Then, first few of $X_n,Y_n$ are obtained as
\begin{align}
X_1&=\frac{1}{2^6\cdot 3^4\cdot 5}
 \left(5E_2^3-3E_2E_4-2E_6\right),\nn\\
X_2&=\frac{1}{2^{17}\cdot 3^8\cdot 5^2}
 \left(-6125E_2^6+10095E_2^4E_4+15280E_2^3E_6\right.\nn\\
 &\hspace{6.5em}\left.
  -12231E_2^2E_4^2-25008E_2E_4E_6+13125E_4^3+4864E_6^2\right),\\
Y_1&=\frac{1}{288}\left(-E_2^2+E_4\right),\nn\\
Y_2&=\frac{1}{2^{13}\cdot 3^6\cdot 5}
 \left(175E_2^5-478E_2^3E_4-232E_2^2E_6+1023E_2E_4^2-488E_4E_6\right).
\quad
\end{align}
We have computed $X_n$ and $Y_n$ up to $n=60$.
Plugging these into \eqref{ZeoinXY}
one obtains corresponding $\Zeven_n$ and $\Zodd_n$.
Once we know $\Zeven_n$ and $\Zodd_n$, the genus $g$ free energies 
$\Feven_g$ and $\Fodd_g$ are obtained
from the relation similar to \eqref{eq:ZtoF}.

\section{Large order behavior \label{sec:large}}

In this section we will study the large order behavior
of the genus expansion coefficients
$\Ztop_n$, $\Zeven_n$ and $\Zodd_n$.
According to the theory of resurgence, non-perturbative corrections
are encoded in the large order behavior of the perturbative series.
This means that one can
``decode'' the non-perturbative effects from the information 
of perturbative computation alone.
We will first perform this analysis using the exact forms
of $\Ztop_n$, $\Zeven_n$ and $\Zodd_n$ up to $n=60$
obtained in the last section.
On the other hand, by adopting a certain analytic continuation
we have obtained in section~\ref{sec:gen}
the all-order instanton corrections to the perturbative partition
functions $\psi^{\text{top}}$, $\Zefull$ and $\Zofull$.
Based on these results,
it is in fact possible to derive analytically
the large order behavior of $\Ztop_n$, $\Zeven_n$ and $\Zodd_n$.
We will also do this and make a comparison
with the results of the former analysis.

Let us first consider the large order behavior of $\Ztop_n$
studied in section~\ref{subsec:genustop}.\footnote{
In this paper we study the large order behavior of
$\Ztop_n$ rather than that of the free energy $F_g$,
simply because the analysis of the former is simpler.
One could study the latter in the same way.}
Following \cite{Marino:2007te},
we write the partition function
with 1-instanton contribution as
\begin{align}
\psi^{\text{top}}(t)\pm \hf \psi^{\text{top}}_{\text{1-inst}}(t).
\label{eq:psi+inst}
\end{align}
For a genus expansion of closed string theory,
it is expected that the 1-instanton correction
takes the form
\begin{align}
 \h{\psi}_{\text{1-inst}}(t)
\equiv\frac{\psi^{\text{top}}_{\text{1-inst}}(t)}{\psi_{01}(t)}
=\pi\ri g_s^{-\btop}\mu(t)e^{-\frac{\Atop(t)}{g_s}}\sum_{n=0}^\infty f_n(t)g_s^n.
\label{eq:Z1-cn}
\end{align}
We can set $f_0(t)=1$ without loss of generality.
Here we have removed the contribution of
genus-zero and genus-one pieces $\psi_{01}(t)$
in \eqref{eq:def-psi01} since we are considering the
asymptotic behavior of $\Ztop_n$ in the $g_s$-expansion of
$\h{\psi}$
in \eqref{eq:Zn-expand}.
As argued in \cite{Marino:2007te}, 1-instanton correction is encoded in
the large order behavior of the perturbative part
\begin{align}
\begin{aligned}
 \Ztop_m(t)
&\sim\frac{1}{2\pi\ri}\int_0^\infty \frac{dz}{z^{m+1}}z^{-\btop/2}
 \pi \ri \mu(t)e^{-\frac{\Atop(t)}{\rt{z}}}\sum_{n=0}^\infty
 f_n(t)z^{n/2}\\
&=\mutop(t) \Atop(t)^{-2m-\btop}\Ga(2m+\btop)\sum_{n=0}^\infty
  \ftop_n(t) \Atop(t)^n\frac{\Ga(2m+\btop-n)}{\Ga(2m+\btop)}.
\end{aligned}
\label{eq:Ztop-asymp}
\end{align}

Following the procedure in \cite{Marino:2007te},
one can extract $\btop,\, \Atop(t),\,\mutop(t),\,\ftop_n(t)$
by constructing some sequence.
In the first step, we consider the following sequence
\begin{align}
 \Atop_m(t):=2m\sqrt{\frac{\Ztop_m(t)}{\Ztop_{m+1}(t)}},\quad (m=1,2,\ldots).
\label{eq:A-seq}
\end{align}
From the asymptotic behavior of $\Ztop_m(t)$ in \eqref{eq:Ztop-asymp},
one can see that $\Atop_m(t)$ approaches $\Atop(t)$ as $m$ increases
\begin{equation}
\begin{aligned}
 \Atop_m(t)=A(t)+\mathcal{O}(m^{-1}),
\end{aligned} 
\end{equation}
and $A(t)$ can be determined from
the large $m$ behavior of $\Atop_m(t)$.
Once we obtain $\Atop(t)$, we next define
the sequence
\begin{align}
 \btop_m(t):=
 m\left(\Atop(t)^2\frac{\Ztop_{m+1}(t)}{4m^2\Ztop_m(t)}-1\right)
 -\frac{1}{2}=b+\mathcal{O}(m^{-1}),
\label{eq:b-seq}
\end{align}
from which we obtain the constant $\btop$.
Then one can extract $\mutop(t)$ 
from the sequence
\begin{align}
\mu_m(t):=\frac{\Atop(t)^{2m+\btop}\Ztop_m(t)}{\Gamma(2m+\btop)}
 &=\mutop(t)+{\cal O}(m^{-1}).
\end{align}
In the same way,
one can  extract $\ftop_n(t)$ by successively defining some sequence.
More specifically, given the forms of $\btop,\, \Atop(t),\,\mutop(t)$
and $\ftop_k(t)$ with $k<n$, one can extract $\ftop_n(t)$
from the sequence
\begin{equation}
\begin{aligned}
f_{n,m}(t):&=\frac{\Atop(t)^{2m+b-n}\Ztop_m(t)}{\Gamma(2m+b-n)\mutop(t)}
-\sum_{k=0}^{n-1}\frac{\Gamma(2m+b-k)}{\Gamma(2m+b-n)}
                 \ftop_k(t)\Atop(t)^{k-n}\\
&=f_n(t)+{\cal O}(m^{-1}). 
\end{aligned} 
\label{eq:muf-seq}
\end{equation}

In the numerical study of the asymptotic behavior of 
a sequence, such
as $A_m(t)$ in \eqref{eq:A-seq},
one can use the 
standard technique
of Richardson extrapolation
which accelerates the convergence of 
sequence towards the leading asymptotics. 
Given a sequence $\{S_m\}_{m=1,2,\ldots}$
\begin{align}
S_m=s_0+\frac{s_1}{m}+\frac{s_2}{m^2}+\cdots,\qquad
\lim_{m\to\infty}S_m=s_0,
\end{align}
its $k$-th Richardson transform is defined as
\begin{align}
S_m^{(k)}:=\sum_{n=0}^k\frac{(-1)^{k+n}(m+n)^nS_{m+n}}{n!(k-n)!}.
\label{eq:Rich}
\end{align}
After this transformation
the subleading terms in $S_m$ are canceled up to $m^{-k}$,
i.e.~$S_m^{(k)}=s_0+{\cal O}(m^{-k-1})$ and hence
the sequence $S_m^{(k)}$ has a much faster convergence to $s_0$.
However, in exchange for a faster convergence we lose some data 
in this transformation:
if we know the original sequence $S_m$ up to $m=m_{\text{max}}$,
the data of $k$-th Richardson transform $S_m^{(k)}$
in \eqref{eq:Rich} 
are available only up to $m=m_{\text{max}}-k$.

By the above described method with the data of $\Ztop_m\ (m\le 60)$,
we find
\begin{align}
 \Atop(t)=\frac{t^2}{2},\quad \btop=\hf,\quad
 \mutop(t)=\rt{\frac{2}{\pi}}e^{-\frac{t}{2}}.
\label{eq:Abmutop}
\end{align}
\begin{figure}[tb]
\begin{center}
\includegraphics[width=5cm]{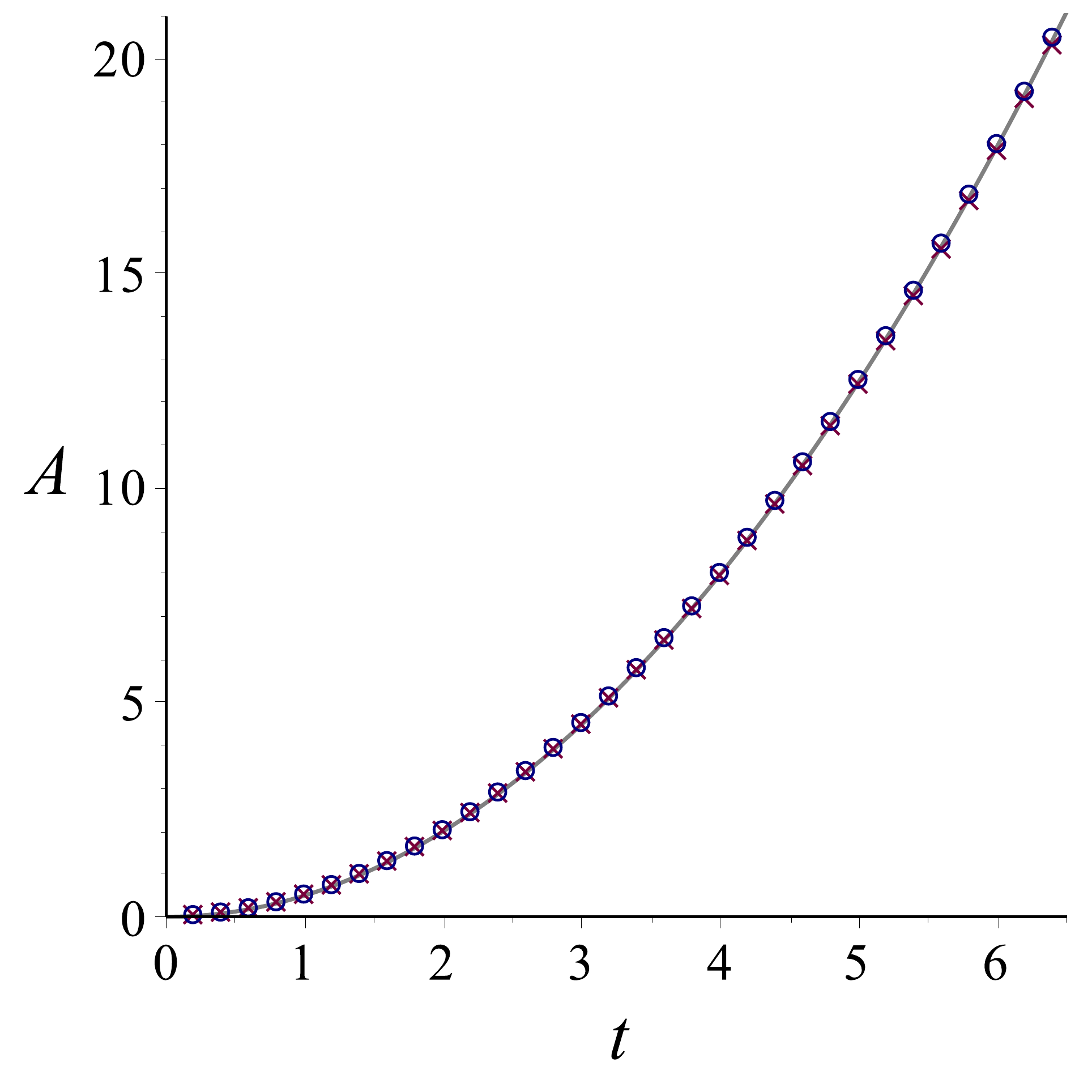}
\includegraphics[width=5cm]{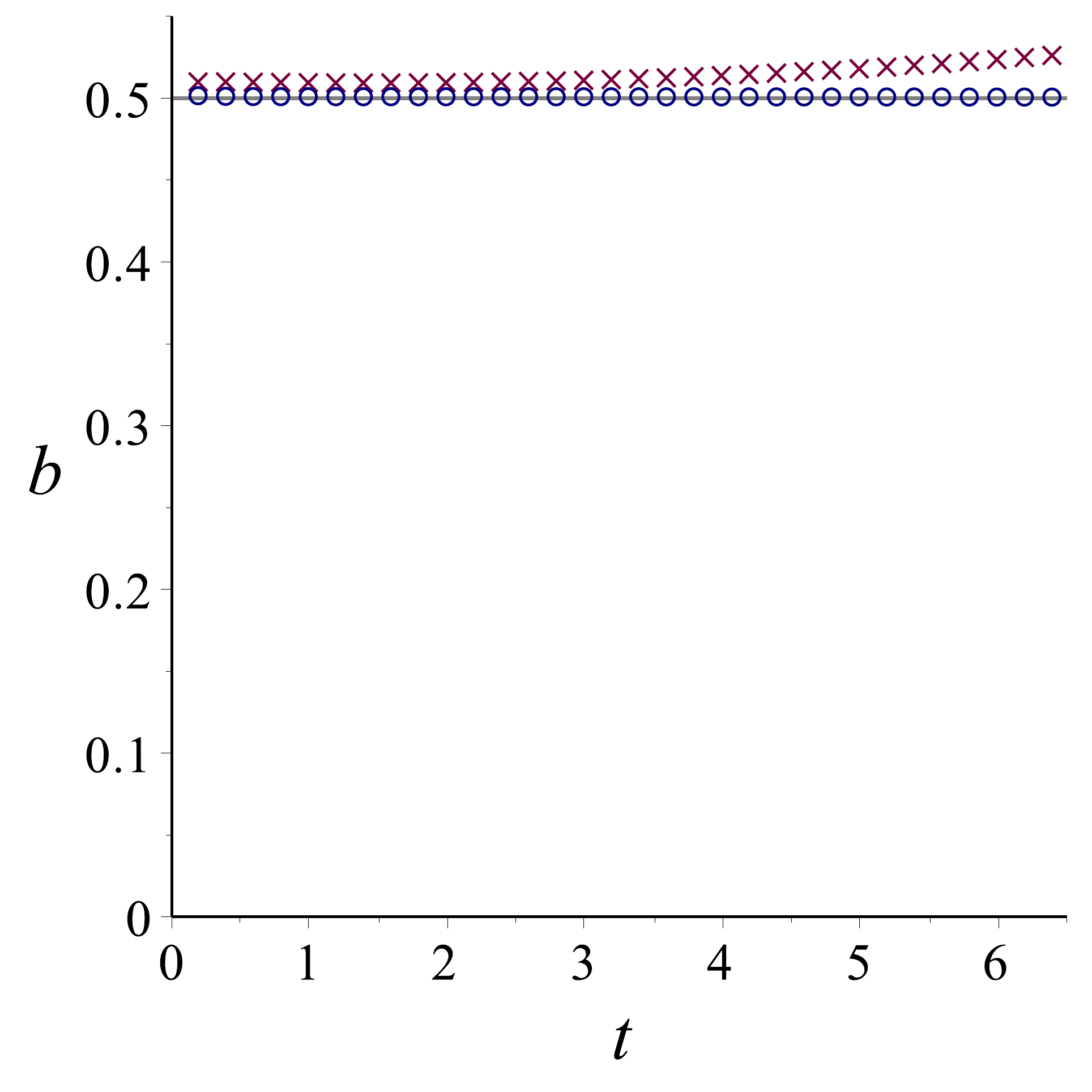}
\includegraphics[width=5cm]{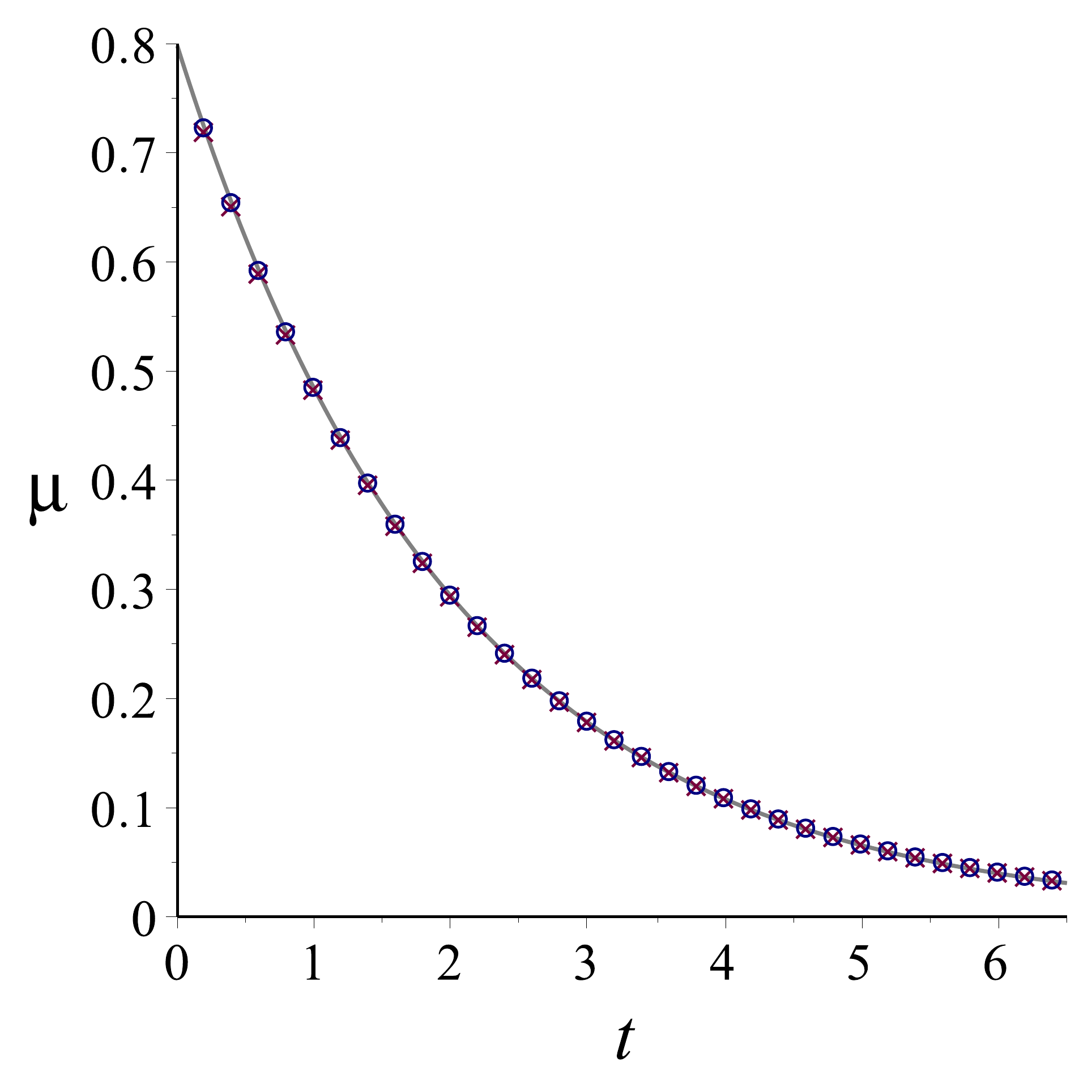}
\end{center}
\vspace{-2ex}
\caption{Numerical estimations of $\Atop(t)$, $\btop$, $\mutop(t)$.
Red diagonal crosses represent
$\Atop_{59}(t)$, $\btop_{59}(t)$, $\mutop_{60}(t)$,
while blue circles represent the first Richardson transforms
$\Atop_{58}^{(1)}(t)$, $\btop_{58}^{(1)}(t)$, $\mutop_{59}^{(1)}(t)$.
Gray solid lines are the plots of analytic expressions
\eqref{eq:Abmutop}.}
\label{fig:curveAbmu}
\end{figure}
As shown in Figure~\ref{fig:curveAbmu},
the data of $\Atop_{59}(t)$ and $\mutop_{60}(t)$
are already accurate enough to estimate the analytic forms.
The value of $\btop$ is also easily determined
by the first Richardson transforms of $\btop_m(t)$.
\begin{figure}[tb]
\vspace{2ex}
\begin{center}
\includegraphics[width=5cm]{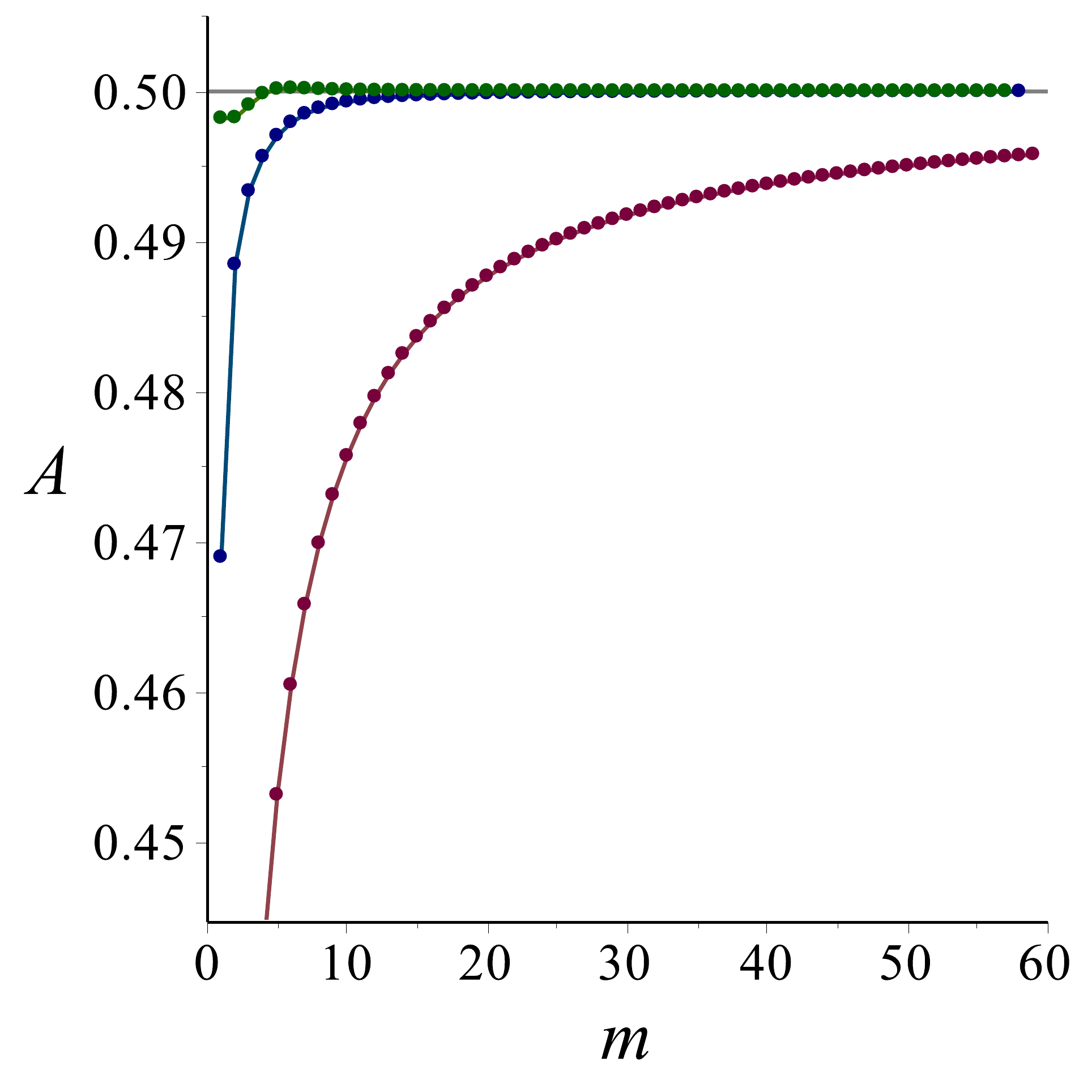}
\includegraphics[width=5cm]{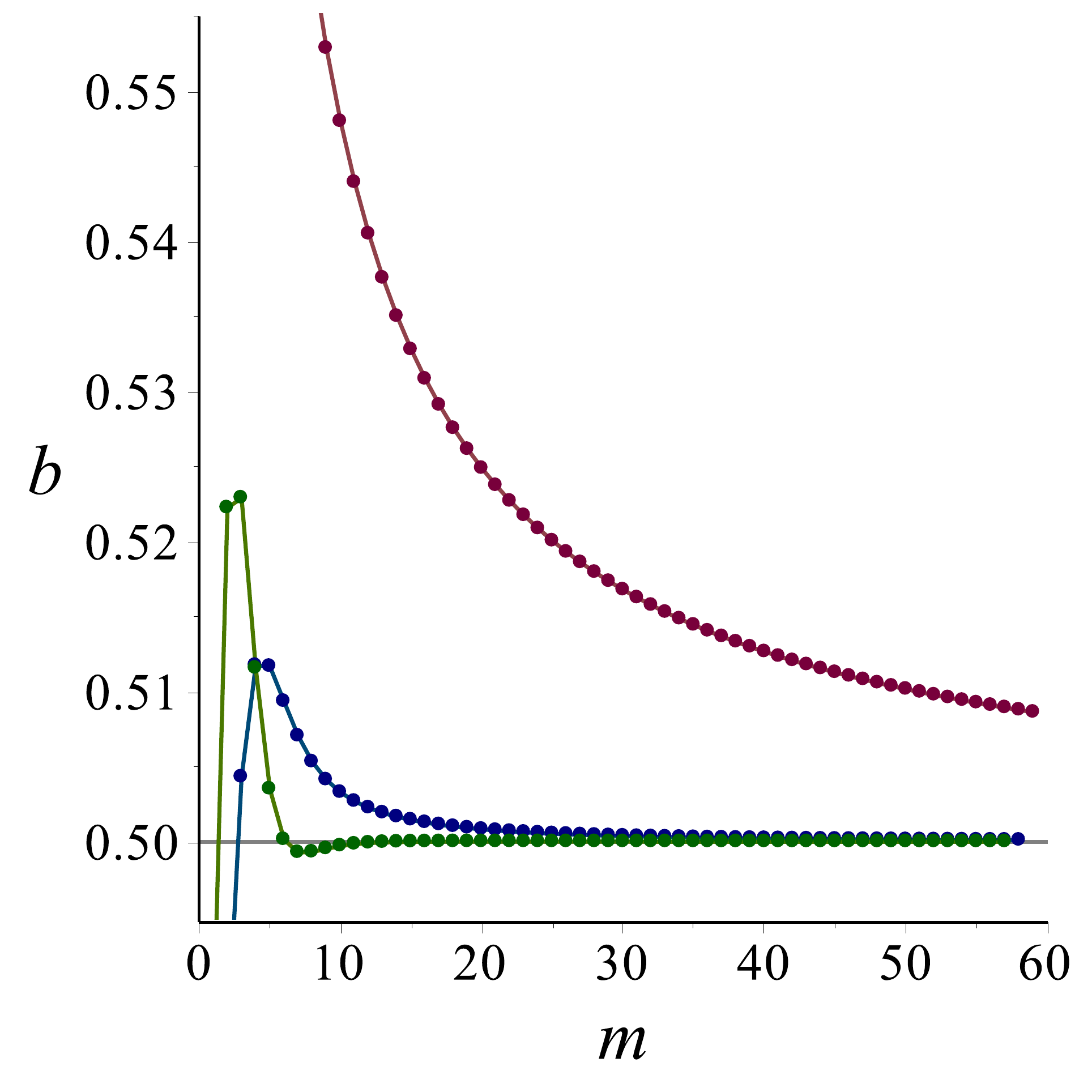}
\includegraphics[width=5cm]{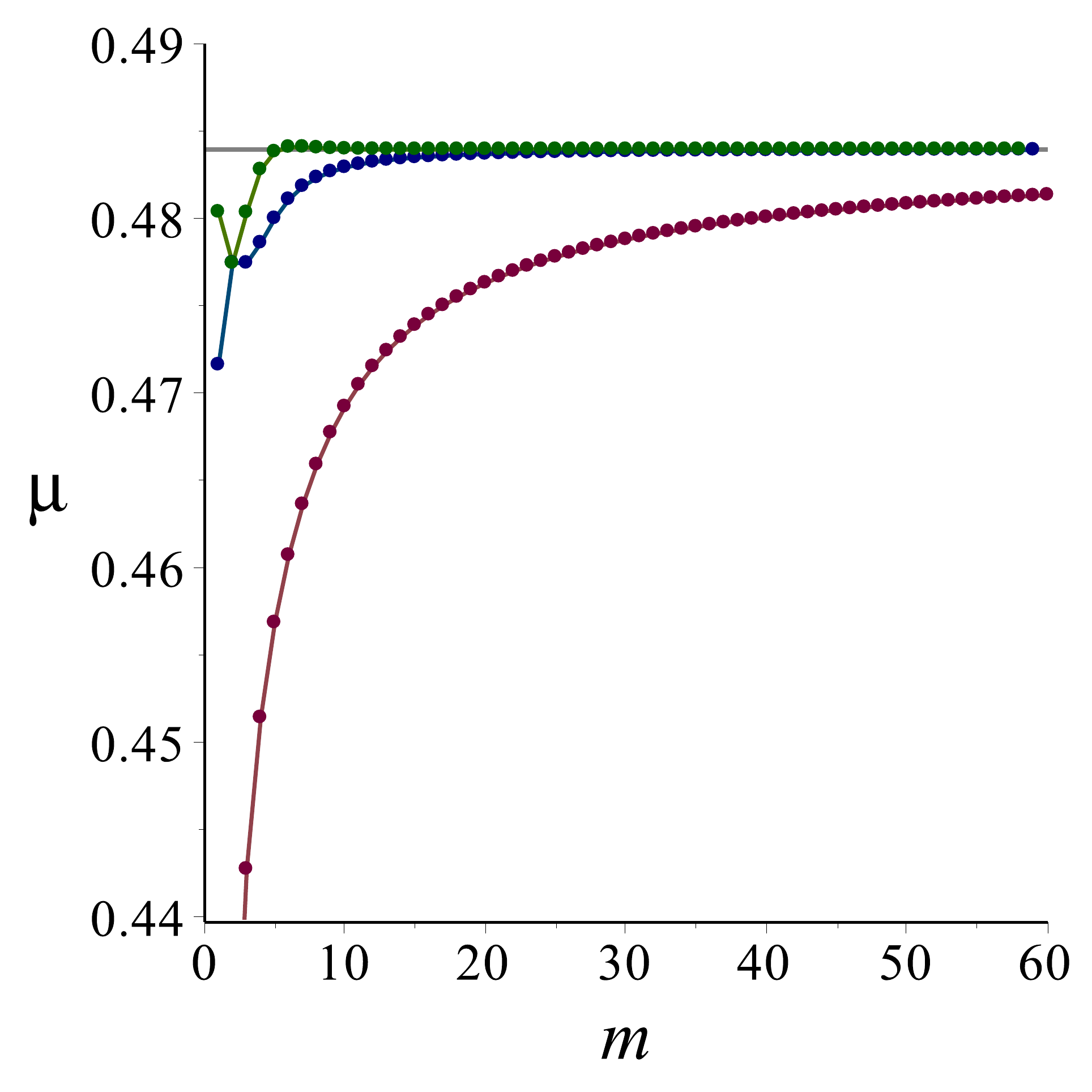}
\end{center}
\vspace{-2ex}
\caption{Plots of sequences
$\Atop_m,\,\btop_m,\,\mutop_m$ at $t=1$.
Red points represent the original sequences,
while blue and green points respectively represent
their first and second Richardson transforms.
The analytic values of $\Atop(1),\,\btop,\,\mutop(1)$ are
expressed by gray solid lines.}
\label{fig:fixedAbmu}
\end{figure}
Figure \ref{fig:fixedAbmu} shows the plots
of sequences $\Atop_m,\,\btop_m,\,\mutop_m$
and their Richardson transforms
at fixed $t$ (we set $t=1$).
As one can see, the Richardson transform drastically
improves the convergence of the sequence.
In fact, we have computed the tenth Richardson transforms
$\Atop_{49}^{(10)}(t)$, $\btop_{49}^{(10)}(t)$,
$\mutop_{50}^{(10)}(t)$ for $t=1,2,3,4,5,6$
and verified that their deviations
from the analytic forms \eqref{eq:Abmutop}
are within $\pm 10^{-10}\%$.

Since $A(t)$ in \eqref{eq:Abmutop} is positive for $t>0$, the asymptotic behavior
of $\Ztop_m(t)$ in \eqref{eq:Ztop-asymp} is non-alternating, i.e. there is no
alternating sign $(-1)^m$ in the large $m$ behavior of $\Ztop_m(t)$.
It follows that the genus expansion of $\psi^{\text{top}}(t)$ is not Borel summable
and the instanton action $A(t)$ appears as a pole on the positive real axis of
the Borel plane.
We can avoid the pole by the so-called the lateral Borel resummation, which will be studied
in the next section.

In section~\ref{sec:gen}
we have obtained
the all-order instanton corrections to 
the perturbative partition function $\psi^{\text{top}}(t)$
by adopting a certain analytic continuation.
That is, we actually know the 1-instanton amplitude
and hence from this we can derive the exact forms of
$\btop,\, \Atop(t),\,\mutop(t),\,\ftop_n(t)$.
Comparing \eqref{eq:psi+inst} with \eqref{eq:psiN-acon} we expect that
1-instanton correction for $\psi^{\text{top}}(t)$ is given by
\begin{align}
 \psi^{\text{top}}_{\text{1-inst}}(t)
 =\ri\rt{\frac{2\pi}{g_s}}\psi^{\text{top}}(t+g_s),
\end{align}
where we have evaluated $\varth_2(q)$ at the leading order
in the small $g_s$ expansion,
discarding the non-perturbative $\mathcal{O}(e^{-2\pi^2/g_s})$
terms
\begin{align}
\varth_2(q)
&=\sqrt{\frac{2\pi}{g_s}}\varth_4\left(e^{-4\pi^2/g_s}\right)
 =\sqrt{\frac{2\pi}{g_s}}\left(1-2e^{-2\pi^2/g_s}+\cdots\right)
 \approx\sqrt{\frac{2\pi}{g_s}}.
\label{eq:th2-approx}
\end{align}
Then $\h{\psi}_{\text{1-inst}}(t)$
in \eqref{eq:Z1-cn} becomes
\begin{equation}
\begin{aligned}
  \h{\psi}_{\text{1-inst}}(t)
=\ri\rt{\frac{2\pi}{g_s}}\frac{\psi^{\text{top}}(t+g_s)}{\psi_{01}(t)}
=\ri\rt{\frac{2\pi}{g_s}}\frac{\psi_{01}(t+g_s)}{\psi_{01}(t)}
  \h{\psi}(t+g_s).
\end{aligned} 
\label{eq:psi-1inst}
\end{equation}
Let us take a closer look at the ratio of $\psi_{01}$
in \eqref{eq:psi-1inst}. For the genus-zero part,
using the expression of 
$F_0(t)=-t^3/6$ in \eqref{eq:free-01} we find
\begin{equation}
\begin{aligned}
 \exp\left[\frac{F_0(t+g_s)-F_0(t)}{g_s^2}\right]
=\exp\left[-\frac{t^2}{2g_s}-\frac{t}{2}-\frac{g_s}{6}\right].
\end{aligned} 
\end{equation}
One can already see the appearance of the instanton factor
$e^{-A(t)/g_s}$ with the instanton action
$A(t)=t^2/2$ obtained numerically in \eqref{eq:Abmutop}. 
Including the contribution of genus-one part $F_1(t)$
in \eqref{eq:free-01}, the ratio of $\psi_{01}$
becomes\footnote{By abusing notation
here we let $\eta(t)$ denote
$\eta\left(Q=e^{-t}\right)
 =Q^{1/24}\prod_{n=1}^\infty(1-Q^n)\big|_{Q=e^{-t}}$.}
\begin{equation}
\begin{aligned}
 \frac{\psi_{01}(t+g_s)}{\psi_{01}(t)}=
e^{-\frac{t^2}{2g_s}-\frac{t}{2}-\frac{g_s}{6}}
\frac{\eta(t)}{\eta(t+g_s)}
=e^{-\frac{t^2}{2g_s}-\frac{t}{2}}
 \frac{\eta_{\frac{1}{6}}(t)}{\eta_{\frac{1}{6}}(t+g_s)},
\end{aligned} 
\end{equation}
where we have introduced $\eta_{\al}(t)$ by
\begin{align}
 \eta_{\al}(t):= e^{\al t}\eta(t).
\label{eq:eta-al}
\end{align}
Finally, plugging the expansion \eqref{eq:Zn-expand}
into the last factor $\h{\psi}(t+g_s)$ of \eqref{eq:psi-1inst},
$\h{\psi}_{\text{1-inst}}(t)$ becomes
\begin{align}
\begin{aligned}
 \h{\psi}_{\text{1-inst}}(t)
&=\ri\rt{\frac{2\pi}{g_s}}
  e^{-\frac{t^2}{2g_s}-\frac{t}{2}}
  \frac{\eta_{\frac{1}{6}}(t)}{\eta_{\frac{1}{6}}(t+g_s)}\sum_{n=0}^\infty 
\Ztop_n(t+g_s)g_s^{2n}.
\end{aligned}
\label{eq:Z1-anal}
\end{align}
Comparing \eqref{eq:Z1-cn} and \eqref{eq:Z1-anal},
one can indeed derive analytically the explicit forms of
$\btop$, $\Atop(t)$ and $\mutop(t)$
that we have previously estimated 
numerically in \eqref{eq:Abmutop}!
Moreover, the analytic form of $\ftop_n(t)$ is found from
\begin{align}
 \sum_{n=0}^\infty \ftop_n(t) g_s^n=\frac{\eta_{\frac{1}{6}}(t)}{\eta_{\frac{1}{6}}(t+g_s)}
\sum_{\ell=0}^\infty \Ztop_\ell(t+g_s)g_s^{2\ell}.
\end{align}
Using the relation
\begin{align}
 \frac{\eta_{\frac{1}{6}}(t)}{\eta_{\frac{1}{6}}(t+g_s)}
=\sum_{m=0}^\infty \frac{(-g_s)^m}{m!}\Big(D_{-1,\frac{1}{6}}\Big)^m 1
\end{align}
with
\begin{equation}
\begin{aligned}
D_{k,\al}=\eta_{\al}(t)^{-k}D\eta_{\al}(t)^{k}=
D+\frac{kE_2}{24}-k\al, 
\end{aligned} 
\label{eq:D-kal}
\end{equation}
we find that $\ftop_n$ is written as
\begin{align}
 \ftop_n=\sum_{m+k+2\ell=n}
  \frac{(-1)^{m+k}}{m!k!}\Big(D_{-1,\frac{1}{6}}\Big)^m 1\cdot D^k\Ztop_{\ell}.
\label{eq:ftop}
\end{align}
Namely, the fluctuation coefficient $f_n$
around the 1-instanton factor $e^{-A(t)/g_s}$
in \eqref{eq:Z1-cn} is completely determined by
the information of perturbative part $\Ztop_n$.
From \eqref{eq:ftop}, one can easily
compute the analytic form of $f_n$
and the first few terms read
\begin{align}
\begin{aligned}
 \ftop_1&=-\frac{1}{6}+\frac{E_2}{24},\\
 \ftop_2&=
 \frac{720-360E_2-45E_2^2+90E_4+5E_2^3-3E_2E_4-2E_6}{51840}.
\end{aligned}
\label{eq:ftop-ex}
\end{align}

\begin{figure}[tb]
\begin{center}
\includegraphics[width=5cm]{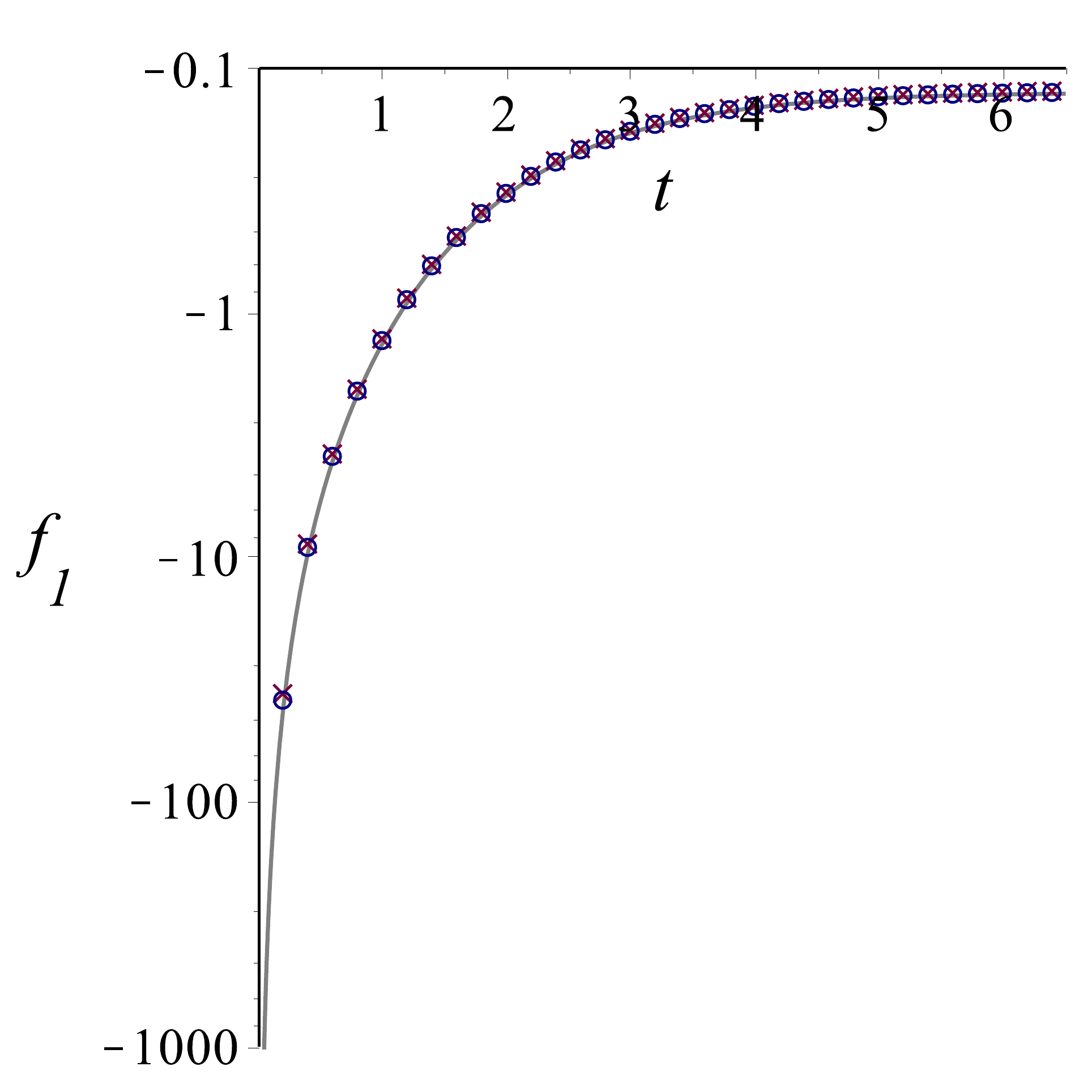}
\includegraphics[width=5cm]{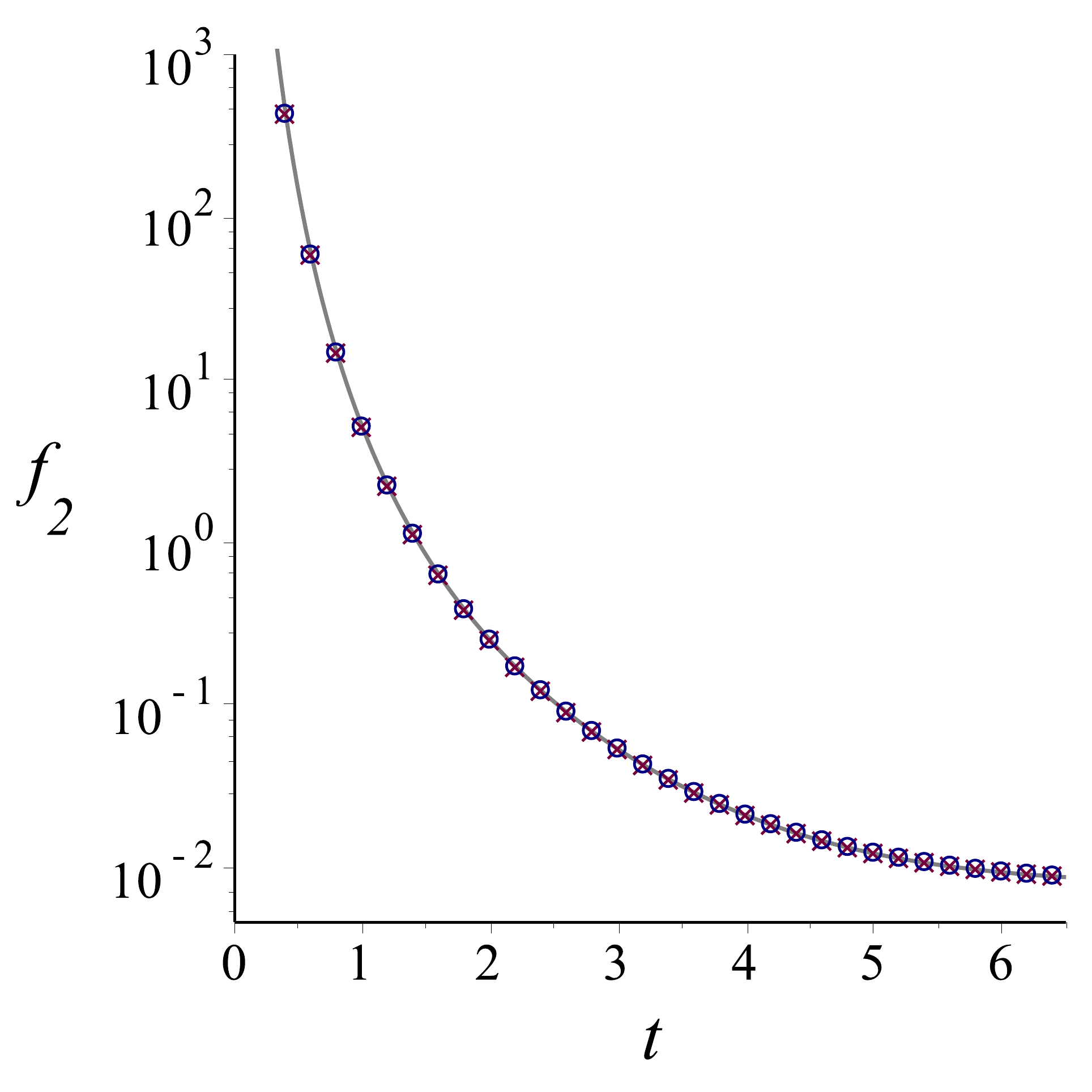}
\includegraphics[width=5cm]{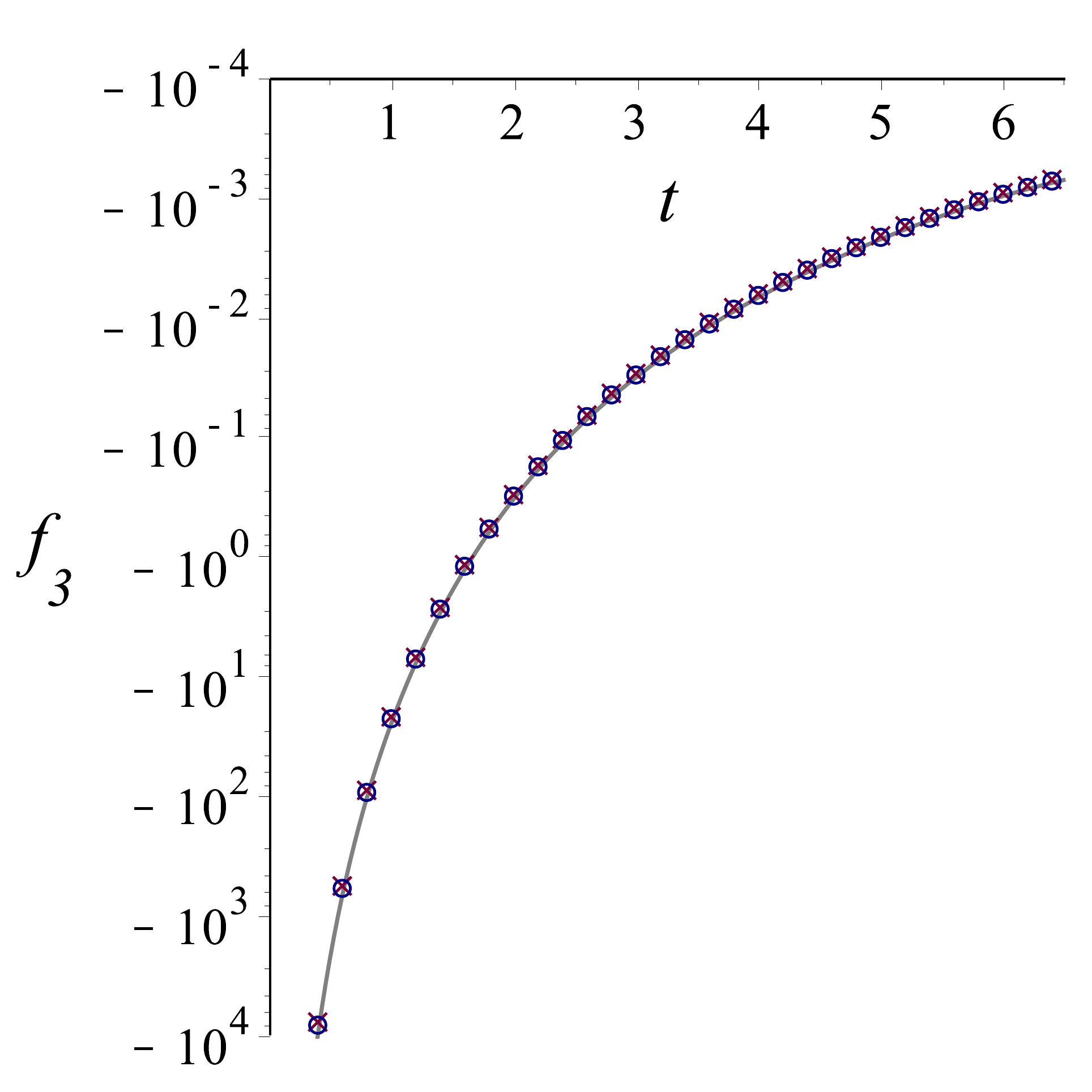}
\end{center}
\vspace{-2ex}
\caption{Asymptotic sequences
versus analytic expressions of $f_n(t),\ n=1,2,3$.
Red diagonal crosses represent $f_{n,60}(t)$ \eqref{eq:muf-seq},
while blue circles represent the first Richardson transforms
$f_{n,59}^{(1)}(t)$.
Gray solid lines are
the plots of analytic expressions \eqref{eq:ftop}.}
\label{fig:curvefn}
\end{figure}

\begin{figure}[tb]
\begin{center}
\includegraphics[width=5cm]{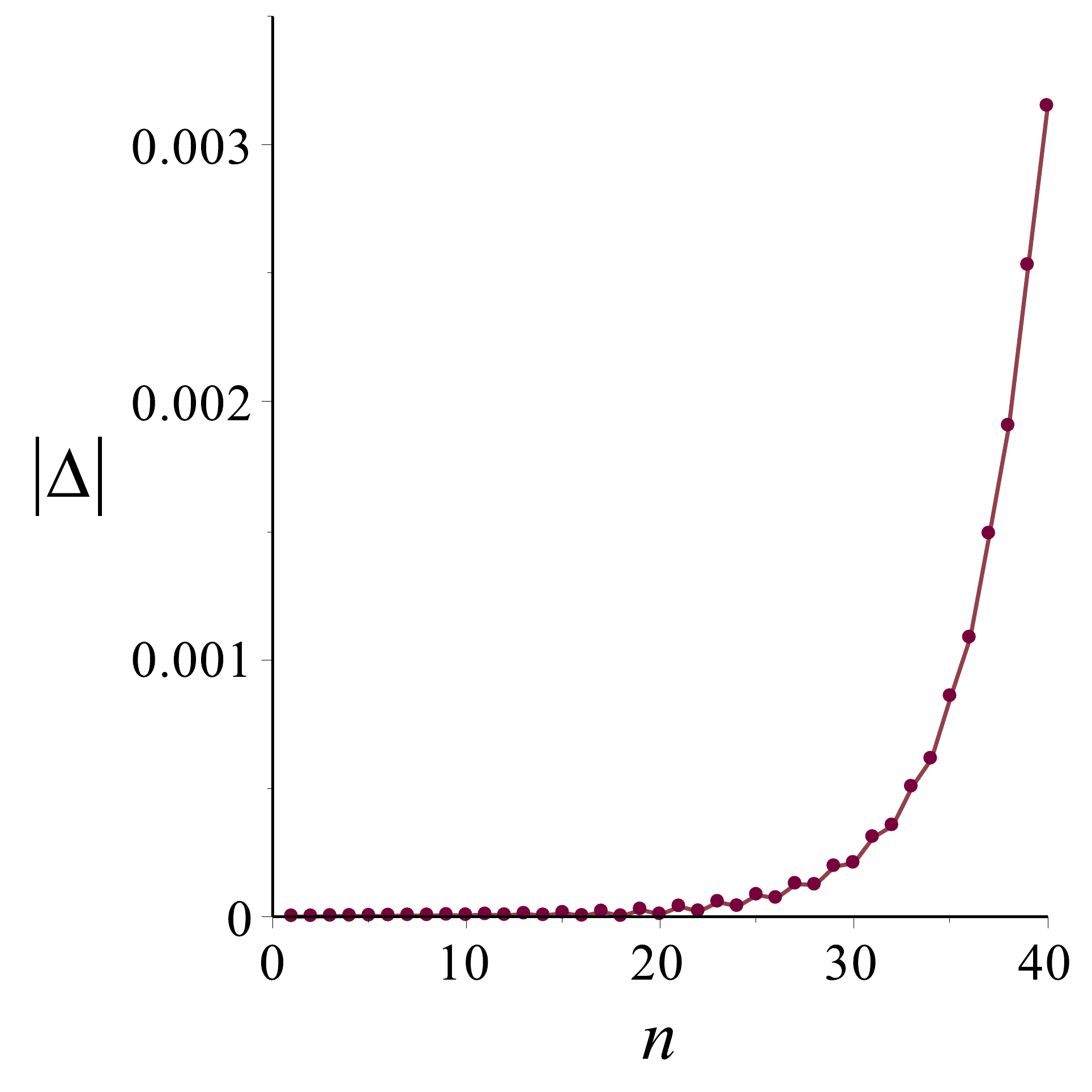}
\hspace{1cm}
\includegraphics[width=5cm]{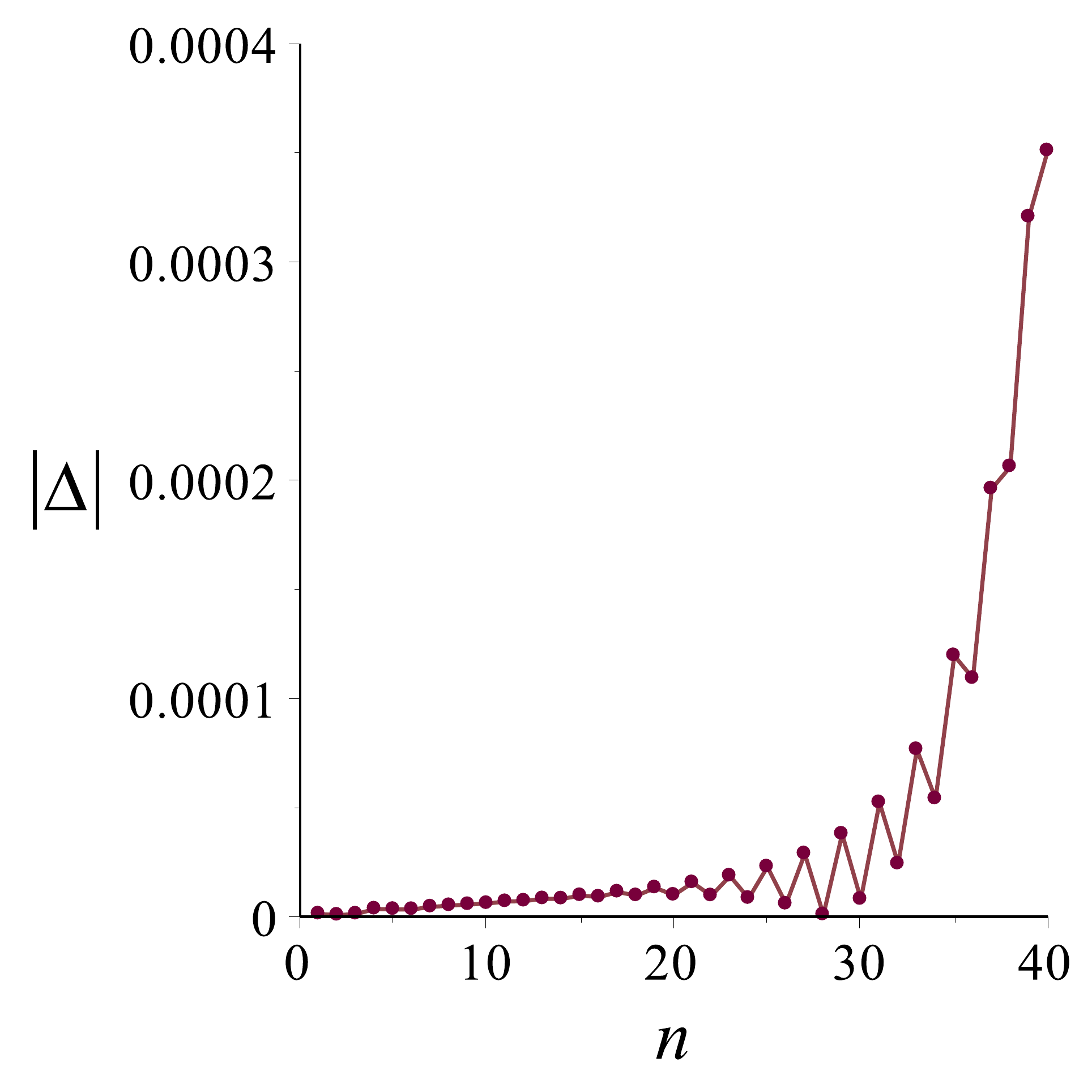}
\end{center}
\vspace{-2ex}
\caption{Relative deviations of $f_{n,m}^{(k)}(t)$ from $f_n(t)$
at $t=4$ (left) and $t=5$ (right).
Here we consider
the second Richardson transforms $k=2,\,m=58$ in \eqref{eq:dev-fn}.}
\label{fig:fdev}
\end{figure}

We have numerically verified
the above obtained exact forms of $\ftop_n$
against the sequences $\ftop_{n,m}$ in \eqref{eq:muf-seq}
based on the data of $\Ztop_m$.
As one can see in Figure~\ref{fig:curvefn},
the analytic expressions of $f_n(t)\ (n=1,2,3)$
are in good agreement with the
asymptotic sequences $f_{n,60}(t)$
and their first Richardson transforms.
In Figure~\ref{fig:fdev}, we plot the absolute value of
the relative deviation
\begin{align}
\Delta := \frac{f_{n,m}^{(k)}(t)-f_n(t)}{f_n(t)}
\label{eq:dev-fn}
\end{align}
for $n=1,2,\ldots,40$ at $t=4$ and $t=5$,
where we consider the second Richardson transform
$f_{n,m}^{(k=2)}$ of $f_{n,m}$ and set $m=58$.
The deviation grows as $n$ increases,
but the error at $n=40$ is still
within $\pm 0.4\%,\, \pm 0.04\%$ for $t=4,\,5$ respectively.
These results give a strong support
for our proposal of the nonperturbative completion
of the topological string partition function
as well as the prescription of analytic continuation
we adopted in section~\ref{sec:gen}.

Next, let us consider the large order behavior of $\Zeven_m(t)$.
We expect that $\Zeven_m(t)$ for large $m$ behave as
\begin{align}
\begin{aligned}
 \Zeven_m(t)
&\sim\mueven(t) \Aeven(t)^{-2m-\beven}\Ga(2m+\beven)\sum_{n=0}^\infty
  \feven_n(t) \Aeven(t)^n\frac{\Ga(2m+\beven-n)}{\Ga(2m+\beven)}.
\end{aligned}
\label{eq:Zeven-asymp}
\end{align}
In the same way as above, we can numerically determine
$\Aeven(t),\beven$ and $\mueven(t)$
from the asymptotic behavior of some sequence, such
as \eqref{eq:A-seq}
with $\Ztop_m(t)$ replaced by $\Zeven_m(t)$. 
The result of this numerical analysis is
\begin{align}
 \Aeven(t)=\frac{t^2}{2},\quad \beven=\hf,\quad
\mueven(t)=2\rt{\frac{2}{\pi}}e^{-\frac{t}{4}}\frac{\Thetaodd}{\Thetaeven}.
\end{align}
This is again derived analytically from 
the expansion \eqref{eq:ZN-acon}
using the approximation of $\vartheta_2(q)$ in \eqref{eq:th2-approx}.
The analytic form of the 1-instanton correction
obtained from \eqref{eq:ZN-acon}
is given by
\begin{equation}
\begin{aligned}
 \Zefull_{\text{1-inst}}(t)=2\ri\rt{\frac{2\pi}{g_s}}
\Zofull(t+g_s/2).
\end{aligned} 
\label{eq:full-inst-anal}
\end{equation} 
For instance, one can see that
the instanton action
$\Aeven(t)$ is reproduced from the genus-zero part 
$\mathcal{F}_0(t)=\til{\mathcal{F}}_0(t)=-t^3/3$
in \eqref{eq:Feo-01}
\begin{equation}
\begin{aligned}
 \exp\left[\frac{\til{\mathcal{F}}_0(t+g_s/2)
-\mathcal{F}_0(t)}{g_s^2}\right]=\exp\left[-\frac{t^2}{2g_s}
-\frac{t}{4}-\frac{g_s}{24}\right].
\end{aligned} 
\label{eq:calF0-shift}
\end{equation}
One can also show that $\mueven(t)$
is reproduced from
\eqref{eq:full-inst-anal} after including the contribution
of genus-one part in \eqref{eq:Feo-01}. 
Moreover, the analytic form of $\feven_n$ is also obtained from 
 \eqref{eq:full-inst-anal}
\begin{align}
 \sum_{n=0}^\infty \feven_n g_s^n
 =\frac{\eta_{\frac{1}{24}}(t)^2}{\eta_{\frac{1}{24}}(t+g_s/2)^2}
  \frac{\Thetaodd(t+g_s/2)}{\Thetaodd(t)}
\sum_{\ell=0}^\infty \Zodd_\ell(t+g_s/2)g_s^{2\ell},
\end{align}
where we have absorbed the last term $-g_s/24$
of \eqref{eq:calF0-shift} into
$\eta_{1/24}$ defined in \eqref{eq:eta-al}, as we did for $\h{\psi}_{\text{1-inst}}$ 
in \eqref{eq:Z1-anal}.
More explicitly, $\feven_n$ is written as
\begin{align}
 \feven_n=\sum_{j+k+m+2\ell=n}\frac{(-1)^{j+k+m}}{2^{j+k+m}j!k!m!}
\Big(D_{-2,\frac{1}{24}}\Big)^j1\cdot \DTodd^k1\cdot D^m \Zodd_{\ell},
\label{eq:Zfull-fn}
\end{align}
where $\DTodd$ is given by \eqref{eq:DTo}
and $D_{-2,1/24}$ is defined in \eqref{eq:D-kal}.
We have also checked numerically 
that \eqref{eq:Zfull-fn}
is consistent with the large order behavior of $\Zeven_m(t)$ in \eqref{eq:Zeven-asymp}.

We can repeat the same analysis for
the large order behavior of $\Zodd_m(t)$
and compare with the analytic expression of 1-instanton in \eqref{eq:tZN-acon}.
The result is similar to the case of $\Zeven_m(t)$ above, so we will be brief.
We find that the large $m$ behavior of $\Zodd_m(t)$ is given by
\begin{align}
\begin{aligned}
 \Zodd_m(t)
&\sim\muodd(t) \Aodd(t)^{-2m-\bodd}\Ga(2m+\bodd)\sum_{n=0}^\infty
  \fodd_n(t) \Aodd(t)^n\frac{\Ga(2m+\bodd-n)}{\Ga(2m+\bodd)},
\end{aligned}
\end{align}
where 
\begin{align}
 \Aodd(t)=\frac{t^2}{2},\quad \bodd=\frac{1}{2},\quad
\muodd(t)=2\rt{\frac{2}{\pi}}e^{-\frac{t}{4}}\frac{\Thetaeven}{\Thetaodd},
\end{align}
and the fluctuation coefficient $\fodd_n$ around 1-instanton
is given by
\begin{align}
 \sum_{n=0}^\infty \fodd_n g_s^n
 =\frac{\eta_{\frac{1}{24}}(t)^2}{\eta_{\frac{1}{24}}(t+g_s/2)^2}
  \frac{\Thetaeven(t+g_s/2)}{\Thetaeven(t)}
\sum_{\ell=0}^\infty \Zeven_\ell(t+g_s/2)g_s^{2\ell}.
\end{align}
From this $\fodd_n$ is written as
\begin{align}
 \fodd_n=\sum_{j+k+m+2\ell=n}\frac{(-1)^{j+k+m}}{2^{j+k+m}j!k!m!}
\Big(D_{-2,\frac{1}{24}}\Big)^j1\cdot \DTeven^k1\cdot D^m \Zeven_{\ell},
\end{align}
where $\DTeven$ 
is defined in \eqref{eq:DTe}.

\section{Borel-Pad\'{e} resummation \label{sec:borel}}

In this section, we consider the 
Borel resummation of the perturbative expansion of
$\psi^{\text{top}}$, $\Zefull$, and $\Zofull$.
We will see that the result is consistent with
our analytic continuation of
$\psi_{N_+}$ in \eqref{eq:psiN-acon},
$Z_N$ in \eqref{eq:ZN-acon}, and $\til{Z}_N$ in \eqref{eq:tZN-acon}.
Let us first consider the genus expansion of
$\psi^{\text{top}}(t)$
\begin{align}
\psi^{\text{top}}(t)
=\frac{e^{-\frac{t^3}{6g_s^2}}}{\eta(Q)}
 \sum_{n=0}^\infty \Ztop_n(t)g_s^{2n}.
\end{align}
As we have seen in the previous section, 
this expansion is not Borel summable and the Borel transform
has a pole on the positive real axis on the Borel plane.
However, we can avoid the pole by deforming the integration
contour slightly above or below the real axis.
This is known as the lateral Borel resummation $\mathcal{S}_{\pm}$
\begin{align}
 \mathcal{S}_{\pm}(\psi^{\text{top}})
=\frac{e^{-\frac{t^3}{6g_s^2}}}{\eta(Q)}\int_0^{\infty\pm\ri 0}
 dx \sum_{n=0}^\infty \frac{\Ztop_nx^{2n}}{\Ga(2n+\hf)}(xg_s)^{-\hf}
 e^{-\frac{x}{g_s}}.
\end{align}
In the numerical analysis, 
the integrand can be approximated by the Pad\'{e} approximation 
\begin{align}
 \sum_{n=0}^{n_{\text{max}}} \frac{\Ztop_nx^{2n}}{\Ga(2n+\hf)}\approx \frac{a_0+a_1x+\cdots 
a_{n_{\text{max}}}x^{n_{\text{max}}}}{1+b_1x+\cdots 
b_{n_{\text{max}}}x^{n_{\text{max}}}}.
\label{eq:Pade}
\end{align}
In the following analysis we set $n_{\text{max}}=60$.

We expect that the lateral Borel resummation of $\psi^{\text{top}}(t=Ng_s)$
is related to its non-perturbative completion $\psi_N$ 
via the relation \eqref{eq:psiN-acon} 
\begin{equation}
\begin{aligned}
 \psi^{\text{top}}(t)=\psi_N+\frac{\ri}{2}\vartheta_2(q)\psi^{\text{top}}(t+g_s)+\cdots
\approx \psi_N+\frac{\ri}{2}\rt{\frac{2\pi}{g_s}}\psi_{N+1},
\end{aligned} 
\end{equation}
where we have used the approximation of
$\vartheta_2(q)$ in \eqref{eq:th2-approx}
in the last step.
It turns out that the two branches of the square-root $(-1)^{\zeta(0)}=\pm\ri$
mentioned below \eqref{eq:acon} should be correlated with the two
choices of the lateral Borel resummation $\mathcal{S}_\pm$
\begin{align}
 \mathcal{S}_{\pm}(\psi^{\text{top}})\approx \psi_N\pm\frac{\ri}{2}\rt{\frac{2\pi}{g_s}}\psi_{N+1}.
\label{eq:BPtop}
\end{align}
This ensures that 
the imaginary part of the RHS of \eqref{eq:psiN-acon} 
is canceled at the 1-instanton level
\begin{equation}
\begin{aligned}
 \textrm{Im}\Bigl[\mathcal{S}_{\pm}\psi^{\text{top}}(t)\mp\frac{\ri}{2}\vartheta_2(q)
\mathcal{S}_{\pm}\psi^{\text{top}}(t+g_s)\Bigr]=0,
\end{aligned} 
\end{equation}
and  in total the RHS of \eqref{eq:psiN-acon} becomes real. 
This should be the case 
since $\psi_N$ on the LHS of \eqref{eq:psiN-acon}
is manifestly real for $g_s>0$.

\begin{figure}[tb]
\begin{center}
\includegraphics[width=5cm]{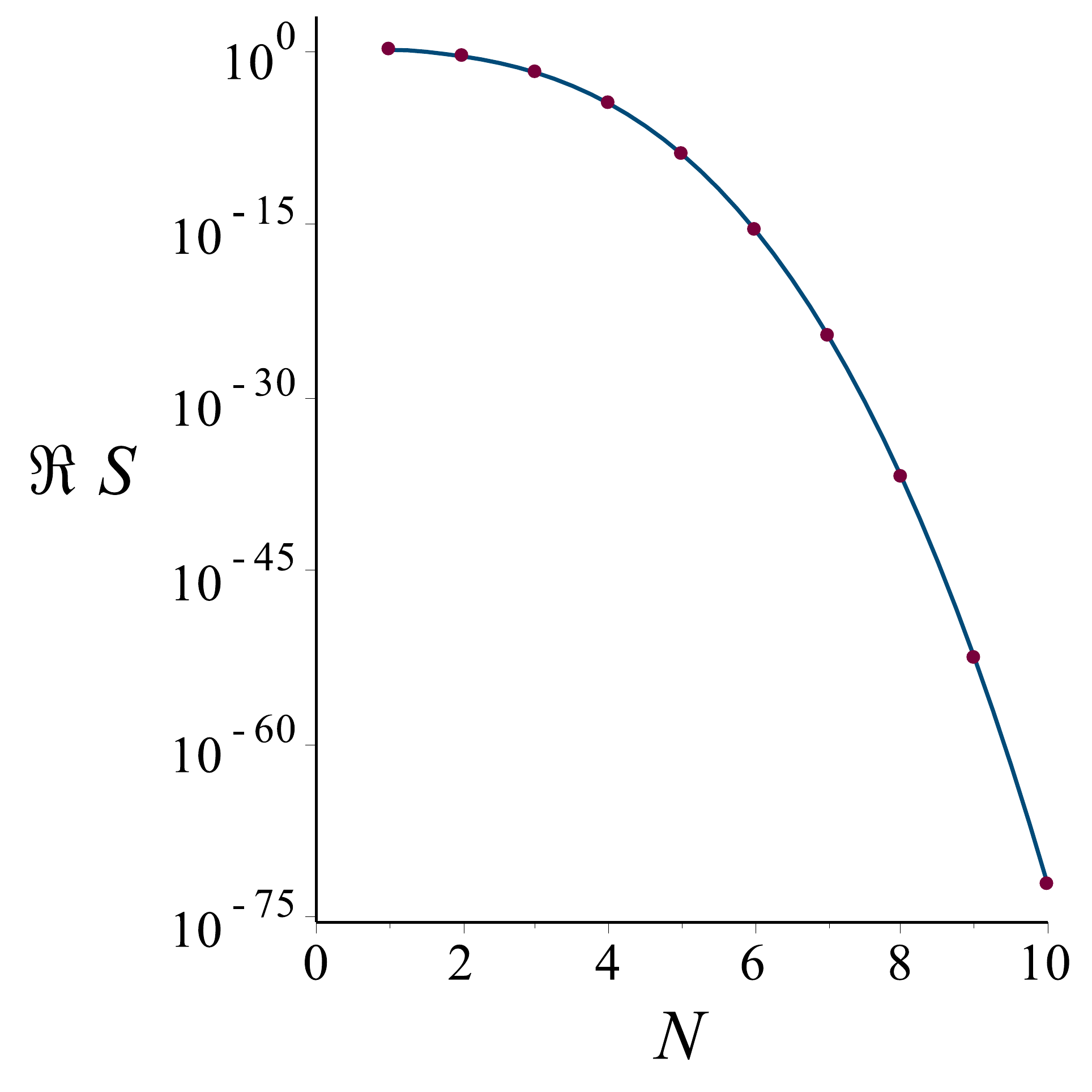}
\hspace{1cm}
\includegraphics[width=5cm]{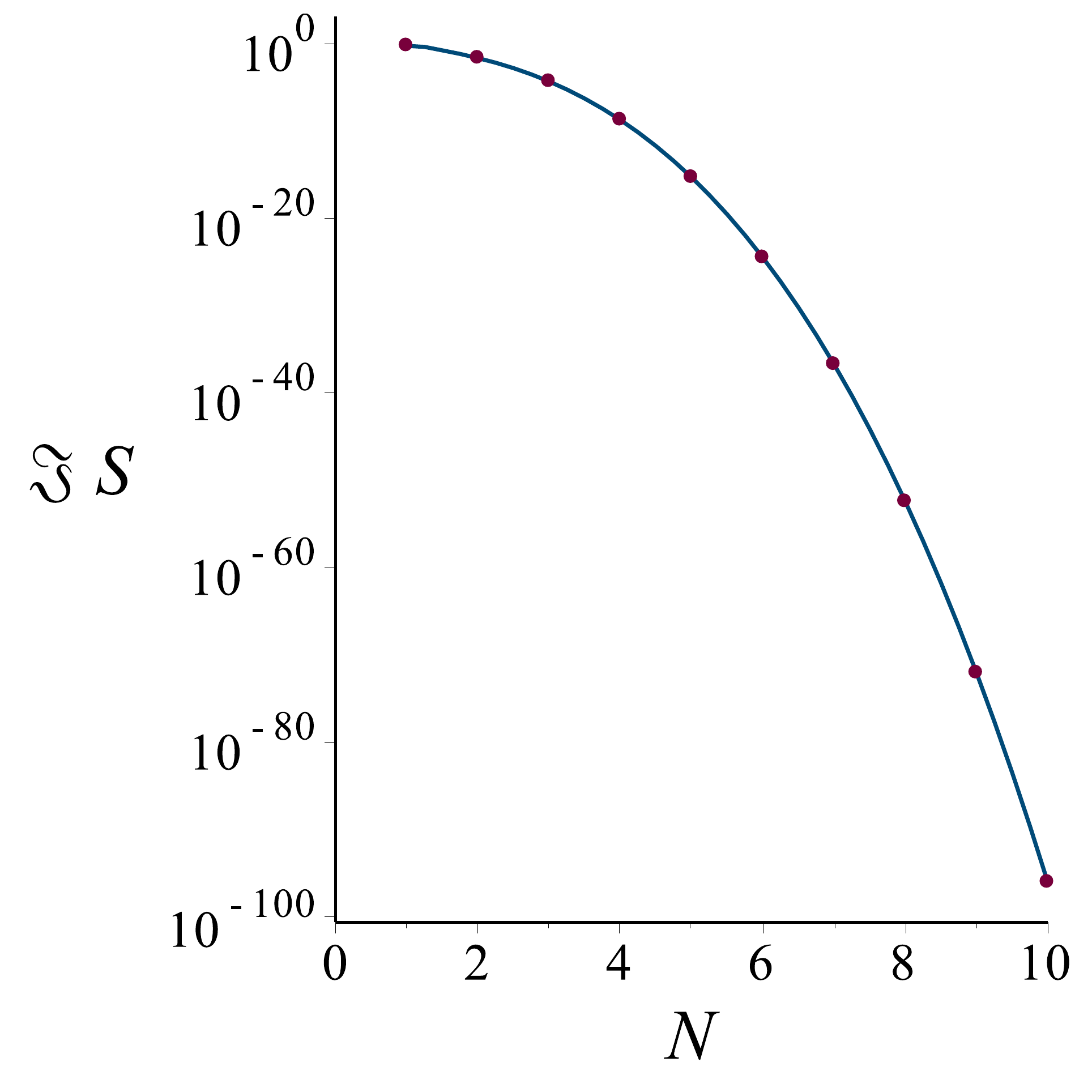}
\end{center}
\vspace{-2ex}
\caption{Comparison of the nonperturbative result
$\psi_N+\frac{\ri}{2}\sqrt{\frac{2\pi}{g_s}}\psi_{N+1}$ (red dots)
with the lateral Borel resummation
$\mathcal{S}_+(\psi^{\text{top}})$ (blue solid line)
at $g_s=1$.
The real part (left) and the imaginary part (right)
are plotted separately.}
\label{fig:bp}
\end{figure}
\begin{figure}[tb]
\begin{center}
\includegraphics[width=5cm]{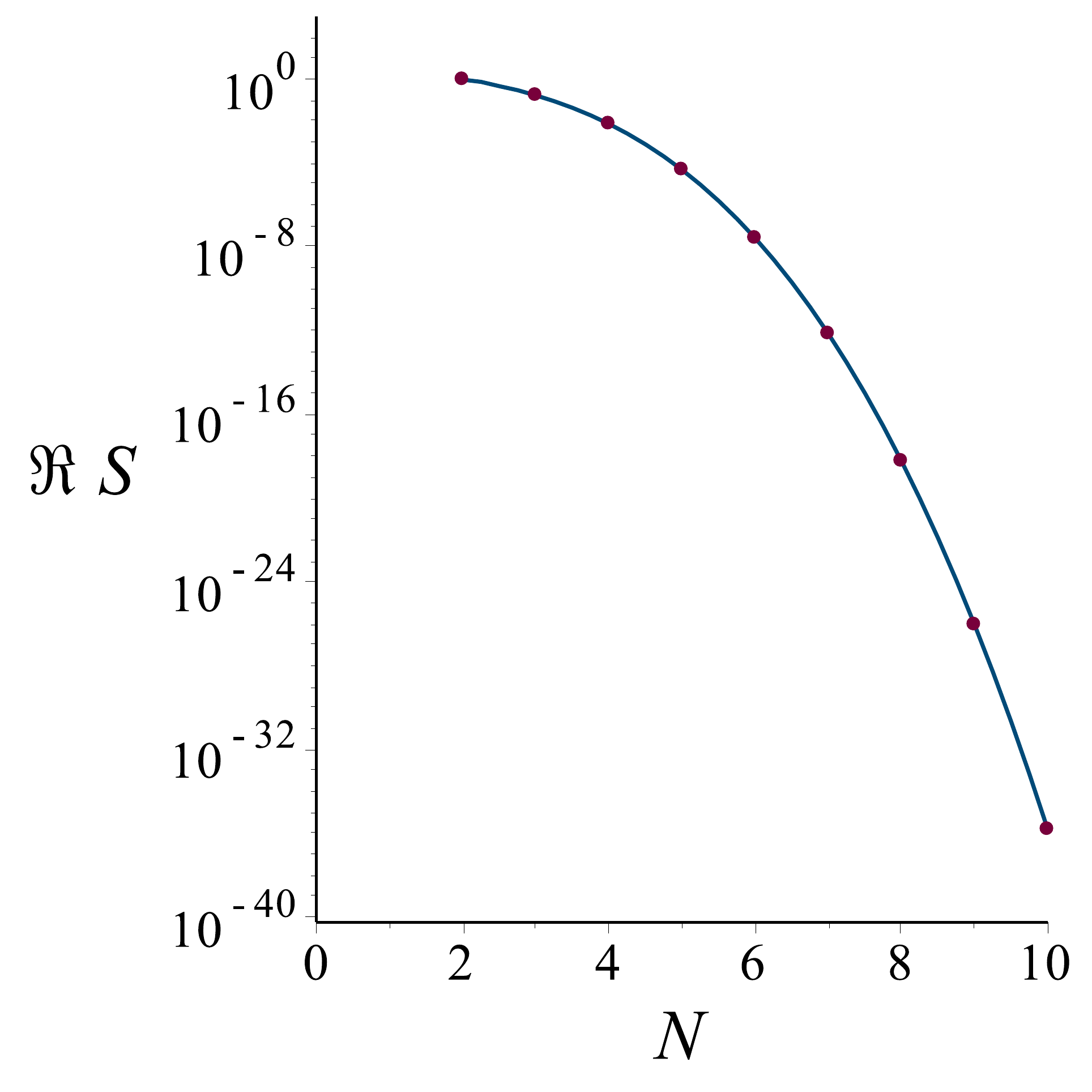}
\hspace{1cm}
\includegraphics[width=5cm]{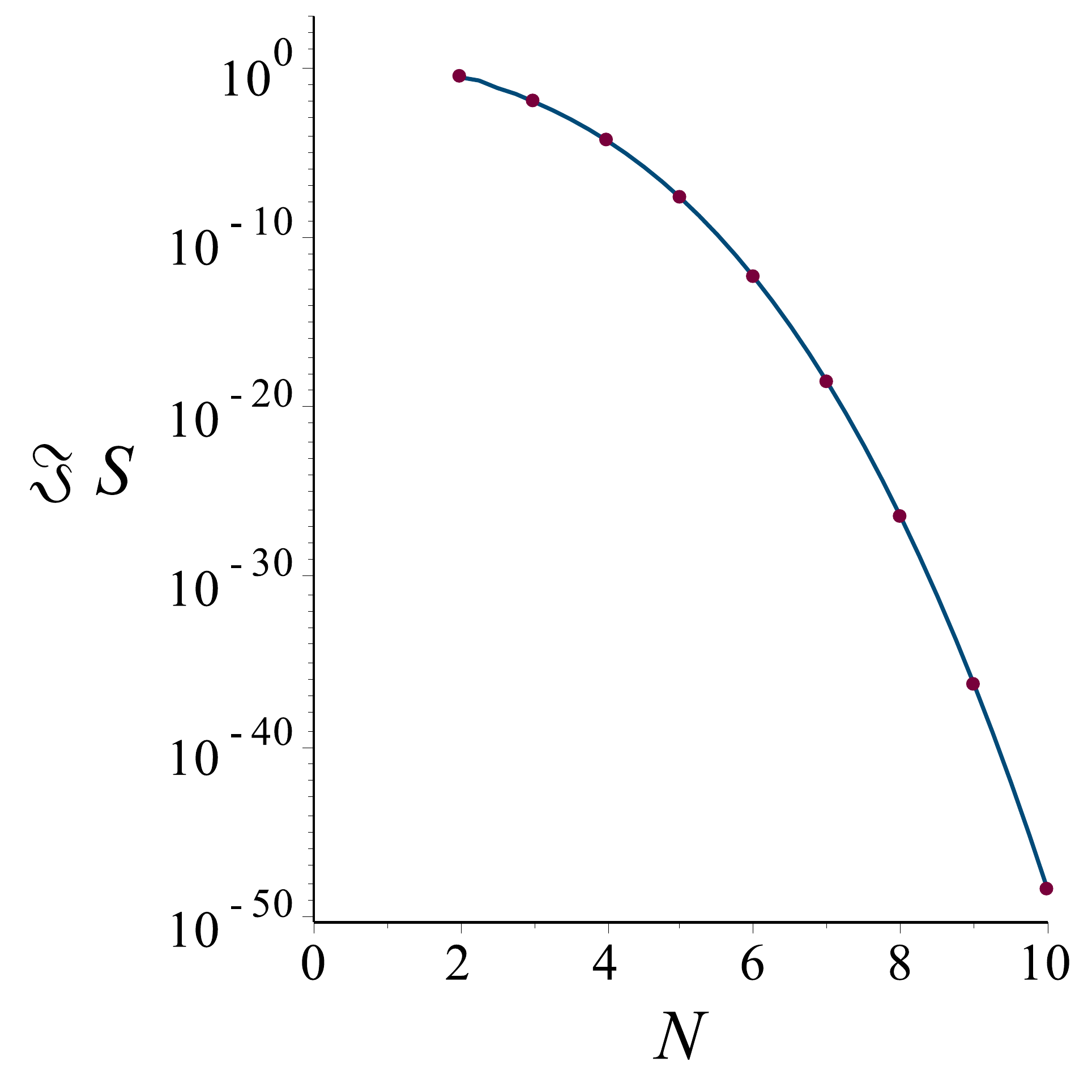}
\end{center}
\vspace{-2ex}
\caption{Comparison of the nonperturbative result
$Z_N+\ri\sqrt{\frac{2\pi}{g_s}}\tZ_{N+1}$ (red dots)
with the lateral Borel resummation
$\mathcal{S}_+(\Zefull)$ (blue solid line)
at $g_s=2$.
The real part (left) and the imaginary part (right)
are plotted separately.}
\label{fig:bpe}
\end{figure}
\begin{figure}[tb]
\begin{center}
\includegraphics[width=5cm]{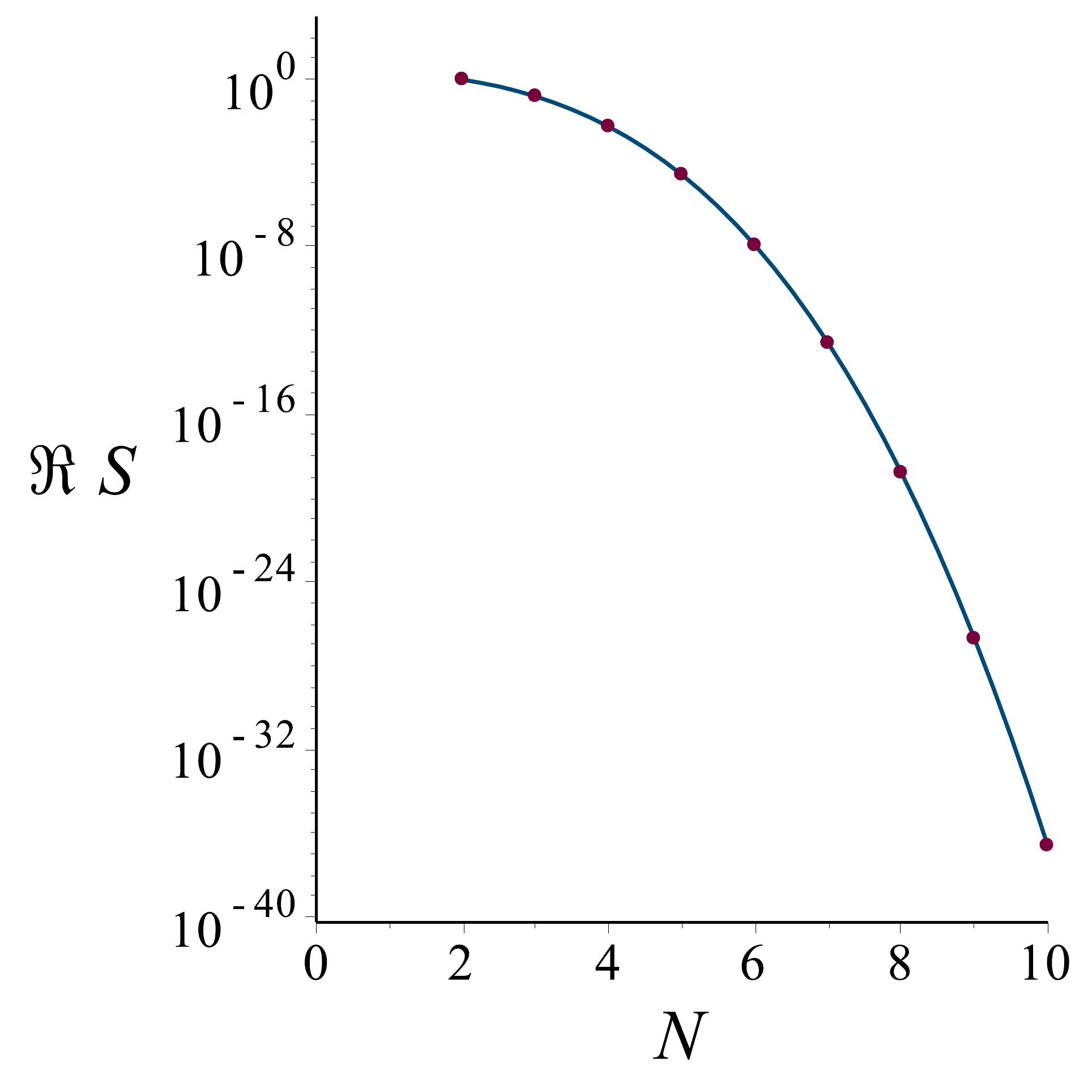}
\hspace{1cm}
\includegraphics[width=5cm]{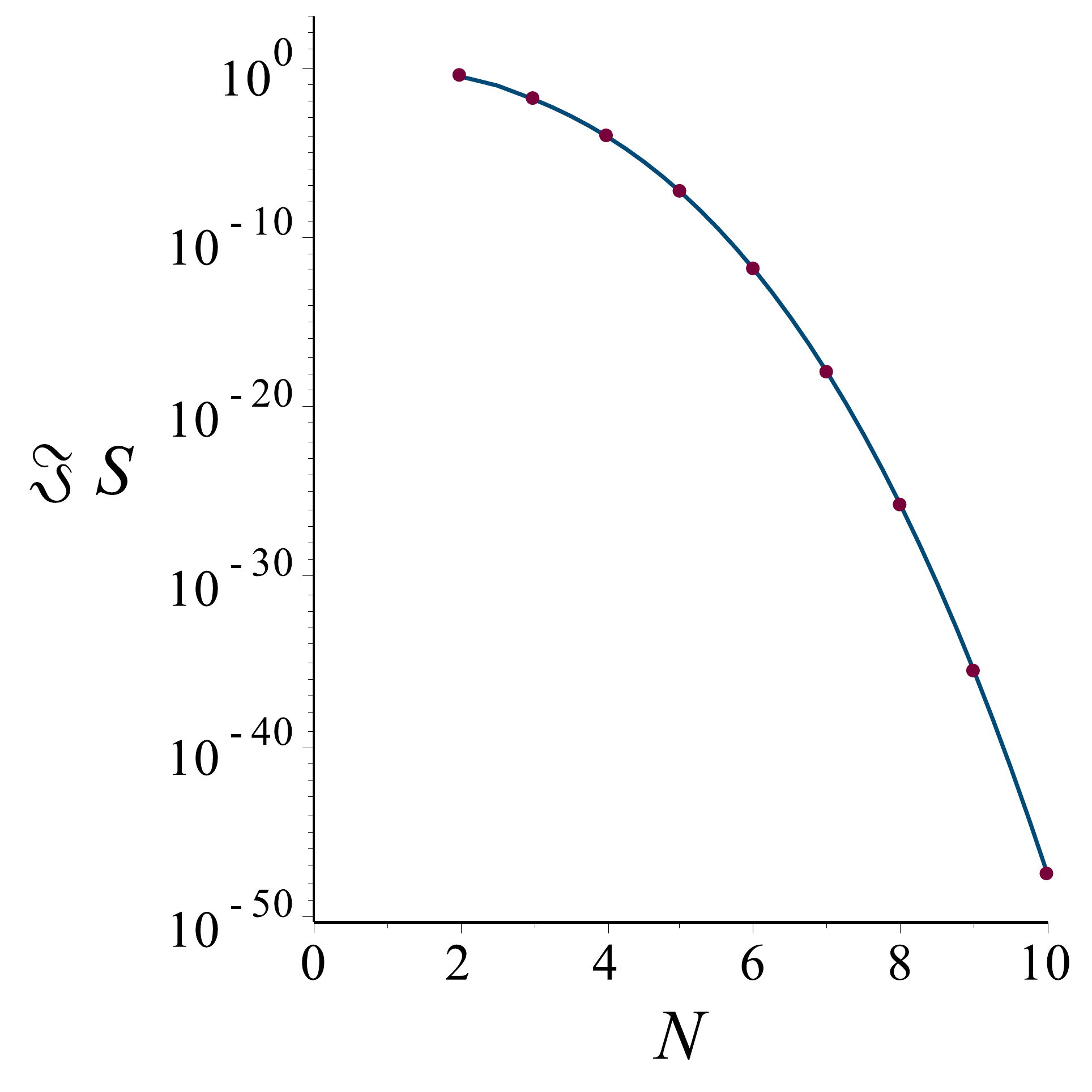}
\end{center}
\vspace{-2ex}
\caption{Comparison of the nonperturbative result
$\tZ_N+\ri\sqrt{\frac{2\pi}{g_s}}Z_{N+1}$ (red dots)
with the lateral Borel resummation
$S_+(\Zofull)$ (blue solid line)
at $g_s=2$.
The real part (left) and the imaginary part (right)
are plotted separately.}
\label{fig:bpo}
\end{figure}

We can numerically evaluate $\mathcal{S}_\pm(\psi^{\text{top}})$ on the LHS of \eqref{eq:BPtop}
by the Borel-Pad\'{e} approximation and see if it agrees with the RHS
of \eqref{eq:BPtop}.
We numerically observed that most of the poles of the Pad\'{e}
approximant \eqref{eq:Pade}
are located on the real axis at $x\gtrsim \Atop(t)=t^2/2$
and there are few other poles away from the real axis.
To avoid the poles on the real axis,
we take the integration contour for $\mathcal{S}_+$ as the union of
two line segments:
$[0,t^2/2+\ri\varepsilon]\cup[t^2/2+\ri\varepsilon,\infty+\ri\varepsilon]$
where $\varepsilon$ is a small positive number.\footnote{We set $\varepsilon=1/50$
in the numerical integration
in Figure \ref{fig:bp}, \ref{fig:bpe} and \ref{fig:bpo}, but
we observe that the results are rather insensitive
to the value of $\varepsilon$ as long as the
integration contour does not hit the poles away from the real axis.}
From Figure \ref{fig:bp}, 
one can see that the lateral Borel resummation
nicely reproduces not only the real part but also
the imaginary part of \eqref{eq:BPtop},
i.e.~the 1-instanton contribution.
\begin{figure}[tb]
\vspace{2ex}
\begin{center}
\includegraphics[width=5cm]{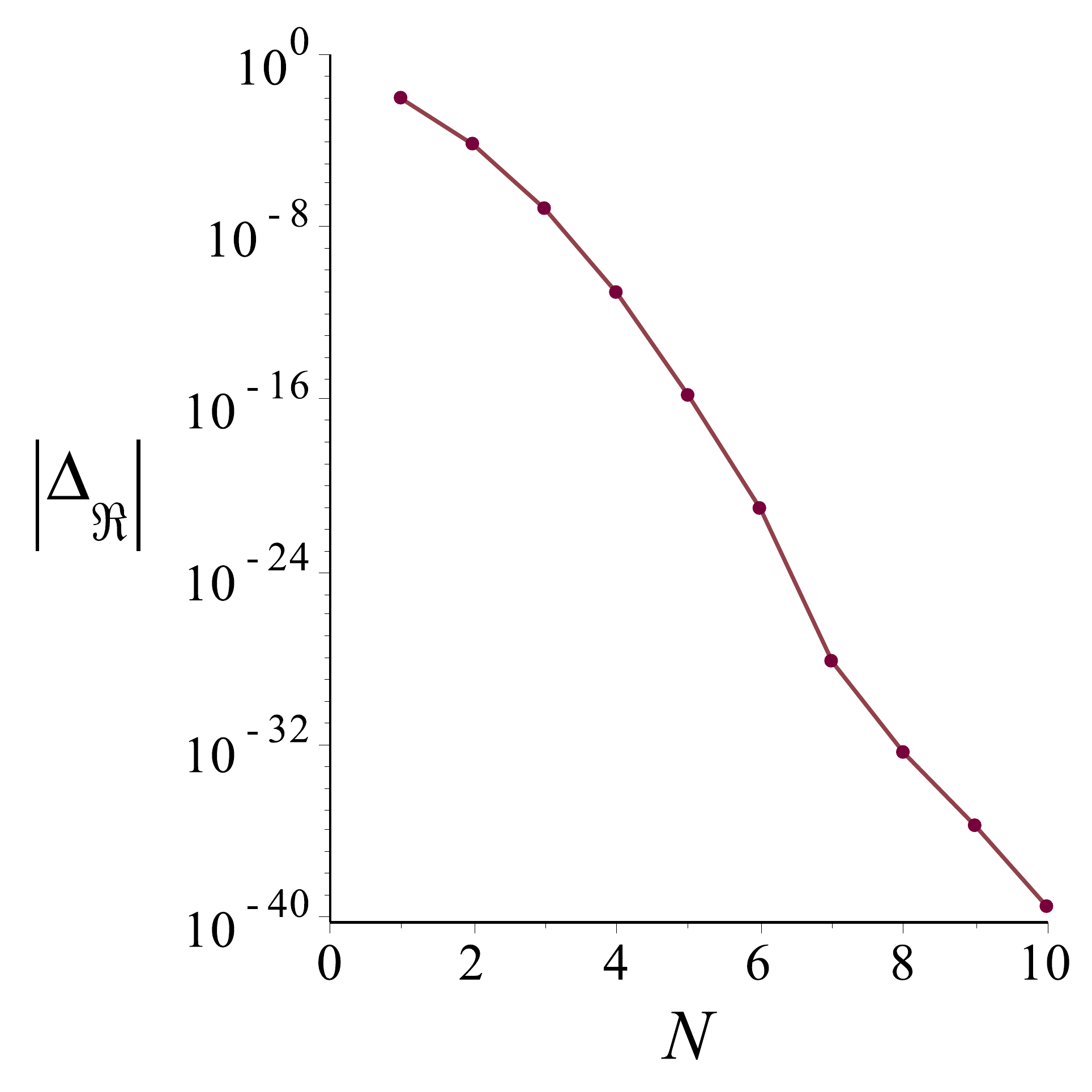}
\hspace{1cm}
\includegraphics[width=5cm]{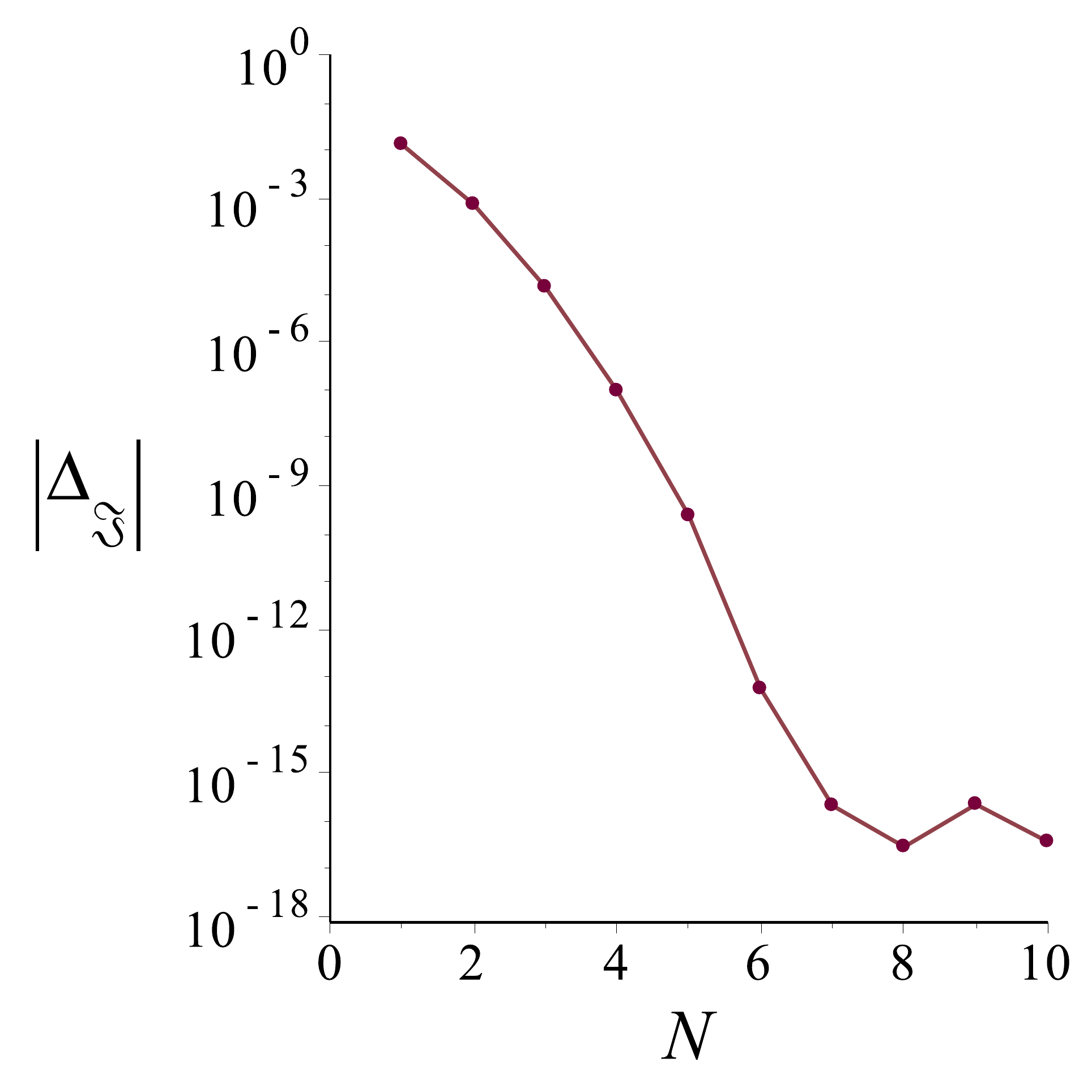}
\end{center}
\vspace{-2ex}
\caption{Relative deviations of
the lateral Borel resummation
$S_+(\psi^{\text{top}})$
from the nonperturbative result
$\psi_N+\frac{\ri}{2}\sqrt{\frac{2\pi}{g_s}}\psi_{N+1}$
at $g_s=1$.
The real part (left) and the imaginary part (right)
are plotted separately.}
\label{fig:bpdev}
\end{figure}
Figure \ref{fig:bpdev} shows the relative deviations
\begin{align}
|\Delta_{\rm Re}|
 =\left|\frac{{\rm Re}\,\mathcal{S}_+}{\psi_N}-1\right|,
\quad
|\Delta_{\rm Im}|
 =\left|\frac{{\rm Im}\,\mathcal{S}_+}
             {\frac{1}{2}\sqrt{\frac{2\pi}{g_s}}\psi_{N+1}}
        -1\right|
\end{align}
at $g_s=1$.
As one can see, the relative deviation
of the real part
$|\Delta_{\rm Re}|$
decreases exponentially as $N$
increases.
On the other hand, the relative deviation
of the imaginary part
$|\Delta_{\rm Im}|$ decreases for small $N$
but it no longer decreases for $N> 7$.
One may also notice an ``inflection point'' at $N=7,6$ in
the plots of $|\Delta_{\rm Re}|,\,|\Delta_{\rm Im}|$ respectively.
Currently we do not understand why this happens,
but we expect that this is merely an artifact of our numerical analysis.

\begin{figure}[tb]
\vspace{2ex}
\begin{center}
\includegraphics[width=5cm]{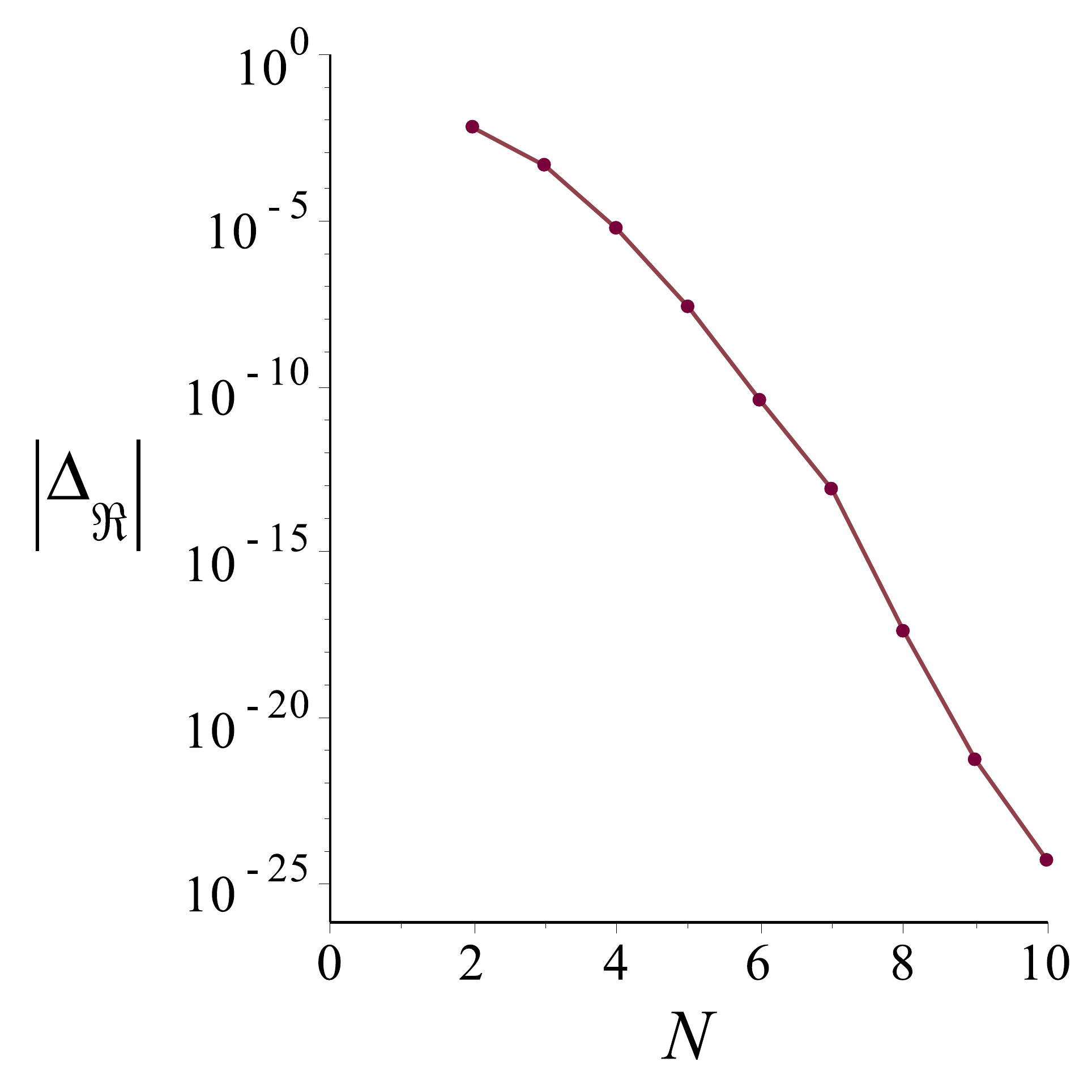}
\hspace{1cm}
\includegraphics[width=5cm]{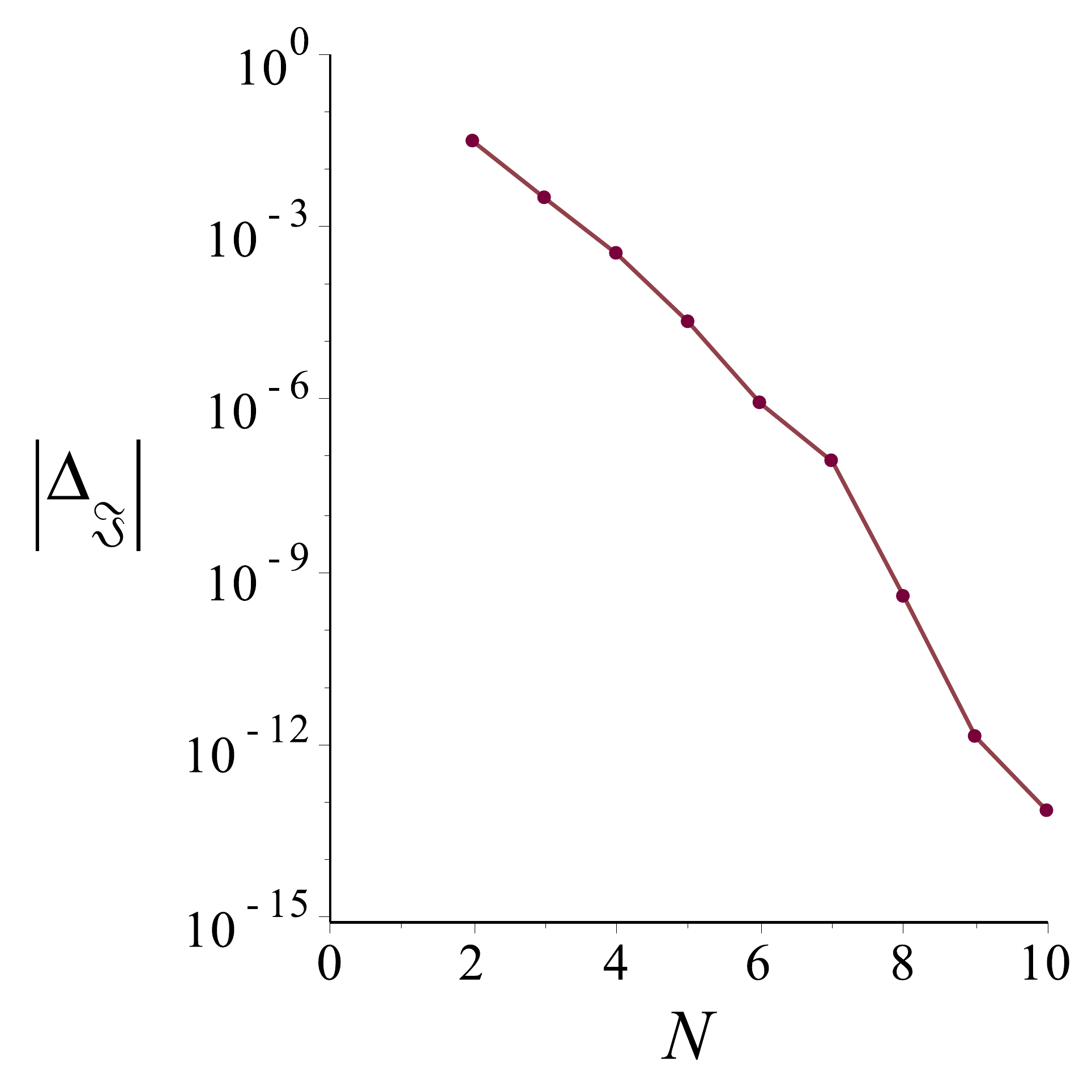}
\end{center}
\vspace{-2ex}
\caption{Relative deviations of
the lateral Borel resummation
$\mathcal{S}_+(\Zefull)$
from the nonperturbative result
$Z_N+\ri\sqrt{\frac{2\pi}{g_s}}\tZ_{N+1}$
at $g_s=2$.
The real part (left) and the imaginary part (right)
are plotted separately.}
\label{fig:bpedev}
\end{figure}
\begin{figure}[tb]
\vspace{2ex}
\begin{center}
\includegraphics[width=5cm]{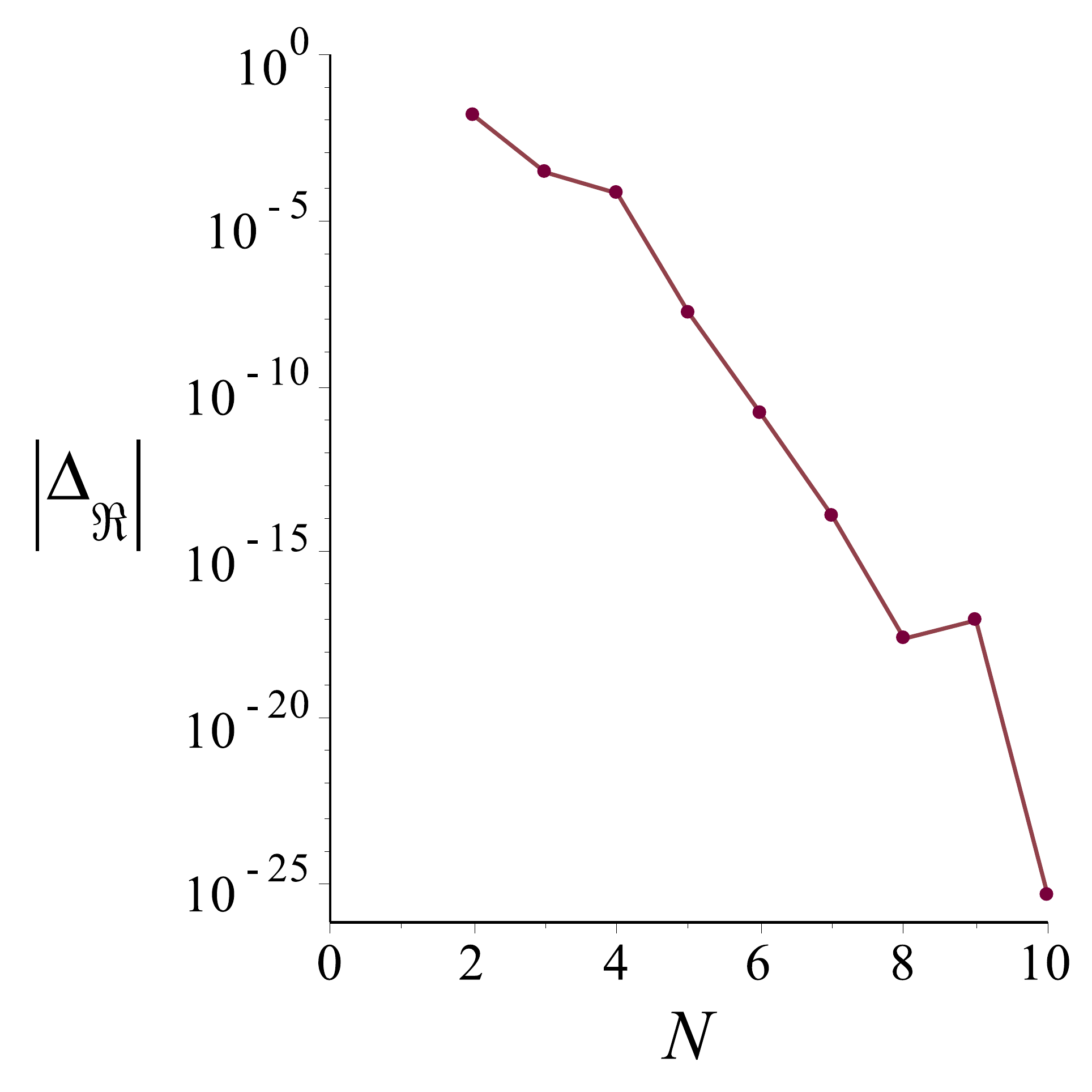}
\hspace{1cm}
\includegraphics[width=5cm]{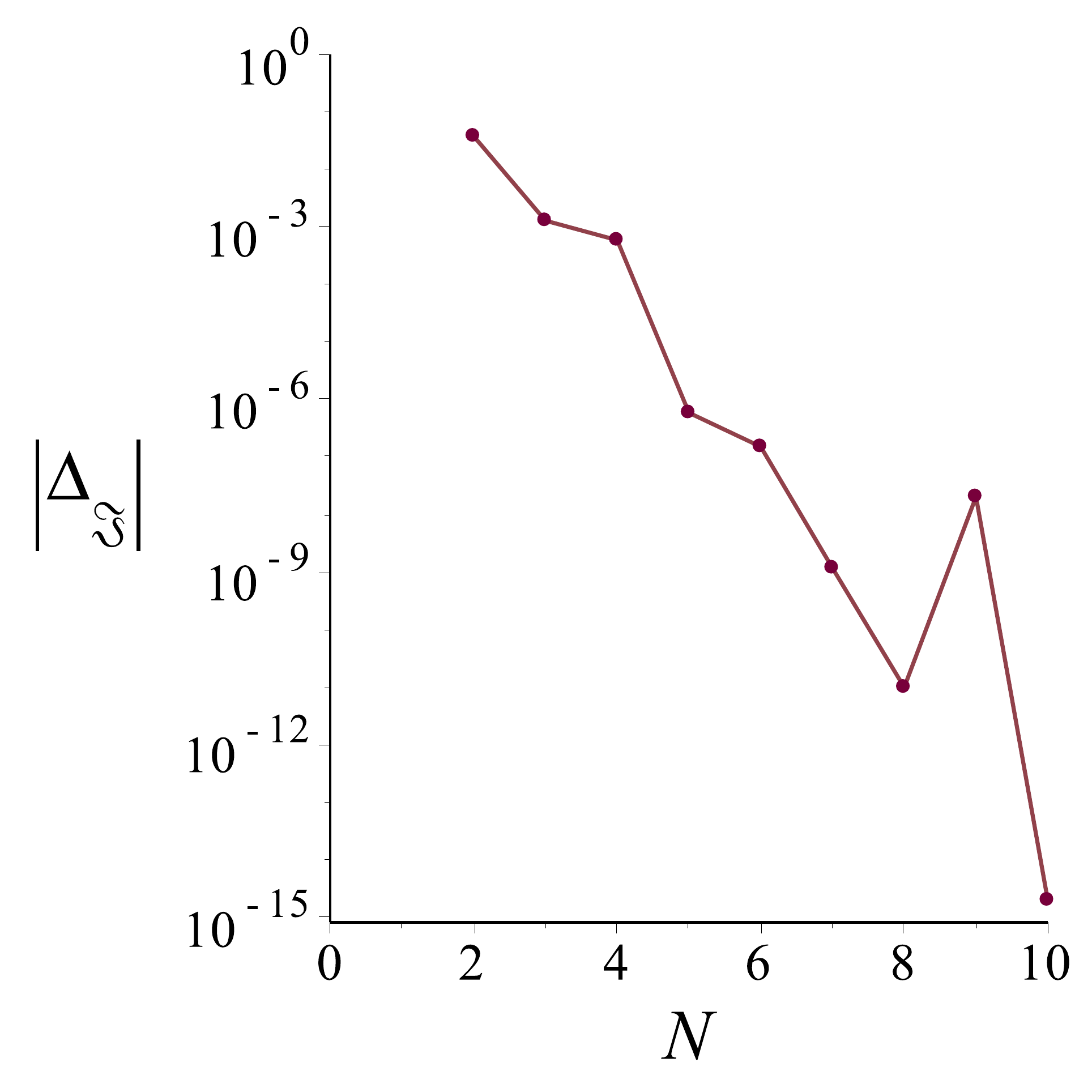}
\end{center}
\vspace{-2ex}
\caption{Relative deviations of
the lateral Borel resummation
$S_+(\Zofull)$
from the nonperturbative result
$\tZ_N+\ri\sqrt{\frac{2\pi}{g_s}}Z_{N+1}$
at $g_s=2$.
The real part (left) and the imaginary part (right)
are plotted separately.}
\label{fig:bpodev}
\end{figure}

We can consider the lateral Borel resummation of 
full partition functions $\Zefull$ and $\Zofull$ as well
\begin{align}
 \begin{aligned}
  \mathcal{S}_{\pm}(\Zefull)&=\frac{e^{-\frac{t^3}{3g_s^2}}\Thetaeven}{\eta^2}\int_0^{\infty\pm\ri 0}
dx \sum_{n=0}^\infty \frac{\Zeven_n x^{2n}}{\Ga(2n+\hf)}(xg_s)^{-\hf} e^{-\frac{x}{g_s}},
\\
\mathcal{S}_{\pm}(\Zofull)&=
\frac{e^{-\frac{t^3}{3g_s^2}}\Thetaodd}{\eta^2}\int_0^{\infty\pm\ri 0}
dx \sum_{n=0}^\infty \frac{\Zodd_n x^{2n}}{\Ga(2n+\hf)}(xg_s)^{-\hf} e^{-\frac{x}{g_s}}.
 \end{aligned}
\end{align}
From the 
expansion of $Z_N$ in \eqref{eq:ZN-acon} and $\til{Z}_N$ in \eqref{eq:tZN-acon},
we expect that at the 1-instanton level
the lateral Borel resummation of full partition function
is approximately given by
\begin{align}
\begin{aligned}
 \mathcal{S}_{\pm}(\Zefull)&\approx Z_N\pm\ri\rt{\frac{2\pi}{g_s}}\til{Z}_{N+1},\\
\mathcal{S}_{\pm}(\Zofull)&\approx\til{Z}_N\pm\ri\rt{\frac{2\pi}{g_s}}Z_{N+1}.
\end{aligned} 
\label{eq:S-full}
\end{align}
Again, we can test this relation by evaluating the LHS numerically 
using the Borel-Pad\'{e} approximation. Figure \ref{fig:bpe} and
Figure \ref{fig:bpo}
show the real and the imaginary parts of $\mathcal{S}_+(\Zefull)$
and $\mathcal{S}_+(\Zofull)$ at $g_s=2$, respectively, while Figure \ref{fig:bpedev} and 
Figure \ref{fig:bpodev}
represent the relative deviation from the expected behavior 
on the RHS of \eqref{eq:S-full}.
From these figures, one can clearly see that 
the lateral Borel resummation $\mathcal{S}_+(\Zefull)$
and $\mathcal{S}_+(\Zofull)$ correctly reproduce the finite $N$
result on the RHS of \eqref{eq:S-full}.
This numerical result strongly supports our prescription 
of analytic continuation \eqref{eq:ZN-acon} and \eqref{eq:tZN-acon}
for the full partition functions.

\section{Comment on $\th\ne0$ \label{sec:theta}}

In the previous sections we have assumed $\th=0$.
In this section we will consider the partition function $Z_N$ with non-zero $\th$
given by \eqref{eq:ZN-int-th}.
As shown in \eqref{eq:thooft}, when $\th$ is non-zero
the 't Hooft coupling $t$ becomes complex and  
$\th$ appears as the imaginary part of $t$.
For the general $\th\ne0$ case,
one can study the large $N$ expansion of $Z_N$ 
using the free fermion picture 
\begin{equation}
\begin{aligned}
 Z_N=\sum_{p_1<\cdots <p_N} q^{E}e^{\ri\th P},
\end{aligned} 
\label{eq:ZN-fermi}
\end{equation}
where $E$ and $P$ denote the total energy and total momentum of $N$ fermions
\begin{equation}
\begin{aligned}
 E=\sum_{i=1}^N\hf p_i^2,\qquad
P=\sum_{i=1}^N p_i.
\end{aligned} 
\end{equation}
From this expression \eqref{eq:ZN-fermi} one can show that 
$Z_N$ is invariant under $\th\to-\th$ and $\th\to\th+2\pi$.

As discussed around \eqref{eq:Egnd}, the ground state corresponds to
the configuration of fermions where the modes between $p=-p_F$ and $p=+p_F$ are occupied,
with the ``Fermi momentum'' $p_F$ being
\begin{equation}
\begin{aligned}
 p_F=\frac{N-1}{2}.
\end{aligned} 
\end{equation}

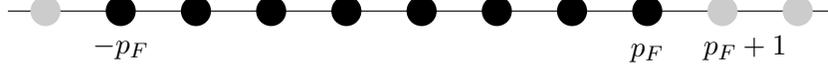
\begin{figure}[t]
\centering
\hskip10mm
\begin{tikzpicture}
\draw (-0.5,6)--(10.5,6);
\fill [gray!40] (0,6) circle [radius=2mm];
\fill [black] (1,6) circle [radius=2mm];
\fill [black] (2,6) circle [radius=2mm];
\fill [black] (3,6) circle [radius=2mm];
\fill [black] (4,6) circle [radius=2mm];
\fill [black] (5,6) circle [radius=2mm];
\fill [black] (6,6) circle [radius=2mm];
\fill [black] (7,6) circle [radius=2mm];
\fill [black] (8,6) circle [radius=2mm];
\fill [gray!40] (9,6) circle [radius=2mm];
\fill [gray!40] (10,6) circle [radius=2mm];
\coordinate (a1) at (1,5.8) node at (a1) [below] {$-p_F$};
\coordinate (a2) at (9.3,5.8) node at (a2) [below] {$p_F+1$};
\coordinate (a3) at (8,5.7) node at (a3) [below] {$p_F$};
\end{tikzpicture}
\caption{Maya diagram for the ground state.
The black nodes ($|p|\leq p_F$) are occupied 
by fermions while the gray nodes ($|p|>p_F$) are empty.} 
\label{fig:gnd}
\end{figure}

\begin{figure}[t]
\vspace{2ex}
\centering
\begin{tikzpicture}
\draw (-0.5,4)--(10.5,4);
\fill [gray!40] (1,4) circle [radius=2mm];
\fill [black] (2,4) circle [radius=2mm];
\fill [black] (3,4) circle [radius=2mm];
\fill [black] (4,4) circle [radius=2mm];
\fill [black] (5,4) circle [radius=2mm];
\fill [black] (6,4) circle [radius=2mm];
\fill [black] (7,4) circle [radius=2mm];
\fill [black] (8,4) circle [radius=2mm];
\fill [black] (9,4) circle [radius=2mm];
\fill [gray!40] (0,4) circle [radius=2mm];
\fill [gray!40] (10,4) circle [radius=2mm];
\coordinate (b1) at (1,3.8) node at (b1) [below] {$-p_F$};
\coordinate (b2) at (9.3,3.8) node at (b2) [below] {$p_F+1$};
\coordinate (b3) at (8,3.7) node at (b3) [below] {$p_F$};
\coordinate (B) at (-1.5,4) node at (B) [right] {(b)}; 
\draw (-0.5,6)--(10.5,6);
\fill [gray!40] (0,6) circle [radius=2mm];
\fill [black] (1,6) circle [radius=2mm];
\fill [black] (2,6) circle [radius=2mm];
\fill [black] (3,6) circle [radius=2mm];
\fill [black] (4,6) circle [radius=2mm];
\fill [black] (5,6) circle [radius=2mm];
\fill [black] (6,6) circle [radius=2mm];
\fill [black] (7,6) circle [radius=2mm];
\fill [gray!40] (8,6) circle [radius=2mm];
\fill [black] (9,6) circle [radius=2mm];
\fill [gray!40] (10,6) circle [radius=2mm];
\coordinate (a1) at (1,5.8) node at (a1) [below] {$-p_F$};
\coordinate (a2) at (9.3,5.8) node at (a2) [below] {$p_F+1$};
\coordinate (a3) at (8,5.7) node at (a3) [below] {$p_F$};
\coordinate (A) at (-1.5,6) node at (A) [right] {(a)};
\end{tikzpicture}
\caption{Examples of excited states: (a) chiral excitation (b) non-chiral excitation.} 
\label{fig:excitation}
\end{figure}
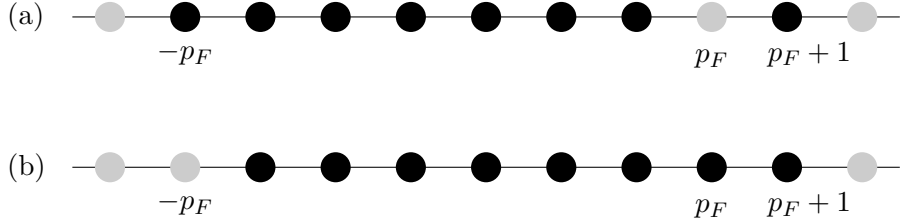
Now it is convenient to use the so-called Maya diagram
to represent the configuration of 
fermions,
as shown in Figure~\ref{fig:gnd} and Figure~\ref{fig:excitation}.
In this diagram, the black nodes are occupied by fermions while the gray nodes are empty.
The configuration in Figure~\ref{fig:gnd} represents the ground state while
Figure~\ref{fig:excitation} is an example of excited states.
The  energy  and the momentum
of the states (a) and (b)  in Figure~\ref{fig:excitation} 
can be easily found as
\begin{align}
 \begin{aligned}
  \mbox{(a)}:&~~E=E_0+\frac{N}{2},
& P&=1,\\
\mbox{(b)}:&~~ E=E_0+\frac{N}{2},
& P&=N,
 \end{aligned}
\end{align}
where $E_0$ is the ground state energy \eqref{eq:Egnd},
and their contributions to the partition function are given by
\begin{align}
  Z_{\text{(a)}}=q^{E_0+\frac{N}{2}}e^{\ri\th},\qquad
Z_{\text{(b)}}=q^{E_0+\frac{N}{2}}e^{\ri N\th}.
\label{eq:Zab}
\end{align}
There are two more states with the same energy,
obtained by changing the sign of momenta $p_i\to-p_i$ in Figure~\ref{fig:excitation}.
In this manner, we can systematically find
the expansion of partition function as
\begin{align}
 Z_N=q^{E_0}\Bigl[1+(e^{\ri\th}+e^{-\ri\th}+e^{\ri N\th}+e^{-\ri N\th})q^{\frac{N}{2}}+\cdots\Bigr].
\label{eq:Z-th}
\end{align}
As discussed by Gross and Taylor \cite{Gross:1992tu,Gross:1993hu,Gross:1993yt},
there is a clear distinction between the excitations (a) and (b) in  Figure~\ref{fig:excitation}:
(a) is chiral while (b) is non-chiral.
This distinction is reflected in the different behavior of
$Z_{\text{(a)}}$ and $Z_{\text{(b)}}$ in the 't Hooft limit \eqref{eq:thooft}. 
In fact, up to the overall factor $q^{E_0}$, $Z_{\text{(a)}}$ is
a holomorphic function of the 't Hooft coupling $t$ in \eqref{eq:thooft}
while $Z_{\text{(b)}}$ is non-holomorphic in $t$
\begin{equation}
\begin{aligned}
 Z_{\text{(a)}}q^{-E_0}=e^{-t},\qquad
Z_{\text{(b)}}q^{-E_0}=e^{-\frac{t+\b{t}}{2}-\frac{t^2-\b{t}^2}{2g_s}}.
\end{aligned} 
\end{equation}
Note that $Z_{\text{(b)}}$ is already ``non-perturbative'' in $g_s$, 
i.e. it behaves as $\mathcal{O}(e^{-1/g_s})$. 
$Z_{\text{(b)}}$ can also be thought of as originating from the sum over RR flux ($l=1$ term in \eqref{eq:RRsum})
\begin{align}
 F^{\text{cl}}(t+ g_s)+ F^{\text{cl}}(\b{t}- g_s)-F^{\text{cl}}(t)
-F^{\text{cl}}(\b{t})=-\frac{t+\b{t}}{2}-\frac{t^2-\b{t}^2}{2g_s},
\label{eq:Fcl-sa}
\end{align}
where $ F^{\text{cl}}(t)$ is given by \eqref{eq:Fcl}.

However, the contribution of $Z_{\text{(b)}}$ was treated as a part of
the perturbative partition function $\Zeven$ in \eqref{eq:Zfull-pert} when 
$\th=0$. Indeed the last term on the RHS of \eqref{eq:Fcl-sa}
vanishes when $t=\b{t}$.
This discussion  suggests that the distinction between the 
perturbative part and the non-perturbative part becomes 
much more complicated when $\th\ne0$ compared to the $\th=0$ case considered in
the previous sections.

Nevertheless, 
it turns out that $Z_N$ has a simple large $N$ expansion for some special value of $\th$.
One can see that $\th=\pi$ is such a special value.
To see this, we first rewrite $Z_{N}$ in \eqref{eq:ZN-int-th} as 
\begin{align}
  Z_{2M}(g_s,\th)=\oint\frac{dx}{2\pi\ri x^{2M+1}}\exp\left[\sum_{\ell=1}^\infty 
\frac{(-1)^{\ell-1}x^\ell}{\ell}\sum_{p\in\mathbb{Z}_{\geq0}+\hf} 2q^{\hf\ell p^2}
\cos(\th\ell p )\right],
\label{eq:Z2N-th}
\end{align}
where we assumed $N=2M$ is an even integer.
When $\th=\pi$,
the summation over $\ell$ is non-vanishing only for even $\ell$. Then, by setting $\ell=2k$ we find
\begin{align}
\begin{aligned}
  Z_{2M}(g_s,\th=\pi)&=\oint\frac{dx}{2\pi\ri x^{2M+1}}\exp\left[\sum_{k=1}^\infty 
\frac{-x^{2k}}{2k}\sum_{p\in\mathbb{Z}_{\geq0}+\hf}  2q^{k p^2}(-1)^k\right]\\
 &=\oint\frac{dx}{2\pi\ri x^{2M+1}}\prod_{p\in\mathbb{Z}_{\geq0}+\hf}(1+x^2 q^{p^2})\\
&=\psi_M(2g_s,\th=0).
\end{aligned}
\end{align}
Namely, the full partition function at $\th=\pi$ is equal to
the chiral partition function at $\th=0$ with rescaled string coupling $g_s\to2g_s$.
More generally, we expect that when $\th/\pi$ is a rational number
the partition function $Z_N(g_s,\th)$ has a simple large $N$ expansion.
We leave the study of rational $\th/\pi$ case as an interesting future problem.

\section{Discussions \label{sec:discussion}}

In this paper we have considered the non-perturbative $\mathcal{O}(e^{-N})$
correction in the $1/N$
expansion of 2d Yang-Mills theory on $T^2$,
which in turn is related to the topological string on a local Calabi-Yau
threefold $X$ \eqref{eq:CY} via the OSV conjecture \eqref{eq:OSV}.
We proposed a non-perturbative completion $\psi_{N_+}$
of the topological string partition function $\psi^{\text{top}}(t)$, 
 as a partition function
of $N_+$ fermions with positive momentum.
We emphasize that our non-perturbative completion $\psi_{N_+}$ of
$\psi^{\text{top}}(t)$ makes sense
at finite $N_+$.
We have also studied the large genus behavior of the $g_s$-expansion
of $\psi^{\text{top}}(t)$ and confirmed that it is consistent with our analytic continuation
of the formal expansion of $\psi_{N_+}$ \eqref{eq:psiN-acon}.
In particular, the 1-instanton coefficient is imaginary and it is 
precisely canceled by the imaginary part coming from the Borel resummation of $\psi^{\text{top}}(t)$
in accord with the theory of resurgence.
We have also studied the genus expansion of
the full partition functions $\Zefull$ in \eqref{eq:Zfull} 
and $\Zofull$
in \eqref{eq:tZfull} 
when $\th=0$. Again, it is consistent 
with our analytic continuation of 
the expansion of $Z_{N}$  in \eqref{eq:ZN-acon}
and $\tilde{Z}_N$ in \eqref{eq:tZN-acon}.
We should stress that our analytic continuation is different from that
in \cite{Dijkgraaf:2005bp} and ours is supported by the resurgence analysis as we mentioned above.
However, our analysis was limited to the 1-instanton
level and it would be very interesting to study the 
higher instanton corrections.

There are several open questions.
Of particular interest is the implication of our findings to the black hole physics.
In \cite{Dijkgraaf:2005bp} the expansion \eqref{eq:baby-expansion} of Yang-Mills partition function
$Z_N$ was considered based on a certain analytic continuation \eqref{eq:acon-baby},
and it was interpreted as the creation of baby universes.
However, our resurgent analysis strongly suggests that we should consider a different analytic continuation.
Moreover, by our definition of non-perturbative completion of $\psi^{\text{top}}(t)$
the chiral factorization is exact \eqref{eq:ZN-decomp}.
From these observations,
it is tempting to conclude that the creation of baby universes
is an artifact of the semi-classical expansion and in the full non-perturbative set-up
such a process is not included in the partition function of 
2d Yang-Mills theory. It is very important to confirm or refute
this conjecture by a further analysis of 2d Yang-Mills theory or other models.
For instance, it would be interesting to study the large $N$ behavior
of 2d Yang-Mills theory on higher genus Riemann
surfaces, where the creation of baby universes is also argued to occur \cite{Aganagic:2006je}. 

Another important problem is the more precise understanding of the
OSV conjecture \eqref{eq:OSV} in the case of 2d Yang-Mills on $T^2$
(see  \cite{Guica:2007wd}
for a review of the status of the OSV conjecture).
It is expected that the black hole partition function in this case has the form
\begin{align}
 Z_{\text{BH}}=\sum_{N_2,N_0}\Om(N,N_2,N_0)\exp\left[
-\frac{2\pi\th}{g_s}N_2-\frac{4\pi^2}{g_s}N_0\right],
\label{eq:BH-Om}
\end{align}
where $\Om(N,N_2,N_0)$ denotes the number (or index) of BPS bound states
with D4, D2, and D0 charges being $(N,N_2,N_0)$. 
We expect that $\log\Om(N,N_2,N_0)$ reproduces the entropy of black hole made of the
D-brane bound states.
However, the exact Yang-Mills partition function $Z_N$ does not
have this form \eqref{eq:BH-Om}. For instance, 
after performing the modular $S$-transformation of
Jacobi theta functions, the exact partition function for $N=2$ in \eqref{eq:Z2}
is rewritten as
\begin{equation}
\begin{aligned}
 Z_2&=\frac{\pi}{g_s}e^{-\frac{\th^2}{g_s}}\sum_{n,m\in\bbZ}(-1)^{n+m}e^{-\frac{2\pi\th}{g_s}(n+m)
-\frac{2\pi^2}{g_s}(n^2+m^2)}-\hf\rt{\frac{\pi}{g_s}}e^{-\frac{\th^2}{g_s}}\sum_{n\in\bbZ}
(-1)^n e^{-\frac{2\pi\th}{g_s}n
-\frac{\pi^2}{g_s}n^2}.
\end{aligned} 
\label{eq:Z2-S} 
\end{equation}
The factor $e^{-\th^2/g_s}$ is common for the two terms and it can be removed by the overall
normalization of the partition function. However,  
the coefficient of the two terms in \eqref{eq:Z2-S} have
different power of $g_s$ which cannot be removed by a simple 
rescaling of $Z_2$.
This is not consistent with the expansion of black hole
partition function \eqref{eq:BH-Om} 
if we assume that $\Om(N,N_2,N_0)$ is a $g_s$-independent pure number.
In \cite{Aganagic:2004js,Caporaso:2006kk,Griguolo:2006kp} it was proposed that only the first term of 
\eqref{eq:Z2-S}, or more generally $\vartheta_2(e^{\ri\th},q)^N$ term in $Z_N$
for general $N$, should be compared with the black hole partition function 
\eqref{eq:BH-Om}. However, it is not clear whether this  definition of 
$Z_{\text{BH}}$ correctly reproduces the BPS degeneracy $\Om(N,N_2,N_0)$.
It would be very interesting to clarify the precise dictionary between
the Yang-Mills partition function $Z_N$
and the black hole partition function \eqref{eq:BH-Om}.

Also, it would be interesting to study the analytic structure of $\psi^{\text{top}}(t)$
as we change the phase of $t$. In this paper we mainly considered the case $t>0$
and analyzed the Borel resummation of $\psi^{\text{top}}(t)$ assuming $t>0$.
In general, it is expected that the complex $t$-plane is divided into several sectors
and the asymptotic expansion of $\psi^{\text{top}}(t)$
takes different form in each sector. Due to the quasi-modularity of $F_g(t)$,
one can restrict $\tau=\ri t/2\pi$ to be in the fundamental region of $SL(2,\bbZ)$
on the upper-half $\tau$-plane. It would be very interesting to understand how
this fundamental region of $SL(2,\bbZ)$ is divided into sectors with different asymptotic expansion
of $\psi^{\text{top}}(t)$.

\vskip8mm
\acknowledgments
We would like to thank Marcos Mari\~{n}o,  Yasuyuki Hastuda, and Masaki Shigemori for correspondence and discussion.
A preliminary result of this work was presented by one of the authors
(KO) at the Chubu Summer School 2017 supported by the Yukawa Institute
for Theoretical Physics, and in the seminar at Rikkyo University on April 25, 2018.
This work was supported in part by JSPS KAKENHI Grant Nos. 
26400257  and 16K05316,
and JSPS Japan-Russia Research Cooperative Program.

\appendix
\section{Convention of Jacobi theta functions \label{app:notation}}

The Jacobi theta functions are defined as
\begin{align}
\varth_1(y,q)&:=
 \ri\sum_{n\in\bbZ}(-1)^n y^{n-1/2}q^{(n-1/2)^2/2},\nn\\
\varth_2(y,q)&:=
  \sum_{n\in\bbZ}y^{n-1/2}q^{(n-1/2)^2/2},\nn\\
\varth_3(y,q)&:=
  \sum_{n\in\bbZ}y^n q^{n^2/2},\nn\\
\varth_4(y,q)&:=
  \sum_{n\in\bbZ}(-1)^n y^n q^{n^2/2}.
\end{align}
We often use the abbreviated notation
\[
\varth_k(q):=\varth_k(1,q).
\]
%

\section{Proof of relations \eqref{eq:WK-cond} and \eqref{eq:tWK-cond}
         \label{app:Zfullproof}}

We first start with $\Zefull_K$ given in \eqref{eq:ZefullN-def}.
It is written as
\begin{align}
\Zefull_K
&=q^{\frac{K^3-K}{24}}\oint\frac{dx}{2\pi\ri x}
 \prod_{p>0}\left(1+xq^{\hf(p^2+Kp)}\right)^2
 \prod_{p>0}\left(1+x^{-1}q^{-\hf(p^2-Kp)}\right)^2,\nn\\
\intertext{where the product is over half-integer $p$.
By replacing $x$ by $q^{K^2/8}x$ we obtain}
\Zefull_K
&=q^{\frac{K^3-K}{24}}\oint\frac{dx}{2\pi\ri x}
 \prod_{p>0}\left(1+xq^{\hf\left(p+\frac{K}{2}\right)^2}\right)^2
 \prod_{p>0}
  \left(1+x^{-1}q^{-\hf\left(p-\frac{K}{2}\right)^2}\right)^2.\nn\\
\intertext{If we write
$p+\frac{K}{2}=r$, $p-\frac{K}{2}=\tilde{r}$ and split the second
 product into two parts,}
\Zefull_K
&=q^{\frac{K^3-K}{24}}\oint\frac{dx}{2\pi\ri x}
 \prod_{r>\frac{K}{2}}\left(1+xq^{\hf r^2}\right)^2
 \prod_{-\frac{K}{2}<\tilde{r}<\hf}
  \left(1+x^{-1}q^{-\hf \tilde{r}^2}\right)^2
 \prod_{\tilde{r}>0}\left(1+x^{-1}q^{-\hf \tilde{r}^2}\right)^2.\nn\\
\intertext{Note that $r,\,\tilde{r}$ are half-integer (integer)
when $K$ is even (odd).
Next, we rewrite the second product by the substitution $\tilde{r}=-r$}
\Zefull_K
&=q^{\frac{K^3-K}{24}}\oint\frac{dx}{2\pi\ri x}
 \prod_{r>\frac{K}{2}}\left(1+xq^{\hf r^2}\right)^2\nn\\
&\phantom{=q^{\frac{K^3-K}{24}}\oint\frac{dx}{2\pi\ri x}}
 \times\hspace{-.5em}\prod_{-\hf<r<\frac{K}{2}}
 \left[x^{-2}q^{-r^2}\left(xq^{\hf r^2}+1\right)^2\right]
 \prod_{\tilde{r}>0}\left(1+x^{-1}q^{-\hf \tilde{r}^2}\right)^2\nn\\
&=\oint\frac{dx}{2\pi\ri x^{K+\epsilon+1}}
 \prod_{r>-\hf}\left(1+xq^{\hf r^2}\right)^2
 \prod_{\tilde{r}>0}\left(1+x^{-1}q^{-\hf \tilde{r}^2}\right)^2,\nn
\end{align}
where $\epsilon=0$ ($\epsilon=1$) for even (odd) $K$.
We thus obtain
\begin{align}
\Zefull_K
&=\left\{\begin{array}{l}
\mathcal{W}_K \qquad \mbox{($K$: even)},\\[1ex]
\til{\mathcal{W}}_K \qquad \mbox{($K$: odd)}.
	 \end{array}\right.
\end{align}

We next start with $\Zofull_K$ given in \eqref{eq:ZofullN-def}.
It is written as
\begin{align}
\Zofull_K
&=q^{\frac{K^3}{24}+\frac{K}{12}}\oint\frac{dx}{2\pi\ri x^2}
 \prod_{n\ge 0}\left(1+xq^{\hf(n^2+Kn)}\right)^2
 \prod_{n\ge 1}\left(1+x^{-1}q^{-\hf(n^2-Kn)}\right)^2,\nn\\
\intertext{where the product is over integer $n$.
By replacing $x$ by $q^{K^2/8}x$ we obtain}
\Zofull_K
&=q^{\frac{K^3}{24}-\frac{K^2}{8}+\frac{K}{12}}\oint\frac{dx}{2\pi\ri x^2}
 \prod_{n\ge 0}\left(1+xq^{\hf\left(n+\frac{K}{2}\right)^2}\right)^2
 \prod_{n\ge 1}
  \left(1+x^{-1}q^{-\hf\left(n-\frac{K}{2}\right)^2}\right)^2.\nn\\
\intertext{If we write
$n+\frac{K}{2}=r$, $n-\frac{K}{2}=\tilde{r}$ and split the second
 product into two parts,}
\Zofull_K
&=q^{\frac{K^3}{24}-\frac{K^2}{8}+\frac{K}{12}}\oint\frac{dx}{2\pi\ri x^2}
 \prod_{r\ge\frac{K}{2}}\left(1+xq^{\hf r^2}\right)^2\nn\\
&\phantom{=q^{\frac{K^3}{24}-\frac{K^2}{8}+\frac{K}{12}}
          \oint\frac{dx}{2\pi\ri x^2}}
 \times\hspace{-.5em}
 \prod_{-\frac{K}{2}+1\le\tilde{r}<\hf}
  \left(1+x^{-1}q^{-\hf \tilde{r}^2}\right)^2
 \prod_{\tilde{r}>0}\left(1+x^{-1}q^{-\hf \tilde{r}^2}\right)^2.\nn\\
\intertext{Note that $r,\,\tilde{r}$ are integer (half-integer)
when $K$ is even (odd).
Next, we rewrite the second product by the substitution $\tilde{r}=-r$}
\Zofull_K
&=q^{\frac{K^3}{24}-\frac{K^2}{8}+\frac{K}{12}}\oint\frac{dx}{2\pi\ri x^2}
 \prod_{r\ge\frac{K}{2}}\left(1+xq^{\hf r^2}\right)^2\nn\\
&\phantom{=q^{\frac{K^3}{24}-\frac{K^2}{8}+\frac{K}{12}}
          \oint\frac{dx}{2\pi\ri x^2}}
 \times\hspace{-.5em}\prod_{-\hf<r\le\frac{K}{2}-1}
 \left[x^{-2}q^{-r^2}\left(xq^{\hf r^2}+1\right)^2\right]
 \prod_{\tilde{r}>0}\left(1+x^{-1}q^{-\hf \tilde{r}^2}\right)^2\nn\\
&=\oint\frac{dx}{2\pi\ri x^{K+\epsilon+1}}
 \prod_{r>-\hf}\left(1+xq^{\hf r^2}\right)^2
 \prod_{\tilde{r}>0}\left(1+x^{-1}q^{-\hf \tilde{r}^2}\right)^2,\nn
\end{align}
where $\epsilon=1$ ($\epsilon=0$) for even (odd) $K$.
We thus obtain
\begin{align}
\Zofull_K
&=\left\{\begin{array}{l}
\til{\mathcal{W}}_K \qquad \mbox{($K$: even)},\\[1ex]
\mathcal{W}_K \qquad \mbox{($K$: odd)}.
	 \end{array}\right.
\end{align}
%


\end{document}